\documentclass[aps,prd,twocolumn,notitlepage,preprintnumbers, showpacs, nofootinbib, superscriptaddress]{revtex4-1}
\usepackage[utf8]{inputenc}
\usepackage{array}
\usepackage{graphicx}
\usepackage{braket}
\usepackage{amsmath,amssymb}
\usepackage{xcolor}
\usepackage{ulem}
\usepackage{etoolbox}
\usepackage{pifont}

\usepackage[hypertexnames=false]{hyperref}
\hypersetup{
    colorlinks=true,       % false: boxed links; true: colored links
    linkcolor=blue,          % color of internal links
    citecolor=blue,        % color of links to bibliography
    filecolor=blue,      % color of file links
    urlcolor=blue           % color of external links
}

\newcommand{\tkeucl}[1]{\ifstrequal{#1}{}{\hat{R}_{E}}{\hat{R}_{E,#1}}}
\newcommand{\tkmink}[1]{\ifstrequal{#1}{}{\hat{R}_{M}}{\hat{R}_{M,#1}}}
\newcommand{\teucl}[1]{\ifstrequal{#1}{}{\hat{T}_{E}}{\hat{T}_{E,#1}}}
\newcommand{\tmink}[1]{\ifstrequal{#1}{}{\hat{T}_{M}}{\hat{T}_{M,#1}}}

\newcommand{\HKbar}[0]{\overline{\text{HK}}}
\newcommand{\wloop}[0]{W}
\DeclareMathOperator{\sign}{sign}
\DeclareMathOperator{\Tr}{Tr}

\renewcommand{\Re}[0]{\text{Re}}
\renewcommand{\Im}[0]{\text{Im}}
\newcommand{\cmark}{\ding{51}}
\newcommand{\xmark}{\ding{55}}

\newcommand{\mbraket}[3]{\left< #1 \vphantom{#2#3} \right|
 #2 \left| #3 \vphantom{#1#2} \right>} % for Dirac matrix elements

\begin{document}

\title{Real-time lattice gauge theory actions: \texorpdfstring{\\}{} unitarity, convergence, and path integral contour deformations}

\author{Gurtej Kanwar}
\affiliation{Center for Theoretical Physics, Massachusetts Institute of Technology, Cambridge, MA 02139, USA}
\affiliation{The NSF AI Institute for Artificial Intelligence and Fundamental Interactions}
\author{Michael L. Wagman}
\affiliation{Fermi National Accelerator Laboratory, Batavia, IL 60510, USA}
\date{\today}

\preprint{FERMILAB-PUB-21-071-T}
\preprint{MIT-CTP/5291}

\begin{abstract}
The Wilson action for Euclidean lattice gauge theory defines a positive-definite transfer matrix that corresponds to a unitary lattice gauge theory time-evolution operator if analytically continued to real time.
Hoshina, Fujii, and Kikukawa (HFK) recently pointed out that applying the Wilson action discretization to continuum real-time gauge theory does not lead to this, or any other, unitary theory and proposed an alternate real-time lattice gauge theory action that does result in a unitary real-time transfer matrix.
The character expansion defining the HFK action is divergent, and in this work we apply a path integral contour deformation to obtain a convergent representation for $U(1)$ HFK path integrals suitable for numerical Monte Carlo calculations.
We also introduce a class of real-time lattice gauge theory actions based on analytic continuation of the Euclidean heat-kernel action.
Similar divergent sums are involved in defining these actions, but for one action in this class this divergence takes a particularly simple form, allowing construction of a path integral contour deformation that provides absolutely convergent representations for $U(1)$ and $SU(N)$ real-time lattice gauge theory path integrals.
We perform proof-of-principle Monte Carlo calculations of real-time $U(1)$ and $SU(3)$ lattice gauge theory and verify that exact results for unitary time evolution of static quark-antiquark pairs in $(1+1)$D are reproduced.
\end{abstract}

\maketitle

\section{Introduction}\label{sec:intro}

Lattice quantum field theory methods have long been used to calculate equilibrium properties of strongly coupled quantum systems. Discretization of quantum field theories on a spacetime lattice provides a UV regulator 
that allows expectation values of observables to be computed after renormalization and extrapolation to the continuum limit. Lattice gauge theory (LGT) in particular allows observables to be computed in strongly coupled gauge theories, for example including observables in quantum chromodynamics (QCD) that are relevant for understanding the dynamics of the strong force in the Standard Model. LGT calculations typically rely on Monte Carlo methods to evaluate Euclidean path integrals by stochastically sampling gauge field configurations from a probability distribution proportional to $e^{-S_E}$, where the Euclidean action $S_E$ is assumed to be real. This approach is suitable for determinations of many equilibrium gauge theory properties, which can be related to expectation values in Euclidean spacetime.

Non-equilibrium properties of gauge theories are more difficult to obtain from Euclidean correlation functions; for example, transport coefficients of the quark-gluon plasma and neutron stars, properties of early universe phase transitions, and inclusive hadron scattering cross-sections involving resonance production have been inaccessible to ab initio Euclidean LGT approaches despite significant theoretical and phenomenological interest.
In principle, out-of-equilibrium gauge theory properties defined by correlation functions of timelike separated operators can instead be accessed using the Schwinger-Keldysh formalism~\cite{Schwinger:1960qe,Keldysh:1964ud}, which represents out-of-equilibrium observables with path integrals involving regions of both Minkowski and Euclidean signature.
In practice, Monte Carlo methods cannot directly be used for efficient nonperturbative calculations of these lattice-regularized real-time or Schwinger-Keldysh LGT path integrals, since the path integral weights involve a pure phase factor $e^{iS_M}$, where $S_M$ is the action for the Minkowski region, and the path integral weights thus cannot be interpreted as a probability distribution to be used for importance sampling.

In some cases, path integral contour deformations have been used to tame this sign problem associated with real-time path integrals; for example, real-time calculations have been performed using path integral deformations in $(0+1)$D quantum-mechanical theories~\cite{Tanizaki:2014xba,Alexandru:2016gsd,Alexandru:2017lqr,Mou:2019tck,Lawrence:2021izu}.
Suitably chosen path integral contour deformations in complexified field space can exactly preserve the total path integral while reducing phase fluctuations of the integrand and introducing fluctuations in the magnitude of $e^{iS_M}$ that are amenable to importance sampling.
Path integral contour deformations have also been applied to sign problems for finite-density systems and other observables in a variety of quantum field theories; see Ref.~\cite{Alexandru:2020wrj} for a recent review. This promising approach to mitigating sign problems has recently been extended to generic observables with signal-to-noise problems in $SU(N)$ gauge theories~\cite{Detmold:2021ulb}, but the construction of contour deformations for real-time LGT path integrals has not yet been addressed.

It was pointed out by Hoshina, Fujii, and Kikukawa (HFK) in Ref.~\cite{Hoshina:2020gdy} that another obstacle facing calculations of real-time LGT path integrals is the determination of a suitable discretized action $S_M$.
In particular, Ref.~\cite{Hoshina:2020gdy} argues that applying the same discretization as the Wilson gauge action~\cite{Wilson:1974sk} to the real-time continuum theory results in a real-time LGT that is non-unitary.
Although it is formally possible to construct a unitary time evolution operator for LGT through analytic continuation of the eigenvalues of the imaginary-time transfer matrix~\cite{Luscher:1976ms}, this formal construction cannot be practically implemented without first solving for the LGT spectrum.
Instead, HFK propose an alternative real-time LGT transfer matrix obtained by analytically continuing the character expansion of the kinetic term in the Wilson action (i.e.~all terms involving timelike plaquettes).
The resulting real-time transfer matrix is unitary and formally allows the definition of a discretized real-time action, called the HFK action below, and of associated real-time LGT observables.

The HFK action, however, is defined by an infinite series and we demonstrate below that it does not converge for some or all values of the gauge field in $U(1)$ and $SU(N)$ gauge theory, making it impossible to calculate the weights necessary for an importance sampling approach.
This divergent representation is a generic feature of real-time LGT actions that are defined by local kinetic and potential energy terms and that give rise to unitary transfer matrices, as discussed below and detailed in Appendix~\ref{sec:divergent-actions}.
In order to remedy this situation, this work introduces path integral contour deformations that explicitly depend on summation indices appearing in the definition of the action.
These path integral contour deformations can change the convergence properties of the action for fixed gauge field values, though singularities still arise for gauge field configurations corresponding to endpoints of the path integral contour that must be kept fixed during deformation.
Rotating the prefactor of the kinetic term in the action starting from $-1$ and approaching $i$ (analogous to a Wick rotation of the continuum theory) regularizes these singularities and makes the path integral absolutely convergent everywhere outside of the limit. By deforming the path integral contour as a function of the prefactor and exactly cancelling contour segments that are related by periodicity, the remaining path integral is made absolutely convergent even after analytically taking the limit to $i$.
A simple contour deformation that renders the $U(1)$ HFK path integral convergent in this sense is introduced below.
Unfortunately, it is challenging to construct an analogous contour deformation that would lead to an absolutely convergent representation of the $SU(N)$ HFK path integral.

\begin{table}[]
    \centering
    \begin{ruledtabular}
    \begin{tabular}{c c c c c}
         Action & Unitary & Convergent & \multicolumn{2}{c}{Convergent deformation} \\
         & & & \hspace{.35cm}$U(1)$\hspace{.35cm} & \hspace{.25cm}$SU(N)$\hspace{.25cm} \\
         \hline
         Wilson & \xmark & \cmark & & \\
         HFK & \cmark & \xmark & \cmark & \xmark \\
         HK & \cmark & \xmark & \xmark & \xmark \\
         $\HKbar$ & \cmark & \xmark & \cmark & \cmark
    \end{tabular}
    \end{ruledtabular}
    \caption{Unitarity in the continuum limit, convergence of the definition of the action, and existence of convergent path integral representations using contour deformations for the four actions considered in this work --- the Wick-rotated real-time Wilson action, the Hoshina-Fujii-Kikukawa (HFK) real-time action~\cite{Hoshina:2020gdy}, the real-time heat-kernel action (HK) derived from the Euclidean action given in Ref.~\cite{Menotti:1981ry}, and the real-time modified heat-kernel ($\HKbar$) action introduced in the main text. A truncated heat-kernel action which is convergent and non-unitary is also studied in Appendix~\ref{sec:truncated}, including details on the continuum limit of this action. Based on the contour deformations introduced in this work, the $\HKbar$ action is the only unitary choice for which a convergent Monte Carlo calculation can be performed in real-time $SU(N)$ LGT, but it is possible that more sophisticated contour deformations could be discovered to provide convergent deformations for the cases marked with \xmark.
    }
    \label{tab:action-comparison}
\end{table}

This challenge motivates the introduction of another class of actions based on analytic continuation of the heat-kernel action of Menotti and Onofri~\cite{Menotti:1981ry}. The key feature of these actions is a kinetic term that is exactly unitary at all values of the lattice spacing, much like the HFK action; in comparison to the HFK action, however, a simpler summation is involved in their definition. The heat-kernel kinetic term can be combined with any real potential without violating unitarity. Possible choices of the potential include either a term with the same heat kernel form applied to the spacelike plaquettes or the Wilson potential term (i.e.~the terms in the Wilson action involving spacelike plaquettes). In the following, these two options are respectively termed the real-time heat-kernel (HK) action and the modified real-time heat-kernel ($\HKbar$) action. The summations defining the HK and $\HKbar$ actions are also divergent, but a simple path integral contour is shown to provide an absolutely convergent representation of the $\HKbar$ action for $U(1)$ and $SU(N)$ gauge theory in Minkowski spacetime with any dimension. This is achieved using a prescription similar to the one applied for the HFK action: parameterizing a ``Wick rotation'' of the kinetic term prefactor, exactly cancelling pieces of the deformed contour related by symmetry, and then analytically taking the limit to prefactor $i$. Thus real-time path integrals using the $\HKbar$ action can be evaluated after contour deformation using Monte Carlo techniques, and real-time LGT observables can be numerically computed. The comparison of the HK and $\HKbar$ actions to the HFK action and the real-time Wilson action is summarized in Table~\ref{tab:action-comparison} and is analyzed in detail below.

Real-time LGT has also been investigated in semi-classical approximations by several previous works.
Real-time evolution of classical lattice gauge fields has been studied in order to gain insight into electroweak sphaleron transitions in the early universe~\cite{Ambjorn:1987qu,Grigoriev:1988bd,Ambjorn:1988gf,Grigoriev:1989ub,Grigoriev:1989je,Ambjorn:1990wn,Ambjorn:1990pu,Ambjorn:1992np,Ambjorn:1995xm,Bodeker:1995pp,Arnold:1996dy,Moore:1997cr} and the evolution of quantum fermions in these classical background gauge fields has been investigated~\cite{Aarts:1998td,Borsanyi:2008eu,Saffin:2011kc,Saffin:2011kn,Mou:2013kca}. 
Real-time LGT calculations of fermion-antifermion pair production in quantum electrodynamics with the $U(1)$ gauge field treated classically have also been performed~\cite{Hebenstreit:2013baa,Hebenstreit:2013qxa,Kasper:2014uaa,Gelis:2015kya,Mueller:2016aao,Shi_2018}.
Semi-classical real-time LGT has been studied in the context of heavy ion collisions~\cite{Krasnitz:1998ns,Gelis:2004jp,Gelis:2005pb,Berges:2007re,Berges:2008mr,Berges:2008zt,Fukushima:2011nq,Berges:2012ev,Schlichting:2012es,Kurkela:2012hp,Berges:2013eia,Berges:2013fga,Mueller:2016ven,Gelfand:2016prm,Tanji:2016dka,Mace:2016shq,Tanji:2017xiw,Boguslavski:2018beu,Boguslavski:2021baf}.
In semi-classical calculations, discretized gauge field equations of motion are derived from the real-time Wilson action and are solved.
Although the real-time Wilson action is not suitable for quantum LGT, the derived equations of motion do give well-defined deterministic evolution of the fields involved and result in a well-formed classical theory. In these approaches, the lattice spacing is a free parameter that is independent of the gauge field coupling, rather than a dynamical quantity whose value in physical units must be determined by tuning the gauge coupling, as described for example in Ref.~\cite{Krasnitz:1998ns}.
The non-existence of a unitary continuum limit for the real-time Wilson action in quantum LGT is thus irrelevant for calculations of solutions to the classical equations of motion associated with the real-time Wilson action.

In the quantum setting, the (non-unitary) real-time Wilson actions for $U(1)$ and $SU(2)$ LGT have previously been used in complex Langevin calculations~\cite{Berges:2006xc,Berges:2007nr}.
Complex Langevin methods are not guaranteed to reproduce exact results for real-time LGT, and Ref.~\cite{Berges:2006xc} finds that complex Langevin results only reproduce analytically calculable results for simple real-time LGT observables for values of the Langevin evolution time that are not too large.
Methods based on reweighting and gauge fixing are found to increase the region in which complex Langevin reproduces exact results in Ref.~\cite{Berges:2007nr}.
The exact results for the one-plaquette model calculated using the real-time Wilson LGT action in this reference agree with the exact results for unit area Wilson loops in $(1+1)$D Minkowski spacetime presented in Sec.~\ref{sec:exact} below.
The non-unitarity of time evolution in the one-plaquette model with the real-time Wilson action implied by these results
is not mentioned in Ref.~\cite{Berges:2007nr}.
The lack of a well-defined continuum limit for the $SU(2)$ one-plaquette model with the real-time Wilson action is discussed in the reference; however, Ref.~\cite{Berges:2007nr} assumes that the corresponding continuum limit of $(3+1)$D LGT with the real-time Wilson action exists and can be used to calculate physical observables in real-time gauge theory.
It is demonstrated below that the real-time Wilson action is not unitary in arbitrary spacetime dimensions, even in the small gauge coupling limit, and that the continuum limit of real-time LGT with the Wilson action in $(3+1)$D either does not exist or is non-unitary, depending on the choice of gauge group $SU(N)$.
In either case, the real-time Wilson action does not provide a suitable starting point for calculating physical observables in real-time gauge theory.

Finally, it is possible to simulate real-time gauge theory dynamics using the Hamiltonian formalism rather than the path integral, as is often considered in the context of tensor network and quantum computing approaches; see Ref.~\cite{Banuls:2019bmf} for a recent review.
Though unitarity of time-evolution in the Hamiltonian formalism is also a key condition, the focus of the present work is on the construction of actions suitable for classical simulation of discrete real-time path integrals, and the discussed actions and contour deformations are not immediately relevant to Hamiltonian quantum simulation. However, the analysis of unitarity explored here may have relevance for similar analyses for quantum computing approaches.
For example, Ref.~\cite{Farrelly:2020ckc} shows how real-time path integrals in lattice field theory can be described using quantum circuits with Trotterization errors described as discretization effects, but asserts that a time-evolution operator for quantum simulation of $SU(N)$ gauge theory obtained using the real-time Wilson action is unitary.\footnote{The real-time transfer matrix elements computed in Ref.~\cite{Farrelly:2020ckc} can be represented as integrals with pure-phase integrands, as shown in Eq.~(H10) of that work. The real-time transfer matrix can also be written as an integral over unitary operators, as given in their Eq.~(18). However, these features do not imply that the eigenvalues of the real-time transfer matrix are pure phases and therefore do not establish unitarity.}
It would be interesting to consider analogous quantum circuit descriptions of real-time transfer matrices for the unitary real-time LGT actions considered in this work.

The remainder of this work is structured as follows. The (non)-unitarity of real-time transfer matrices defining discretized path integrals of compact and non-compact variables, including the non-unitarity of the real-time Wilson action, is given in Sec.~\ref{sec:theory}.
The unitary real-time LGT actions shown in Table~\ref{tab:action-comparison} are introduced in detail and discussed in Sec.~\ref{sec:actions}.
In Sec.~\ref{sec:contour}, path integral contour deformation techniques are used to construct absolutely convergent representations of path integrals involving the HFK action for $U(1)$ LGT, the $\HKbar$ action for $U(1)$ and $SU(N)$ LGT, and the associated Schwinger-Keldysh action; proof-of-principle Monte Carlo calculations in $(1+1)$D are also discussed.
Our conclusions and outlook are summarized in Sec.~\ref{sec:concl}.

\section{Real-time evolution of compact and non-compact variables}\label{sec:theory}

In a quantum field theory (QFT) defined on a Minkowski spacetime background, the continuum action $S_M$ provides a useful starting point for understanding and perturbatively calculating expectation values of quantum operators and corresponding physical observables. The continuum action is manifestly Lorentz and translation invariant, ensuring that the physics encoded in the action satisfies the Poincar{\'e} symmetry observed in nature. 
A Hamiltonian and related quantum operators can be defined using the construction of a Hilbert space on a co-dimension-one submanifold of spacetime, which superficially breaks Poincar{\'e} invariance~\cite{Dirac:1949cp}. The path integral formalism~\cite{Feynman:1948ur}, however, provides a way to relate expectation values of quantum operators to integrals over fluctuations of classical fields weighted by the manifestly Poincar{\'e}-invariant function $e^{iS_M}$.
Path integrals in imaginary time can be used to calculate expectation values for systems in thermal equilibrium when suitable temporal boundary conditions are used for bosons and fermions
and the inverse temperature $\beta$ is set by the length of the imaginary-time direction.

Formally, the imaginary-time path integral can be related to a real-time path integral by Wick rotation $x^0 \leftrightarrow i x^0$ of the time coordinate~\cite{Wick:1954eu}. The action $S_M$ to be used in real (Minkowski) time consists of the kinetic minus potential energy, whereas the action $S_E$ in imaginary (Euclidean) time consists of the kinetic plus potential energy, with the corresponding path integral weights related by Wick rotation as $e^{i S_M} \leftrightarrow e^{-S_E}$. Although these path integral relations are straightforward in continuum QFT, subtleties arise in the connections between real- and imaginary-time path integrals with lattice-regularized actions involving compact variables as discussed below.

\subsection{The SHO and quantum rotator}\label{sec:trotter}
The continuum path integral is ill-defined without a prescription for regularizing UV divergences and extrapolating to remove the regulator. Discretizing spacetime provides such a regulator with the desirable properties of being non-perturbative, gauge-invariant, and amenable to numerical simulation. This discretization is well-understood for gauge fields in imaginary-time lattice field theory, but as discussed in Ref.~\cite{Hoshina:2020gdy} and below subtleties arise when considering real-time path integrals or path integrals involving real-time components, such as the path integrals required in the Schwinger-Keldysh approach to computing out-of-equilibrium observables~\cite{Schwinger:1960qe,Keldysh:1964ud}. These difficulties can be simply demonstrated even in $(0+1)$D quantum mechanical systems. Though these systems do not possess a Lorentz symmetry, the challenges in these $(0+1)$D path integrals are investigated as simple analogues to the challenges that arise in $(3+1)$D LGT.

We first consider the simple harmonic oscillator (SHO), the $(0+1)$D theory of a single noncompact variable $x \in \mathbb{R}$ constrained by a quadratic potential. The continuum actions for the SHO in real and imaginary time can be written respectively as
\begin{equation}
\begin{aligned} \label{eq:sho-ctm-action}
    S_M[x(t)] &= \int dt \, \frac{1}{2} (\partial_t x(t))^2 - \frac{\omega^2}{2} x(t)^2, \\
    S_E[x(\tau)] &= \int d\tau \, \frac{1}{2} (\partial_\tau x(\tau))^2 + \frac{\omega^2}{2} x(\tau)^2,
\end{aligned}
\end{equation}
where $x(t)$ and $x(\tau)$ are the position histories in real and imaginary time respectively, and we work in units for which the mass is set to $1$. Transition amplitudes between states $\ket{x}$ and $\ket{x'}$ after time evolution by $L_T$ are given in real time by $\braket{x' | e^{-i \hat{H} L_T} | x}$ and in imaginary time by $\braket{x' | e^{-\hat{H} L_T} | x}$. These transition amplitudes can be used to extract the energy spectrum and other physical properties. A discretized path integral can be used to write the transition amplitudes using time steps of size $a$ as
\begin{equation}
\begin{aligned}
    &\braket{x' | e^{s \hat{H} L_T} | x}
    = \int dx_a \dots dx_{L_T-a} \Big[ \\
    &\;\; \prod_{n=0}^{N_T-1} e^{s \frac{a \, \omega^2}{4} x_{na+a}^2 + \frac{(x_{na+a}-x_{na})^2}{2 s\, a} + s \frac{a\, \omega^2}{4} x_{na}^2} \Big] + O(a^2),
\end{aligned}
\end{equation}
where $x_0 \equiv x$, $x_{L_T} \equiv x'$, $N_T = L_T/a$, and $s \in \{-i, -1\}$ gives the prefactor associated with time evolution on the real- and imaginary-time contours, respectively. Factoring out $1/s$ in the exponent allows one to collect the exponentials into a temporally-discretized version of the continuum action in Eq.~\eqref{eq:sho-ctm-action},
\begin{equation} \label{eq:sho-path-integral}
    \braket{x' | e^{s \hat{H} L_T} | x} = \int_{x_0 = x}^{x_{L_T} = x'} \mathcal{D}x\,
    e^{\frac{1}{s}S(x;s)} + O(a^2),
\end{equation}
where $\int \mathcal{D}x \equiv \int dx_a \dots dx_{L_T-a}$ and the discretized action is given by
\begin{equation}
    S(x;s) = a \sum_{n=0}^{N_T-1} \left[ K(x_{na+a},x_{na}) + \frac{1}{s^2}V(x_{na}) \right],
    \label{eq:SHOS}
\end{equation}
with the kinetic and potential energy terms $K$ and $V$ given by
\begin{equation}
    \begin{split}
        K(x_{na+a},x_{na}) &= \frac{(x_{na+a}-x_{na})^2}{2 a^2}, \\
        V(x_{na}) &= \frac{\omega^2}{2}x_{na}^2.
    \end{split}
\end{equation}
Noting that $s^2 = -1$ for real time and $s^2 = +1$ for imaginary time, we arrive at the usual conclusion that the discretized action for real or imaginary time can be related by simply replacing the sign in front of the potential term and using the appropriate prefactor of $1/s$ in the path integral weights in Eq.~\eqref{eq:sho-path-integral}.

In either real or imaginary time, the discretized path integral $\int_{x_0 = x}^{x_{L_T} = x'} \mathcal{D}x \, e^{\frac{1}{s} S(x;s)}$ gives an approximation to $\braket{x' | e^{s\hat{H}L_T} | x}$ which is accurate to $O(a^2)$ under the assumption that the integral exists as $a \rightarrow 0$ and can be expanded in powers of $a$. The existence of such a limit is critical to extrapolating physical quantities of interest to the continuum. 
For the SHO, the $a \rightarrow 0$ limit is well-defined and can be studied using the transfer matrix.
The matrix elements of the real- and imaginary-time transfer matrices respectively are given by one factor of the path integrand,
\begin{equation}
\begin{aligned}
    \tmink{}(x_{t+a}, x_{t}) &= e^{-i \frac{a \omega^2}{4} x_{t+a}^2 + i \frac{(x_{t+a} - x_t)^2}{2 a} -i \frac{a \omega^2}{4}x_t^2}, \\
    \teucl{}(x_{\tau+a}, x_{\tau}) &= e^{- \frac{a \omega^2}{4} x_{\tau+a}^2 - \frac{(x_{\tau+a} - x_\tau)^2}{2 a} - \frac{a \omega^2}{4}x_\tau^2},
\end{aligned}
\label{eq:SHOTM1}
\end{equation}
where $t = na$ and $\tau = na$ in real and imaginary time.
The corresponding SHO Hamiltonian is given by $\hat{H} = \hat{K} + \hat{V}$, where $\braket{x'|\hat{V}|x} = \frac{\omega^2 x^2}{2} \delta(x'-x)$ and $\hat{K} = \hat{p}^2/2$ is defined in term of the momentum operator $\hat{p}$ satisfying $[\hat{x},\hat{p}]= i$.
Noting that $\braket{x'|e^{s a \hat{p}^2 /2}|x} \propto e^{(x'-x)^2/(2a s)}$, the real- and imaginary-time transfer matrices can be expressed as
\begin{equation}
\begin{aligned}
    \tmink{} &\propto e^{-i a \hat{V}/2} e^{-i a \hat{K}} e^{-ia \hat{V}/2}, \\
    \teucl{} &\propto e^{-a \hat{V}/2} e^{-a \hat{K}} e^{-a \hat{V}/2}.
\end{aligned}
\label{eq:SHOTM2}
\end{equation}
By the Lie-Trotter product formula~\cite{Trotter:1959}, products of the real- and imaginary-time transfer matrices approximate products of the corresponding time evolution operators in real and imaginary time, $e^{s \hat{H} a}$, with errors that vanish as $a\rightarrow 0$.

The continuum real- and imaginary-time transfer matrices $e^{-ia \hat{H}}$  and $e^{- a \hat{H}}$ are unitary and positive definite respectively for Hermitian $\hat{H}$.
These properties allow a spectral decomposition in either case and are desirable to maintain in the discretized theory.
Unitarity of the SHO real-time transfer matrix (up to a constant normalization factor suppressed in Eq.~\eqref{eq:SHOTM2}) can be demonstrated by direct computation,
\begin{equation} \label{eq:sho-unitary-real-transfer}
\begin{aligned}
&\int dy\, \tmink{}(x,y) \tmink{}^{\dagger}(y,x') \\
&\;\;= \int dy e^{-i a\,\omega^2 x^2 / 4} e^{i \frac{(x-y)^2}{2\,a} - i \frac{(y-x')^2}{2\,a}} e^{i a\,\omega^2 (x')^2 / 4} \\
    &\;\;= \frac{e^{i a\,\omega^2 (x')^2 / 4}}{e^{i a\,\omega^2 x^2 / 4}} \frac{e^{i \frac{x^2}{2a}}}{e^{i \frac{(x')^2}{2a}}} \int dy e^{- i y (x - x')/a} \\
    &\;\;\propto \delta(x-x'),
\end{aligned}
\end{equation}
with an analogous calculation for $\hat{T}_M^\dagger \hat{T}_M$ giving $\hat{T}_M^\dagger \hat{T}_M = \hat{T}_M \hat{T}_M^\dagger$.
The imaginary-time transfer matrix is related to the Gaussian integral kernel $e^{-\frac{(x-y)^2}{2\,a}}$ and can be shown to be positive definite by the fact that the Gaussian integral kernel itself is positive definite~\cite{fasshauer2011positive},
\begin{equation} \label{eq:sho-pos-imag-transfer}
\begin{aligned}
    &\int dx\,dy\, f(x) \teucl{}(x,y) f(y) \\
    &\;\;= \int dx\,dy\, [f(x) e^{-a\, \omega^2 x^2 /4}] e^{-\frac{(x-y)^2}{2\,a}} [f(y) e^{-a\, \omega^2 y^2/4}] \\
    &\;\;\geq 0, \quad \forall f(x) \in L^2(\mathbb{R}).
\end{aligned}
\end{equation}
Fundamentally, these desired properties of the transfer matrix emerge for the SHO because the kinetic portion of the discretized action is respectively a unitary integral kernel or a positive-definite integral kernel when the prefactor is $i$ or $-1$. Specifically, in Eq.~\eqref{eq:sho-unitary-real-transfer}, it is exactly unitarity of the kinetic integral kernel that produces the term $\delta(x-x')$. The potential factors multiplied by this delta function are inverses which cancel, regardless of the specifics of the potential. Similarly, in Eq.~\eqref{eq:sho-pos-imag-transfer} only the positive-definiteness of the Gaussian kinetic term is required to show overall positive definiteness, given that the potential appeared in the same way on each side of the transfer matrix (as long as the potential is real and bounded from below so that $f(x) e^{-aV(x)} \in L^2(\mathbb{R})$). For other theories of noncompact variables, the kinetic term in the action can also generically be discretized as a Gaussian integral kernel that satisfies these properties, and these properties therefore extend to lattice field theories of noncompact variables.

We next consider the planar quantum rotator in $(0+1)$D in both real and imaginary time. This system can be physically interpreted as the SHO on a compact domain, which can be chosen for example to be the circular domain $x \in [0,2\pi]$, with $x=0$ identified with $x=2\pi$. This choice normalizes the length of the compact domain such that $x$ can be considered as the angular variable of the quantum rotator. Since the microscopic description of the theory is identical to the SHO, the continuum action given for the SHO in Eq.~\eqref{eq:sho-ctm-action} also describes the physics of the quantum rotator. We are also free to set $\omega = 0$ in the potential because the compact domain ensures convergence of path integrals in the free theory, which simplifies the analytical manipulations below.

Although the continuum actions for the $\omega=0$ SHO and the quantum rotator are the same, the definitions of the discretized action and path integral for the quantum rotator require care. In particular, the derivatives used in the kinetic operator should be compatible with the identification of $x=0$ and $x=2\pi$.
A typical choice is to write the path integral using a cosine for the discrete kinetic term,
\begin{equation} \label{eq:qrot-path-integral}
\begin{aligned}
   \braket{x' | e^{s \hat{H} t} | x}
    &= \int_0^{2\pi} dx_a \dots dx_{L_T-a} \Big[ \\
    &\quad \times \prod_{n=0}^{N_T-1} e^{ \frac{1}{sa}[1 - \cos(x_{na+a}-x_{na})]} \Big] + O(a^2),
\end{aligned}
\end{equation}
where as above $s \in \{-i,-1\}$ gives the appropriate prefactor for real or imaginary time, respectively. The corresponding discretized action for the quantum rotator is
\begin{equation}
\begin{aligned}
    S(x;s) &= a \sum_{n=0}^{N_T-1} \left[ \frac{1}{a^2} [ 1 -\cos(x_{na+a} - x_{na} ) ] \right] \\
    &\equiv a \sum_{n=0}^{N_T-1} K(x_{na+a},x_{na}).
    \label{eq:rotorS}
\end{aligned}
\end{equation}
The Taylor expansion of $1 - \cos(x_{na+a} - x_{na}) = \frac{1}{2}(x_{na+a} - x_{na})^2 + O((x_{na+a} - x_{na})^4)$ demonstrates that Eq.~\eqref{eq:rotorS} is equivalent to the free theory SHO action ($\omega = 0$) for small fluctuations of the position in lattice units.
The two actions should therefore be perturbatively equivalent in the continuum limit.

The differences in behavior at non-zero lattice spacing become apparent when considering the transfer matrix description. The transfer matrices in real and imaginary time associated with the path integral in Eq.~\eqref{eq:qrot-path-integral} are defined respectively by
\begin{equation}
\begin{aligned}
    \tmink{}(x_{t+a},x_t) &= e^{\frac{i}{a} - \frac{i}{a} \cos(x_{t+a} - x_t)}, \\
    \teucl{}(x_{\tau+a},x_\tau) &= e^{\frac{-1}{a} + \frac{1}{a} \cos(x_{\tau+a} - x_\tau)}.
\end{aligned}
\end{equation}
The lack of any potential terms reflects our choice of working with the free theory. As argued above, the unitarity and positive definiteness of the transfer matrix in real and imaginary time depends only on the behavior of this kinetic integral kernel, as any (real) potential terms will cancel from the relations in Eq.~\eqref{eq:sho-unitary-real-transfer} and \eqref{eq:sho-pos-imag-transfer} if the kinetic term satisfies the desired properties. For this choice of discretization, the imaginary-time transfer matrix does satisfy positivity, and can be expanded in terms of Fourier eigenfunctions using the Jacobi-Anger expansion~\cite{olver2010nist1035}
\begin{equation}
    \teucl{}(x,y) = \sum_k e^{-1/a} I_k(1/a) e^{i k (x-y)},
\end{equation}
where $I_k$ is the modified Bessel function of the first kind with rank $k$. The eigenvalues $e^{-1/a} I_k(1/a)$ all vanish as $a \rightarrow 0$, but physical observables are determined by ratios $I_k(1/a) / I_0(1/a)$ which converge to $1$ in the continuum limit, allowing renormalization and extraction of quantities of interest in the continuum. On the other hand, the non-Gaussian nature of the kinetic integral kernel results in non-unitarity of the real-time transfer matrix, which can be seen by direct calculation,
\begin{equation}
\begin{aligned}
    &\tmink{}(x,y) \tmink{}^\dagger(y,x') \\
    &\;\; = \int_0^{2\pi} \frac{dy}{2\pi} e^{-\frac{i}{a} \cos(x-y)}e^{\frac{i}{a}\cos(y-x')} \\
    &\;\; = I_0\left(\frac{i}{a}\left|1 - e^{i(x-x')}\right|\right) \not\propto \delta(x - x').
\end{aligned}
\end{equation}
The breakdown of unitarity at non-zero $a$ is an undesirable feature, but could be considered acceptable if the ratios of transfer matrix eigenvalues converged to unit-norm values in the continuum limit. Instead, many ratios of eigenvalues simply do not have a continuum limit, as can be seen by performing a similar Jacobi-Anger expansion in real time, 
\begin{equation}
    \tmink{}(x,y) = \sum_k e^{i/a} I_k(-i/a) e^{i k (x-y)}.
\end{equation}
As $a\rightarrow0$, the ratio $I_k(-i/a) / I_l(-i/a) \rightarrow 1$ for $k \equiv l \mod 2$ but the limit does not exist for $k \equiv l+1 \mod 2$, demonstrating that observables that depend on these ratios of eigenvalues do not have either a well-defined continuum limit or a spectral representation consistent with unitary time evolution. If a potential is included, these free theory eigenfunctions can generically be expected to mix, and the non-existence of a unitary continuum limit for certain eigenstates of $e^{-ia\hat{K}}$ can be expected to spoil the existence of continuum limits for generic observables.

This simple exploration of the SHO and quantum rotator highlights a concern that must be addressed if attempting to work with discretized real-time path integrals in a position-space representation. For compact variables, a position-space kinetic term in the action that satisfies periodicity may not be compatible with the replacement $K + V \rightarrow K - V$ in moving from the imaginary-time action to the real-time action, as this simple replacement may result in non-unitarity of the transfer matrix and prevent extrapolating to continuum physics.

\begin{figure}[t]
    \centering
    \includegraphics[width=0.8\linewidth]{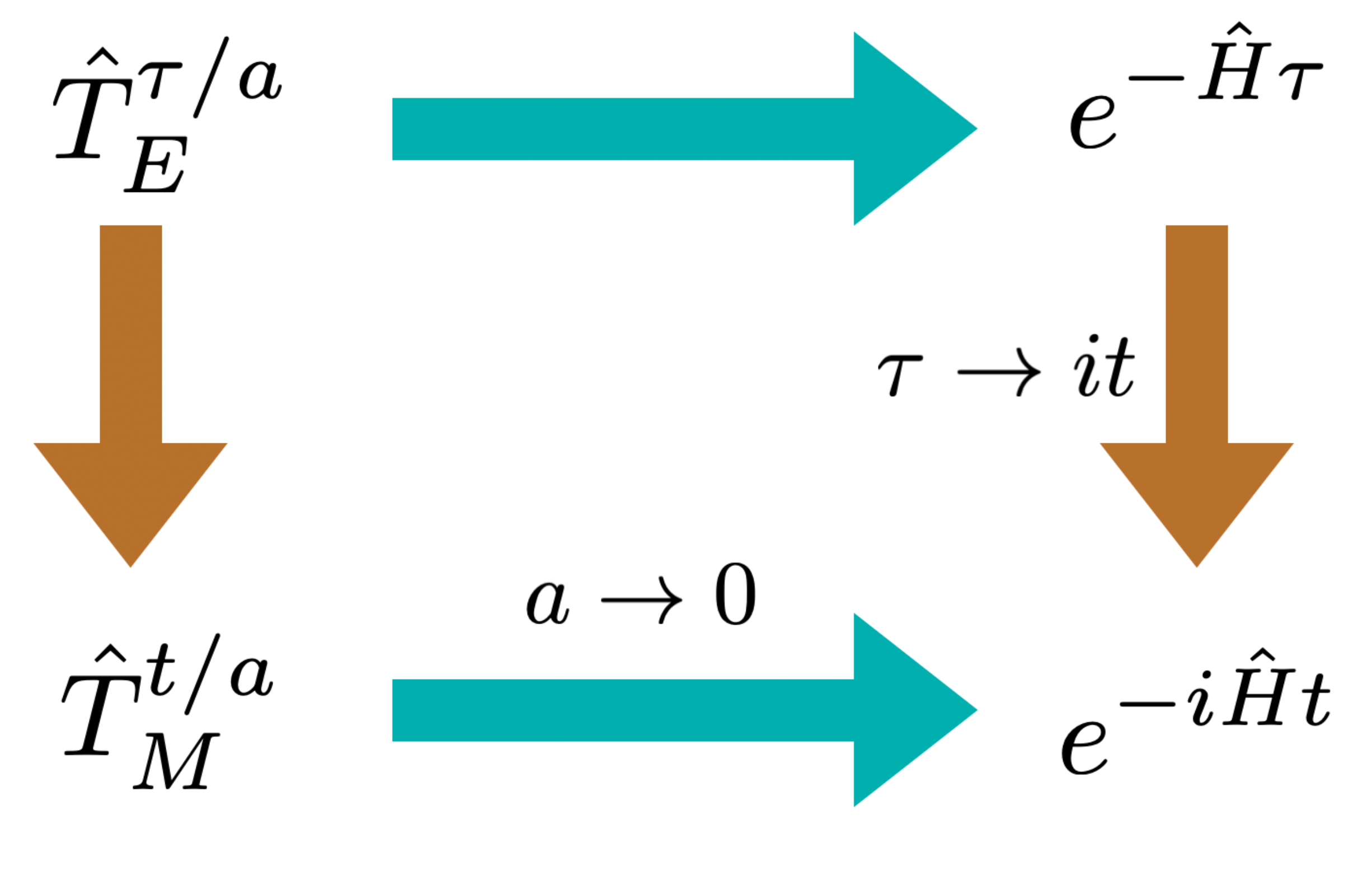}
    \caption{Commutative diagram expressing the desired relationships between real- and imaginary-time transfer matrices $\tmink{}$ and $\teucl{}$ and the operators $e^{-\hat{H}\tau}$ and $e^{-i\hat{H} t}$ describing real- and imaginary-time evolution in the continuum limit. The diagram is satisfied for generic actions involving non-compact variables but is only sometimes satisfied for actions involving compact variables as discussed in the main text. \label{fig:limits}}
\end{figure}

\subsection{Lattice gauge theory in real and imaginary time}\label{sec:transfer}

The Standard Model of particle physics involves the gauge groups $U(1)$ and $SU(N)$, and thus we focus on these two groups in explorations of suitable real-time actions for lattice gauge theory.
As an Abelian group, the continuum limit of $U(1)$ gauge theory can be accessed by extrapolating to zero lattice spacing using either the compact gauge group $U(1)$ or the non-compact gauge group $\mathbb{R}$; see for example Ref.~\cite{Fiore:2005ps}. Using the compact gauge group is susceptible to the subtleties discussed above and is the focus of our $U(1)$ studies.
In order to describe continuum Euclidean $SU(N)$ gauge theory, standard formulations of LGT~\cite{Wilson:1974sk,Kogut:1974ag} use $SU(N)$ variables in the fundamental group representation, which consists of $N \times N$ unitary matrices with unit determinant. These formulations are therefore also susceptible to challenges associated with path integrals involving compact variables when moving to real time.

A lattice gauge theory for gauge group $G$ in $D$ spacetime dimensions is defined in terms of a set of gauge fields $U_{x,\mu} \in G$, where $x = (x^0, x^1,\ldots x^{D-1})$ are the spacetime lattice points, and $\mu=0,\ldots, D-1$ labels the lattice axes with the (real or imaginary) time direction specified by $\mu=0$. We assume a lattice with an extent of $L/a$ sites in each spatial direction and $L_T/a$ sites in the temporal direction, where $a$ is the lattice spacing in physical units. When applying the Schwinger-Keldysh formalism for out-of-equilibrium observables, the lattice extent in the temporal direction is divided into regions of Euclidean, forward Minkowski, and reverse Minkowski time evolution; discussion on extending real-time actions to this setting is deferred to Sec.~\ref{sec:SK} and in other sections the spacetime signature is assumed to be uniform throughout the lattice. Each component $U_{x,\mu}$ of the gauge field is associated with an edge connecting neighboring sites $x$ and $x + \hat{\mu}$. For all actions under consideration an exact gauge symmetry holds: transforming all gauge field components by $U_{x,\mu} \rightarrow \Omega_x U_{x,\mu} \Omega_{x+a\hat{\mu}}^{\dagger}$, for any field $\Omega_x \in G$, does not modify the value of the action.

Equilibrium properties of gauge theories can be determined from the continuum limits of expectation values of ``observables'' in LGT, defined as generic functions of the gauge field, $\mathcal{O}(U)$, with Euclidean path integral representations
\begin{equation}
   \left< \mathcal{O} \right>_E = \frac{1}{Z_E} \int \mathcal{D} U \; \mathcal{O}(U) e^{-S_E(U)},
\end{equation}
where $\mathcal{D}U \equiv \prod_{x,\mu} d U_{x,\mu}$ is the product of the Haar measure $d U_{x,\mu}$ for each gauge field degree of freedom, $S_E$ is the Euclidean action, and $Z_E = \int \mathcal{D} U\; e^{-S_E(U)}$ is the partition function.
We restrict to considering Euclidean actions that can be expressed as a sum of potential and kinetic energy functions involving the lattice gauge fields on individual timeslices and pairs of adjacent timeslices respectively,
\begin{equation}
    S_E(U) = a \sum_{\tau/a=0}^{N_T-1}\left[ K(U_{\tau+a},U_{\tau}) + V(U_{\tau}) \right],
       \label{eq:SEdef}
\end{equation}
where $U_\tau = \{U_{x,\mu} \, | \, x^0 = \tau\}$ and $N_T = L_T/a$ is the number of lattice sites in the (imaginary) time direction. Restricting to actions of this form allows the construction of transfer matrices and explicit analyses of unitarity.
Constructing unitary actions that violate this form, e.g.~due to Symanzik improvement~\cite{Symanzik:1983dc} or inclusion of matter fields, is left as the subject of future work.

The Hilbert space for pure gauge theory on a fixed timeslice can be represented as a product of Hilbert spaces for group-valued quantum rotators~\cite{Kogut:1974ag}.
Gauge field operators $\hat{U}_{\vec{x},k}$ can be defined for $k=1,\ldots,D-1$, i.e.~they are associated with spacelike links, and are analogous to position operators $\hat{x}_{na}$ for the quantum rotator discussed in Sec.~\ref{sec:trotter}.
Assuming temporal gauge, the Hilbert space for pure gauge theory is spanned by polynomial functions of these operators.
States $\ket{U_{\vec{x},k}}$ are defined by the eigenvalue relation $\hat{U}_{\vec{x},k} \ket{U_{\vec{x},k}} = U_{\vec{x},k} \ket{U_{\vec{x},k}}$ and are normalized to satisfy $\braket{U_{\vec{x},k}|U_{\vec{x},k}'} = \delta(U_{\vec{x},k},U'_{\vec{x},k})$.
This Hilbert space can equivalently be described using a basis of $L^2$-normalizable complex-valued functions $f(U)$ as detailed in Ref.~\cite{Luscher:1976ms}.
Hilbert space states $\ket{f}$ can be associated with this function basis by $\left(\otimes_{\vec{x},k} \bra{U_{\vec{x},k}}\right) \ket{f} = f(U)$,
and the actions of Hilbert space operators in the function basis can be represented using integral kernels.

The imaginary-time transfer matrix $\teucl{}$ can be concretely defined by an integral kernel $\teucl{}(U_{\tau+a},U_\tau) = \braket{U_{\tau+a}|\teucl{}|U_{\tau}}$, where $\ket{U_\tau}$ and $\ket{U_{\tau+a}}$ are arbitrary tensor product basis states given by $\otimes_{\vec{x},k} \ket{U_{\vec{x},k}}$ for particular gauge field configurations $U_{(\tau,\vec{x}),k}$ and $U_{(\tau+a,\vec{x}),k}$.
The integral form of the action of this operator on a function-basis state $\ket{f}$ is then
\begin{equation}
    \teucl{} \ket{f} = \int \mathcal{D}U_{\tau} \mathcal{D}U_{\tau+a}\ \teucl{}(U_{\tau+a},U_\tau)\ f(U_{\tau}) \ \ket{U_{\tau+a}}.
\end{equation}
The integral kernel $\teucl{}(U_{\tau+a},U_\tau)$ is analogous to the coordinate space matrix elements $\teucl{}(x_{\tau+a},x_\tau)$ for the quantum rotator.
For a LGT action of the form in Eq.~\eqref{eq:SEdef}, the imaginary-time transfer matrix is defined in a general gauge by the integral kernel~\cite{Montvay:1994cy}
\begin{equation}
\begin{split}
    \teucl{}(U_{\tau+a},U_\tau) &= \int \mathcal{D}U_{\partial(\tau+a,\tau)} \; e^{-a V(U_{\tau+a})/2} \\
    &\hspace{40pt} \times e^{-a K(U_{\tau+a},U_\tau)} e^{-a V(U_\tau)/2} ,
    \label{eq:Tdef}
    \end{split}
\end{equation}
where $U_{\partial(\tau+a,\tau)} = \{ U_{x,0} \, | \, x^0 = \tau \}$.
The imaginary-time transfer matrix describes discretized imaginary-time evolution in LGT corresponding to the generic action in Eq.~\eqref{eq:SEdef}.
For example, Euclidean correlation functions involving a pair of temporally separated operators $\mathcal{A}(U_\tau)$ and  $\mathcal{B}(U_0)$ have a transfer-matrix representation
\begin{equation}
\begin{split}
    &\left< \mathcal{A}(U_\tau) \mathcal{B}(U_0) \right>_E = \frac{1}{Z_E} \int \prod_{\tau' \geq \tau} \left[ \mathcal{D}U_{\tau'} \; \teucl{}(U_{\tau'+a}, U_{\tau'})  \right] \\
    &\hspace{30pt} \times \mathcal{A}(U_\tau) \prod_{0 \leq \tau' < \tau}  \left[  \mathcal{D}U_{\tau'} \;  \teucl{}(U_{\tau'+a}, U_{\tau'}) \right] \mathcal{B}(U_0).
    \label{eq:Taction}
    \end{split}
\end{equation}

It is possible to formally (although not practically) construct a Hermitian Hamiltonian $\hat{H}$ and unitary real-time evolution operator $e^{-i\hat{H}t}$ directly from the imaginary-time transfer matrix. 
Assuming that the transfer matrix for a particular choice of Euclidean action is positive definite, then it is possible to construct the Hamiltonian operator $\hat{H}$ defined by
\begin{equation}
    \hat{H} = -\frac{1}{a}\ln \teucl{}
\end{equation}
without encountering singularities of the logarithm~\cite{Luscher:1976ms}.
Formally, a perfect real-time transfer matrix can then be constructed,
\begin{equation}
    \hat{U} = e^{-i \hat{H} a} = e^{i \ln\teucl{}} = \teucl{}^i.
    \label{eq:Udef}
\end{equation}
By construction $\hat{U}$ is unitary, with eigenvalues $e^{- i E_n a}$ related to the eigenvalues $e^{-E_n a}$ of $\teucl{}$.
The existence of a positive-definite transfer matrix therefore guarantees the existence of a unitary time-evolution operator with the same energy spectrum.
The positivity of the transfer matrix associated with the Wilson action~\cite{Wilson:1974sk} for Euclidean LGT was established early on in the study of lattice QFT through proofs of reflection positivity~\cite{Osterwalder:1973dx,Osterwalder:1974tc}, and the transfer matrix was then explicitly constructed~\cite{Creutz:1976ch} and explicitly demonstrated to be positive~\cite{Luscher:1976ms}.
These results and their generalizations to other actions crucially allow the energy spectra of Euclidean gauge theories obtained from the continuum limits of LGT results to be identified with the energy spectra of the corresponding Minkowski continuum gauge theories relevant for experiments involving real-time dynamics.
This approach to determining the energy spectra of continuum gauge theories corresponds to first taking $a\rightarrow 0$ and then subsequently taking $\tau\rightarrow it$ in Fig.~\ref{fig:limits} and is expected to be valid for any Euclidean LGT with a positive-definite $\teucl{}$.
However, the calculation of correlation functions with timelike separated operators in Minkowski spacetime, relevant for example for inclusive scattering cross-sections and transport coefficients, is an ill-posed and practically challenging problem when using analytic continuation of numerical results for Euclidean correlation functions.
For these and other applications, it may be advantageous to consider the opposite order of limits shown in Fig.~\ref{fig:limits}, in which a real-time transfer matrix is constructed for Minkowski LGT and physical results are obtained by subsequently taking the continuum limits of real-time observables obtained in Minkowski LGT.

The operator $\hat{U}$ defined in Eq.~\eqref{eq:Udef} is not a suitable starting point for practically computing Minkowski LGT observables because it cannot be constructed without explicitly diagonalizing $\teucl{}$ and working in the energy eigenbasis. A seemingly promising alternative approach is to replace the sum of kinetic and potential terms $K + V$ with a difference $K - V$ to move from the LGT Euclidean action to the LGT Minkowski action, with the goal of recovering the physics encoded in $\hat{U}$ in the continuum limit.
The Minkowski action obtained by such a replacement is
\begin{equation}
    S_M(U) = a \sum_{t/a=0}^{N_T-1} \left[ K(U_{t+a},U_t) - V(U_t) \right].
    \label{eq:SMdef}
\end{equation}
Minkowski expectation values described by this action are defined by
\begin{equation}
   \left< \mathcal{O} \right>_M = \frac{1}{Z_M} \int \mathcal{D} U \; \mathcal{O}(U) e^{i S_M(U)},
\end{equation}
where $Z_M = \int \mathcal{D} U\; e^{i S_M(U)}$.
A real-time transfer matrix $\tmink{}$ can be defined for this action in analogy to Eq.~\eqref{eq:Tdef}
\begin{equation}
\begin{split}
    \tmink{}(U_{t+a},U_t) &= \int \mathcal{D}U_{\partial(t+a,t)} \; e^{-i a V(U_{t+a})/2}  \\
    &\hspace{20pt} \times e^{i a K(U_{t+a},U_t)} e^{-i a V(U_t)/2}.
    \label{eq:TMWdef}
    \end{split}
\end{equation}
This real-time transfer matrix can be used to equivalently write matrix elements of products of operators separated in Minkowski time, otherwise given by real-time LGT path integrals involving products of operators.
For example, Minkowski correlation functions involving a pair of temporally separated operators $\mathcal{A}(U_t)$ and  $\mathcal{B}(U_0)$ are given analogously to Eq.~\eqref{eq:Taction} by
\begin{equation}
\begin{split}
    &\left< \mathcal{A}(U_t) \mathcal{B}(U_0) \right>_M = \frac{1}{Z_M} \int \prod_{t' \geq t} \left[ \mathcal{D}U_{t'} \; \tmink{}(U_{t'+a}, U_{t'})  \right] \\
    &\hspace{30pt} \times \mathcal{A}(U_t) \prod_{0 \leq t' < t}  \left[  \mathcal{D}U_{t'} \;  \tmink{}(U_{t'+a}, U_{t'}) \right] \mathcal{B}(U_0).
    \label{eq:Waction}
    \end{split}
\end{equation}
The energy spectrum for Minkowski LGT with action $S_M$ can therefore be obtained from the eigenvalues of $\tmink{}$.

The real- and imaginary-time transfer matrices can be decomposed into products of potential-energy evolution operators that are diagonal in the coordinate basis and kinetic-energy evolution operators $\tkeucl{}$ and $\tkmink{}$,
\begin{equation}
\begin{split}
        \tmink{}(U_{t+a},U_t) &= e^{-iaV(U_{t+a})/2} \tkmink{}(U_{t+a},U_t) \\ 
        &\hspace{20pt}\times e^{-iaV(U_t)/2}, \\
        \teucl{}(U_{\tau+a},U_\tau) &= e^{-aV(U_{\tau+a})/2} \tkeucl{}(U_{\tau+a},U_\tau) \\ &\hspace{20pt} \times e^{-aV(U_\tau)/2}. \\
        \label{eq:EMdef}
        \end{split}
\end{equation}
The real- and imaginary-time kinetic-energy evolution operators $\tkmink{}$ and $\tkeucl{}$ are defined by
\begin{equation}
\begin{split}
    \tkmink{}(U_{t+a},U_t) &= \int \mathcal{D}U_{\partial(t+a,t)} \; e^{ia K(U_{t+a},U_t)},\\ 
    \tkeucl{}(U_{\tau+a},U_\tau) &= \int \mathcal{D}U_{\partial(\tau+a,\tau)} \; e^{-a K(U_{\tau+a},U_\tau)}.
    \label{eq:Mdef}
    \end{split}
\end{equation}
In temporal gauge the integrals over $U_{\partial(\tau+a,\tau)}$ appearing in Eq.~\eqref{eq:Mdef} are trivial, and the kinetic-energy evolution operators can be simplified to
\begin{equation}
\begin{split}
    \tkmink{}(U_{t+a},U_t) &=  e^{i aK(U_{t+a},U_t)}, \\
    \tkeucl{}(U_{\tau+a},U_\tau) &=  e^{-aK(U_{\tau+a},U_\tau)}.
          \label{eq:EMtemporal}
    \end{split}
\end{equation}
The real- and imaginary-time kinetic-energy evolution operators satisfy a relation $\tkmink{}(U_{\tau+a},U_\tau) = \tkeucl{}(U_{\tau+a},U_\tau)^{-i}$ with superficial similarities to Eq.~\eqref{eq:Udef}; however, this relation between coordinate-basis matrix elements of $\tkmink{}$ and $\tkeucl{}$ does not imply that the eigenvalues of $\tkmink{}$ and $\tkeucl{}$ satisfy analogous relations.

In general the real-time transfer matrix $\tmink{}$ in Eq.~\eqref{eq:EMdef} is distinct from the unitary time-evolution operator $\hat{U}$ defined by Eq.~\eqref{eq:Udef}.
The fact that $\tmink{} \neq \hat{U}$ is not by itself problematic, as the validity of results obtained using real-time LGT calculations only requires that a continuum limit exists in which the energy spectra associated with $\tmink{}$ and $\hat{U}$ agree.
However, if $\tmink{}$ is non-unitary for all limits of the action parameters then it is not possible to achieve a continuum limit in which $\tmink{}$ coincides with $\hat{U}$ and the real- and imaginary-time energy spectra will contain unphysical differences.
This undesirable scenario corresponds to the non-commutation of limits shown in Fig.~\ref{fig:limits} and can arise in practice for theories with compact variables such as the quantum rotator discussed in Sec.~\ref{sec:trotter}.
For non-compact scalar field theory, this scenario does not arise and the unitarity of time-evolution associated with the standard finite difference discretization of the continuum Minkowski action has been demonstrated in 0+1D calculations in Refs.~\cite{Alexandru:2016gsd,Alexandru:2017lqr,Mou:2019tck,Lawrence:2021izu}.
A proof of the unitarity of the real-time transfer matrix $\tmink{}$ for scalar field theory proceeds analogously to the SHO case in Sec.~\ref{sec:trotter} and is detailed in Appendix~\ref{sec:scalar}.

Below, the unitarity and continuum limits of $\tmink{}$ will be studied for specific LGT actions.
Proofs of (non\nobreakdash-)unitarity of $\tmink{}$ are facilitated by the observation that
$\braket{U|\tmink{} \tmink{}^\dagger|U'}$,
which must be proportional to a delta function $\delta(U,U')$ for $\tmink{}$ to give rise to unitary physics, satisfies
\begin{equation}
\begin{split}
    \braket{U|\tmink{} \tmink{}^\dagger|U'}
    &= \int \mathcal{D}U^{\prime\prime} \tmink{}(U,U^{\prime\prime})\tmink{}^\dagger(U^{\prime\prime},U') \\
    &= e^{-ia[ V(U) - V(U')^*]}
    \braket{U|\tkmink{} \tkmink{}^\dagger|U'}.
    \end{split}
    \label{eq:WWdag}
\end{equation}
Similarly,
\begin{equation}
    \braket{U|\tmink{}^\dagger \tmink{}|U'} = e^{ia[V(U)^* - V(U')]} \braket{U|\tkmink{}^\dagger \tkmink{}|U'}.
\end{equation}
Assuming that $V(U)$ is real here and below, it follows that $\braket{U|\tmink{} \tmink{}^\dagger|U'} = \braket{U|\tmink{}^\dagger\tmink{}|U'} \propto \delta(U,U')$ 
will hold if and only if $\braket{U|\tkmink{} \tkmink{}^\dagger|U'} = \braket{U|\tkmink{}^\dagger\tkmink{}|U'} \propto \delta(U,U')$,
i.e.~$\tmink{}$ is proportional to unitary if and only if $\tkmink{}$ is proportional to unitary.

\subsection{The real-time Wilson action}\label{sec:wilson}

For the gauge groups $U(N)$ or $SU(N)$, the (Euclidean) Wilson action is defined by~\cite{Wilson:1974sk}
\begin{equation}
\begin{split}
    S_{E,W}(U) &= \frac{2}{g^2}\sum_{x}  \sum_{\mu < \nu} \Re\Tr\left(1- P_{x,\mu\nu} \right), \\
    P_{x,\mu\nu} &\equiv U_{x,\mu} U_{x+a\hat{\mu},\nu} U^\dagger_{x+a\hat{\nu},\mu} U^\dagger_{x,\nu}.
    \label{eq:SWilson}
\end{split}
\end{equation}
For the case of $G=SU(N)$, the bare coupling $g$ is normalized so that the ``naive continuum limit'' obtained by defining $U_{x,\mu} = e^{i a A_\mu(x)}$ and taking the $a\rightarrow 0$ limit of  Eq.~\eqref{eq:SWilson} agrees with the continuum Yang-Mills action for $A_{\mu}(x) \in \mathfrak{g}$ with the same gauge coupling $g$.
For the case of $G=U(1)$, the rescaled coupling $e = g / \sqrt{2}$ has the correct normalization for the naive continuum limit of Eq.~\eqref{eq:SWilson} to match the continuum $U(1)$ gauge action with gauge coupling $e$.
The eigenvalues of the ``plaquette'' $P_{x,\mu\nu}$ can be represented as $e^{i\phi_{x,\mu\nu}^A}$ where $A \in \{1\}$ for $G=U(1)$ and $A \in \{1,\ldots,N\}$ for $SU(N)$, with an additional constraint $\sum_A \phi_{x,\mu\nu}^A=0 \mod 2\pi$.
The Wilson action can be represented in terms of these eigenvalues as
\begin{equation}
    S_{E,W}(U) =  \frac{2}{g^2}  \sum_{x,A} \sum_{\mu < \nu} \left[ 1 - \cos(\phi_{x,\mu\nu}^A)\right].
\end{equation}

The quantum rotator discussed in Sec.~\ref{sec:trotter} is in fact equivalent to $U(1)$ LGT using the Wilson action in $(1+1)$D, and the arguments in that section explicitly demonstrate that $\tmink{}$ is non-unitary in this case.
As discussed in Ref~\cite{Hoshina:2020gdy} and reviewed here, character expansion methods can be used to demonstrate non-unitarity of the Wilson gauge action more generally for $G=U(1)$ or $G=SU(N)$ in arbitrary spacetime dimensions.
A similar check for (non\nobreakdash-)unitarity can be applied to any LGT action that only depends locally on the $P_{x,\mu\nu}$ in a way that can be decomposed into a kinetic energy piece that is a function of ``timelike plaquettes'' $P_{x,0k}$ on each timeslice and a potential energy piece that is a function of ``spacelike plaquettes'' $P_{x,ij}$ on each timeslice.
Since timelike plaquettes are given in temporal gauge by $U_{x,k} U_{x+a\hat{0},k}^\dagger$, the corresponding kinetic-energy evolution operator $\tkmink{}(U,U')$ only depends on the products $U_{\vec{x},k}' U_{\vec{x},k}^\dagger$.
The kinetic-energy evolution operator associated with any plaquette LGT action can therefore be represented using a character expansion of the form
\begin{equation}
    \tkmink{}(U,U') = \prod_{\vec{x},k} \left[ \sum_{r} d_r c^M_r(g^2) \chi_r(U_{\vec{x},k}' U_{\vec{x},k}^\dagger) \right],
    \label{eq:Mchar}
\end{equation}
where $r$ labels the representations of the gauge group $G$, $d_r$ denotes the dimension of representation $r$, and $\chi_r(U)$ is the character of the group element $U \in G$ in representation $r$; see for example Refs.~\cite{Drouffe:1983fv,Montvay:1994cy} for further discussion in the context of LGT.
The coefficients of the character expansion $c^M_r(g^2)$ are functions of only the gauge coupling $g$.
Using the character orthogonality properties
\begin{equation}
     \int dU \chi_r(U)^* \chi_{r'}(U) = \delta_{rr'},
\end{equation}
and
\begin{equation}
    \int dU' \ \chi_r(U' U^\dagger) \chi_{r'}(U^{\prime \dagger})^* = \delta_{rr'} \frac{1}{d_r}\chi_r(U^\dagger),
\end{equation}
the characters $\chi_r(U) \equiv \prod_{\vec{x},k} \chi_{r_{\vec{x},k}}(U_{\vec{x},k})$ can be seen to be eigenfunctions of $\tkmink{}$,
\begin{equation}
\begin{split}
    \tkmink{} \ket{\chi_r(U)} &= \int \mathcal{D}U' \mathcal{D}U \tkmink{}(U',U) \prod_{\vec{x},k} \chi_{r_{\vec{x},k}}(U_{\vec{x},k}) \ket{U'_{\vec{x},k}} \\
    &= \prod_{\vec{x},k} \left[ c_{r_{\vec{x},k}}^M(g^2)   \int dU'_{\vec{x},k} \chi_{r_{\vec{x},k}}(U'_{\vec{x},k}) \ket{U'_{\vec{x},k}} \right] \\
    &= \left[ \prod_{\vec{x},k} c_{r_{\vec{x},k}}^M(g^2) \right] \ket{\chi_r(U)}.
    \end{split}
\end{equation}
The eigenvalue associated to each eigenfunction $\chi_r(U)$ is therefore the product of the character coefficients $c_{r_{\vec{x},k}}^M(g^2)$ associated with the representation $r_{\vec{x},k}$ specified by $\chi_r(U)$ for each gauge link.  
This implies that the kinetic-energy evolution operator $\hat{R}_M$ will be unitary if and only if all $c_r^M(g^2)$ are unit-norm complex numbers. The character expansion coefficients can be obtained using character orthogonality,
\begin{equation}
    c^M_r(g^2) = \left( \int \mathcal{D} U \ \prod_{\vec{x},k} \left[ \frac{1}{d_r} \chi_r(U_{\vec{x},k})^* \right] \tkmink{}(1,U) \right)^{1/N_{L}},
    \label{eq:crdef}
\end{equation}
where $N_L = (D-1)(L / a)^{D-1}$ is the number of spatial links $U_{\vec{x},k}$ at fixed $t$. Unitarity can then in principle be checked for particular actions.

The kinetic and potential energy terms $K_W$ and $V_W$ for the Wilson action are given by
\begin{equation}
\begin{split}
    K_W(U_{\tau+a},U_{\tau}) &= \frac{2}{g^2 a}  \sum_{\vec{x}} \sum_{k} \Re\Tr\left(1 - P_{(\vec{x},\tau),0k} \right), \\
    V_W(U_{\tau}) &= \frac{2}{g^2 a}  \sum_{\vec{x}} \sum_{k < l} \Re\Tr\left(1  - P_{(\vec{x},\tau),kl} \right),
\end{split}
\end{equation}
where $k,l=1,\ldots,D-1$ denote spatial indices.
The Minkowski Wilson action obtained by combining $K_W$ and $V_W$ with a relative minus sign and multiplying by $a$ to remove the factors of $1/a$ above is given by
\begin{equation}
\begin{split}
   S_{M,W} &= a \sum_{\tau} [K_W(U_{\tau+a}, U_{\tau}) - V_W(U_{\tau})] \\
   &= \frac{2}{g^2}  \sum_{x} \sum_{k} \Re\Tr\left(1 - P_{x,0k} \right) \\
   &\hspace{20pt} - \frac{2}{g^2} \sum_{x} \sum_{i < j}  \Re\Tr\left(1 - P_{x,ij} \right).
   \label{eq:SMWdef}
   \end{split}
\end{equation}
The kinetic-energy evolution operator $\tkmink{W}$ associated with the Minkowski Wilson action is then given in temporal gauge by
\begin{equation}
\begin{split}
    \tkmink{W}(U,U') &=  e^{\frac{2i}{g^2} \sum_{\vec{x}}  \sum_k \Re\Tr(1 - U_{\vec{x},k}' U_{\vec{x},k}^\dagger)  } .
    \end{split}
\end{equation}
Applying the character expansion in Eq.~\eqref{eq:Mchar} to $\tkmink{W}$ gives
\begin{equation}
\begin{split}
    \tkmink{W}(U,U') &= \prod_{\vec{x},k}  \left[ \sum_r d_r c^{M,W}_r(g^2) \chi_r ( U_{\vec{x},k}' U_{\vec{x},k}^\dagger )   \right],
    \end{split}
\end{equation}
where the character expansion coefficients for the Wilson action are given by
\begin{equation}
    \begin{split}
        c^{M,W}_r(g^2) &= \left( \int \mathcal{D}U \prod_{\vec{x},k}  \left[ \frac{1}{d_r} \chi_r(U_{\vec{x},k})^* \right] \tkmink{W}(1,U) \right)^{1/N_{L}} \\
        &= \frac{1}{d_r} \int  dU \ \chi_r(U)^* \  e^{\frac{2i}{g^2}  \Re\Tr(1 - U) }.
    \end{split}
    \label{eq:mrWilson}
\end{equation}
The kinetic-energy evolution operator $\tkmink{W}$ is exactly unitary if and only if $|c^{M,W}_r(g^2)|=1$ for all $r$.
However, a change in the overall normalization of the path integral uniformly rescales the magnitudes of all character expansion coefficients while leaving all operator expectation values invariant.
Therefore if there is a particular choice of overall path integral normalization that leads to a unitary real-time transfer matrix then the corresponding action is compatible with unitary real-time evolution.\footnote{We thank Henry Lamm for bringing this point to our attention.}

\begin{figure}[t!]
    \centering
    \includegraphics[width=0.47\textwidth]{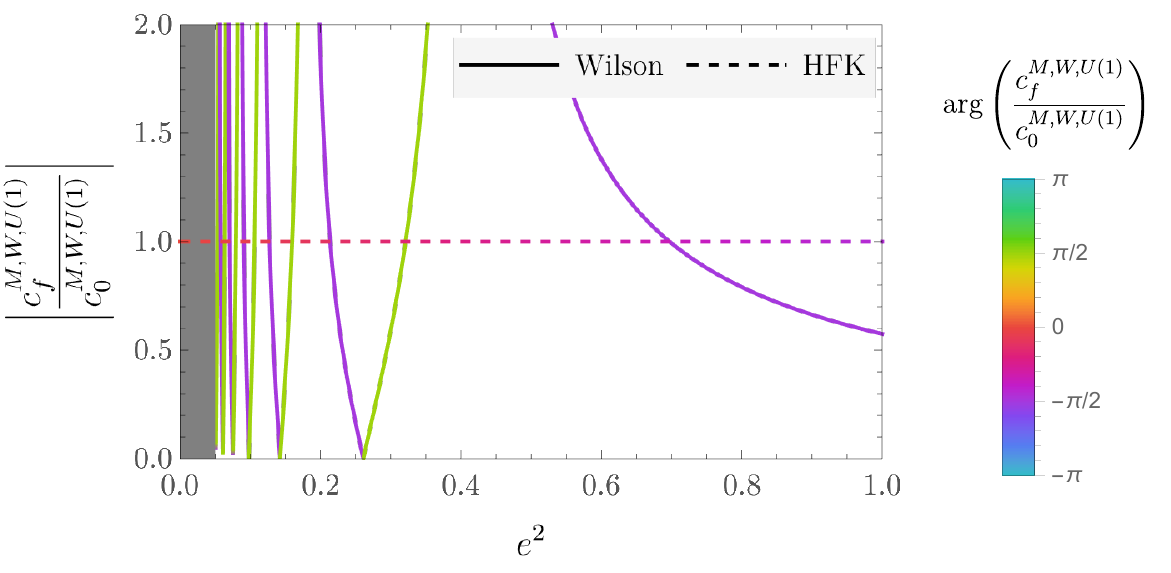}
    \caption{Ratios of the fundamental and trivial irrep character expansion coefficients for the Wilson and HFK actions for $U(1)$ gauge theory. The Wilson action result includes an infinite number of singularities that accumulate as $e^2 \rightarrow 0$ and is replaced by a gray background for $e^2 < 0.05$. The Wilson action can be seen to have a non-unitary coefficient ratio almost everywhere in $e^2$, with an ill-defined $e^2 \rightarrow 0$ limit due to the repeated singularities, while the HFK action has a unitary ratio for all choices of $e^2$.
    \label{fig:Wchar1} }
\end{figure}

To demonstrate the non-unitarity of the Wilson gauge action, we therefore explicitly analyze ratios of the Wilson action character expansion coefficients for the groups $U(1)$ and $SU(N)$ with $N \in \{2, \dots, 9\}$ below. These ratios are independent of the overall path integral normalization.
In all cases, the ratios of Wilson action character coefficients are found to have non-unit magnitudes almost everywhere in $g^2$ and in particular in the would-be continuum limit $g^2 \rightarrow 0$,
which is sufficient to establish that $\tkmink{W}$ is non-unitary and cannot be made unitary by adjusting the path integral normalization.
As discussed below Eq.~\eqref{eq:WWdag}, it follows that $\tmink{W}$ is non-unitary and further cannot be made unitary by adjusting the path integral normalization.

\begin{figure*}[t!]
    \centering
    \includegraphics[width=0.47\textwidth]{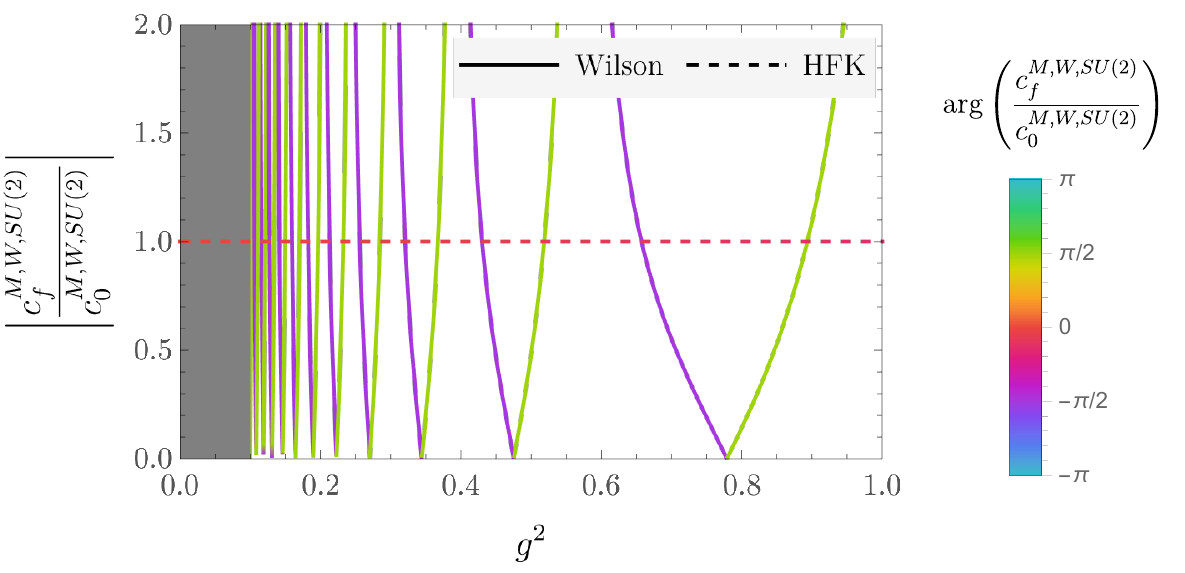} \hspace{25pt}
    \includegraphics[width=0.47\textwidth]{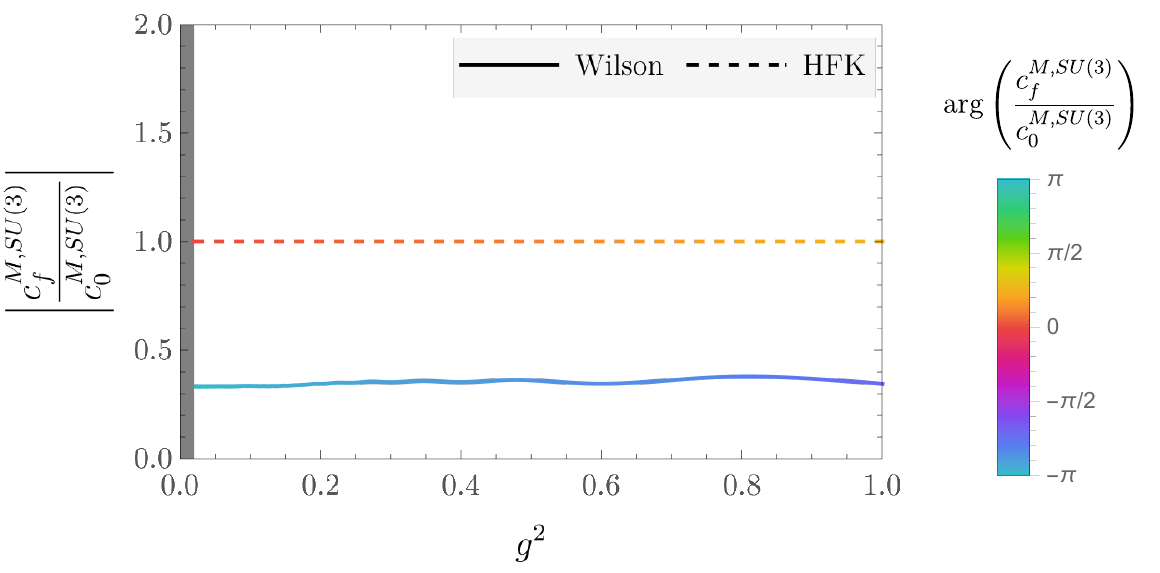}
    \caption{Ratios of the fundamental and trivial irrep character expansion coefficients for the Wilson and HFK actions for $SU(2)$ and $SU(3)$ gauge theory. The Wilson action result for $SU(2)$ includes an infinite number of singularities that accumulate as $g^2 \rightarrow 0$ and is replaced by a gray background for $g^2 < 0.1$. The Wilson action for both $SU(2)$ and $SU(3)$ can be seen to have a non-unitary coefficient ratio almost everywhere in $g^2$. In the $SU(2)$ case, the limit $g^2 \rightarrow 0$ is ill-defined due to the repeated singularities, while for $SU(3)$ there is an apparently well-defined limit (explained by a stationary phase expansion in Appendix~\ref{sec:nonunitary-Wilson}) which is non-unitary. The HFK action for both $SU(2)$ and $SU(3)$ gauge theory has a unitary ratio for all choices of $g^2$.
    \label{fig:Wchar23} }
\end{figure*}

In the simplest case of $G=U(1)$, group elements $U \in U(1)$ can be represented as $U = e^{i\phi}$, the Haar measure is simply $\frac{d\phi}{2\pi}$, and there are only one-dimensional representations with characters $\chi_r(e^{i\phi}) = e^{i r \phi}$.
Eq.~\eqref{eq:mrWilson} therefore gives
\begin{equation}
\begin{split}
    c_r^{M,W,U(1)}(e^2) &=  \int_{-\pi}^{\pi} \frac{d\phi}{2\pi} e^{-ir\phi} e^{-\frac{i}{e^2}\cos(\phi)} e^{\frac{i}{e^2}} \\
    &=  I_{r}\left(-\frac{i}{e^2}\right)e^{\frac{i}{e^2}} .
    \end{split}
        \label{eq:mrWilsonU1}
\end{equation}
The modified Bessel functions $I_r(-i/e^2)$ are oscillating functions of $e^2$ with vanishing magnitude and increasingly rapid oscillations in the $e^2 \rightarrow 0$ limit associated with the continuum limit of Euclidean $U(1)$ LGT.
The form of these functions immediately gives that $|c_r^{M,W,U(1)}(e^2) / c_0^{M,W,U(1)}(e^2)| \neq 1$ almost everywhere in $e^2$ and for $r > 0$ -- these ratios do have unit norm for particular choices of $e^2$ and $r$ as seen for the fundamental representation in Fig.~\ref{fig:Wchar1}, but not for all $r$ at any $e^2$).
It is also possible to directly confirm non-unitarity in the continuum limit $e^2 \rightarrow 0$,
\begin{equation}
\begin{split}
    &\braket{e^{i\phi}|\tkmink{W}^{U(1)} \tkmink{W}^{U(1) \dagger}|e^{i\phi'}} \\ 
    &= \prod_{\vec{x},k}  \int_{-\pi}^\pi \frac{d\phi^{\prime\prime}}{2\pi} \ e^{-\frac{i}{e^2}\cos(\phi_{\vec{x},k} - \phi^{\prime\prime})}\ e^{\frac{i}{e^2}\cos(\phi^{\prime\prime} - \phi'_{\vec{x},k})} \\
    &= \prod_{\vec{x},k}  I_0\left(-\frac{i}{e^2}\left| 1- e^{i(\phi_{\vec{x},k}-\phi'_{\vec{x},k})} \right| \right),
    \end{split}
\end{equation}
which is not proportional to the identity integral kernel as $e^2 \rightarrow 0$.

Analogous explicit results can be obtained for $SU(2)$ gauge theory, where the irreps are labeled by half-integers $j$ or equivalently integers $r = 2j$ and have dimension $d_r = r+1$. Denoting the eigenvalues of group elements $U$ by $e^{i\phi}$ and $e^{-i\phi}$,
the Weyl character formula gives $\chi_r(U) = \sin((r+1)\phi)/\sin(\phi)$. For functions of the eigenvalues, the Haar measure corresponds to $\frac{d\phi}{\pi}\sin^2(\phi)$, and the character expansion coefficients are therefore given by
\begin{equation}
\begin{split}
  &  c_r^{M,W,SU(2)}(g^2) \\
    &= \frac{1}{r+1}\int_{-\pi}^{\pi} \frac{d\phi}{\pi} \sin(\phi)\sin\left( (r+1)\phi\right) e^{-\frac{4i}{g^2}\cos(\phi)} e^{\frac{4i}{g^2}} \\
    &= \frac{1}{r+1}\left[ I_{r}\left(-\frac{4i}{g^2}\right) - I_{r+2}\left(-\frac{4i}{g^2}\right) \right] e^{\frac{4i}{g^2}}  \\
    &= \frac{ig^2}{2} I_{r+1}\left(-\frac{4i}{g^2}\right) e^{\frac{4i}{g^2}} .
    \end{split}
    \label{eq:mrWilsonSU2}
\end{equation}
An explicit calculation of the ratios $|c_r^{M,W,SU(2)}(g^2)/c_0^{M,W,SU(2)}(g^2)|$ for $r \neq 0$ gives non-unitarity almost everywhere in $g^2$. The limit $\lim_{g^2 \rightarrow 0} |c_r^{M,W,SU(2)}(g^2)/c_0^{M,W,SU(2)}(g^2)| $ does not exist and the $SU(2)$ real-time Wilson action is also non-unitary in the would-be continuum limit.
Results for $c_f^{M,W,SU(2)} / c_0^{M,W,SU(2)}$ are shown in Fig.~\ref{fig:Wchar23} and show qualitatively similar features to the $U(1)$ case, including an accumulation of divergences as $g^2 \rightarrow 0$.
It is further demonstrated in Sec.~\ref{sec:exact} below that the $g^2 \rightarrow 0$ limits of simple observables such as $(1+1)$D Wilson loops do not exist using the $SU(2)$ real-time Wilson LGT action.

Explicit results for the $SU(N)$ character expansion coefficients for the imaginary-time Wilson action are known~\cite{Drouffe:1983fv} and can be used to derive analogous results for the general $SU(N)$ real-time Wilson action.
For the imaginary-time Wilson action, the kinetic-energy evolution operator $\tkeucl{W}$ has a character expansion given by
\begin{equation}
\begin{split}
    \tkeucl{W}(U,U') &= \prod_{\vec{x},k}  e^{\frac{2}{g^2}\Re\Tr\left( 1 - U'_{\vec{x},k} U_{\vec{x},k}^\dagger \right)} \\
    &= \prod_{\vec{x},k} \left[ \sum_r d_r c^{E,W}_r(g^2) \chi_r\left(U'_{\vec{x},k} U_{\vec{x},k}^\dagger \right)  \right].
    \label{eq:EW}
    \end{split}
\end{equation}
The coefficients of this expansion are given analogously to Eq.~\eqref{eq:crdef} by
\begin{equation}
\begin{split}
    c^{E,W}_r(g^2) &= \left( \int \mathcal{D}U  \prod_{\vec{x},k} \left[ \frac{1}{d_r}  \chi_r(U_{\vec{x},k})^* \right] \tkeucl{}(1,U) \right)^{1/N_L} \\
    &= \frac{1}{d_r} \int dU \, \chi_r(U)^* \, e^{ \frac{2}{g^2} \Re\Tr (1-U) } .
    \end{split}
    \label{eq:erWilson}
\end{equation}
The character identity $\chi(U) = \chi(U^\dagger)^*$ along with invariance of the trace and Haar measure under the transformation $U \rightarrow U^\dagger$ gives $c^{E,W}_r(g^2) = c^{E,W}_r(g^2)^*$.
It is proven in Ref.~\cite{Luscher:1976ms} that $c^{E,W}_r(g^2) > 0$ by expanding the exponential in Eq.~\eqref{eq:EW} and observing that $c^{E,W}_r(g^2)$ is equal to a positive number (for $g\in \mathbb{R}$) times a counting factor related to the number of times the irrep $r$ appears at a given order of the expansion.
Comparing the definition of $c^{M,W}_r$ in Eq.~\eqref{eq:mrWilson} and the coefficients $c^{E,W}_r$, one finds
\begin{equation}
    c^{E,W}_r(g^2) = c^{M,W}_r(-ig^2).
    \label{eq:ermrWilson}
\end{equation}
The explicit forms of $c^{E,W}_r(g^2)$ for generic $SU(N)$ gauge groups presented in Ref.~\cite{Drouffe:1983fv} can thus be used with Eq.~\eqref{eq:ermrWilson} to obtain results for $c^{M,W}_r(g^2)$.
We numerically compute results for $c^{M,W}_f(g^2)/ c^{M,W}_0(g^2)$ for the cases of $N \in \{2,\ldots,9\}$ as detailed in  Appendix~\ref{sec:nonunitary-Wilson}.
For $SU(3)$, 
these numerical results are shown in Fig.~\ref{fig:Wchar23} and indicate that, although $c^{M,W,SU(3)}_f(g^2)/ c^{M,W,SU(3)}_0(g^2)$ is finite for all $g^2$, $|c^{M,W,SU(3)}_f(g^2)/ c^{M,W,SU(3)}_0(g^2)|\neq 1$ for all $g^2$ and for the limit $g^2 \rightarrow 0$, and therefore the real-time $SU(3)$ Wilson action is non-unitary.
The analogous results for $N \in \{4,\ldots,9\}$ are shown in Appendix~\ref{sec:nonunitary-Wilson} and indicate similarly that the real-time Wilson action is non-unitary almost everywhere in $g^2$ and in particular is non-unitary in the $g^2 \rightarrow 0$ limit.
The appendix further discusses observed similarities between results for choices of $N$ that are equivalent mod 4 which suggest a pattern in the behavior of $c^{M,W,SU(N)}_f / c^{M,W,SU(N)}_0$ and non-unitarity for general $N$. In the appendix the $g^2 \rightarrow 0$ limit is also analyzed using the stationary phase approximation.

\section{Unitary real-time LGT actions}\label{sec:actions}

Ref.~\cite{Hoshina:2020gdy} directly constructs a unitary real-time evolution matrix, here denoted $\tmink{HFK}$, whose spectrum in the naive continuum limit is related to the spectrum of the usual Euclidean Wilson transfer matrix. Unitarity is guaranteed by defining the kinetic energy evolution operator, $\tkmink{HFK}$, in terms of an explicitly unitary spectrum. The spectrum is determined by the character expansion with coefficients denoted $c^{M,HFK}_r(g^2)$ that are constructed to satisfy $|c^{M,HFK}_r(g^2)| = 1$. Unitarity for any choice of potential, including the Wilson potential $V_W$, then follows from Eq.~\eqref{eq:WWdag}.

An action $S_{HFK}(U)$ can be formally defined using the character expansion for $\tkmink{HFK}$~\cite{Hoshina:2020gdy}, as reviewed below in Sec.~\ref{sec:HFK}. The resulting series, however, is not absolutely convergent in all gauge field configurations as discussed below and cannot be numerically evaluated using either systematically improvable truncations or Monte Carlo sampling techniques.
For $U(1)$ gauge theory, a simple path integral contour deformation can be found that provides an absolutely convergent representation of path integrals involving the HFK action (see Sec.~\ref{sec:contour}), but for $SU(N)$ gauge theory it is challenging to find an analogous contour deformation that gives convergence.

In this work, an alternative real-time LGT action based on analytic continuation of the Euclidean heat-kernel LGT action~\cite{Menotti:1981ry} is therefore studied; the Euclidean construction is reviewed in Sec.~\ref{sec:heatkernelE}, and the Minkowski version is introduced in Sec.~\ref{sec:heatkernelM}.
This Minkowski heat-kernel, or HK, action is also formally defined by a divergent series, and it is difficult to find path integral contour deformations that result in absolutely convergent representations of this action. 
However, a real-time modified heat-kernel, or $\HKbar$, action is introduced in Sec.~\ref{sec:heatkernelbarM} for which path integral contour deformations can be used to construct absolutely convergent representations of the $U(1)$ and $SU(N)$ unitary real-time LGT actions. The real-time $\HKbar$ action is thus in principle suitable for evaluation of unitary real-time LGT dynamics for these gauge groups.
Analytic results for unitary and non-unitary actions computable in $(1+1)$D are discussed in Sec.~\ref{sec:exact}.

\subsection{The real-time HFK action}\label{sec:HFK}

We begin by reviewing in this section the construction of the HFK action given in Ref.~\cite{Hoshina:2020gdy}.
Positivity of $c^{E,W}_r(g^2)$ permits the definition of a set of character coefficients
\begin{equation}
    c^{M,HFK}_r(g^2) = [c^{E,W}_r(g^2)]^i = e^{i \ln(c^{E,W}_r(g^2))},
\end{equation} 
that satisfy $|c^{M,HFK}_r(g^2)| = 1$ by construction.
Explicitly, $\tkmink{HFK}$ is then defined from the character expansion in temporal gauge by
\begin{equation}
    \tkmink{HFK}(U,U') = \prod_{\vec{x},k} \left[ \sum_r d_r [c^{E,W}_r(g^2)]^i \chi_r\left(U'_{\vec{x},k} U_{\vec{x},k}^\dagger \right)  \right].
\end{equation}
A gauge-invariant action whose kinetic energy term is related to $\tkmink{HFK}$ by Eq.~\eqref{eq:EMtemporal} can then be defined as
\begin{equation}
\begin{split}
    S_{M,HFK}(U) &= -\sum_{x,k} \ln \left[ \sum_r d_r [c^{E,W}_r(g^2)]^i \chi_r(P_{x,0k}) \right] \\
    &\hspace{20pt} + \frac{2}{g^2} \sum_{x} \sum_{i<j} \Re\Tr(1 - P_{x,ij}).
    \end{split}
    \label{eq:SMHFK}
\end{equation}
The real-time transfer matrix associated with the HFK action is
\begin{equation}
    \tmink{HFK}(U,U') = e^{-iV_W(U)/2}\tkmink{HFK}(U,U')e^{-iV_W(U')/2},
\end{equation}
and by unitarity of $\tkmink{HFK}$, the entire transfer matrix is unitary.

Positivity of the Euclidean coefficients $c^{E,W}_r(g^2)$ guarantees the existence of an operator $\hat{K}_W$ satisfying $\tkeucl{W} = e^{-a \hat{K}_W}$.
The eigenvalue relation $c^{M,HFK}_r(g^2) = [c^{E,W}_r(g^2)]^i$ then gives $\tkmink{HFK} = e^{-i a \hat{K}_W}$.
It follows that
\begin{equation}
    \tmink{HFK} = e^{-ia\hat{V}_W/2} e^{-ia\hat{K}_W} e^{-ia\hat{V}_W/2}, \label{eq:WHFKhat}
\end{equation}
where $\hat{V}_W(U,U') \equiv V_W(U) \delta(U,U')$ and the imaginary-time Wilson transfer matrix is given similarly by
\begin{equation}
    \teucl{W} = e^{-a\hat{V}_W/2} e^{-a\hat{K}_W} e^{-a\hat{V}_W/2}. \label{eq:TWhat}
\end{equation}
The relationship between $\tmink{HFK}$ and $\teucl{W}$ in Eqs.~\eqref{eq:WHFKhat}--\eqref{eq:TWhat} is the expected analytic continuation relating discretized real- and imaginary-time evolution operators. Proving the existence of the continuum limit for LGT in either real or imaginary time is outside the scope of this work, but from the forms of $\tmink{HFK}$ and $\teucl{W}$ in Eqs.~\eqref{eq:WHFKhat}--\eqref{eq:TWhat} and the Lie-Trotter product formula one can expect that $\teucl{W}$ and $\tmink{HFK}$ will satisfy the commutative diagram in Fig.~\ref{fig:limits} in the continuous time limit.
The HFK action therefore provides a theoretically suitable action for real-time LGT.

A practical complication associated with the HFK action is that the character expansion in the definition of the action does not provide an absolutely convergent function of the gauge field $U$ across the entire group domain.
In particular, the constraint $|c^{M,HFK}_r(g^2)|=1$ and the fact that each character is normalized by $\int dU |\chi_r(U)|^2 = 1$ and $\chi_r(U)$ cannot vanish for all $U \in G$ implies that the $r$-th term in the sum does not vanish as $r\rightarrow \infty$ using any enumeration of the representations of $G$, at least for a set of non-zero measure in $G$.\footnote{ Stronger statements can be proven for particular choices of the gauge group. For $G=U(1)$ the characters $\chi_r(U) = e^{i r \phi}$ satisfy $|\chi_r(U)|=1$ and therefore the sum in Eq.~\eqref{eq:SMHFK} diverges for all $U$. For $G=SU(2)$, there are combinations of $r$ and $U$ satisfying $\chi_r(U) = 0$; however, the Weyl character formula can be used to explicitly show that $\lim_{r\rightarrow \infty} \chi_r(U)$ is oscillatory and not equal to zero for any $U$ and therefore that the sum in Eq.~\eqref{eq:SMHFK} diverges for all $U$. The  explicit $SU(3)$ character formula in Ref.~\cite{Baaquie_1988} can be used to analogously prove that $\lim_{p,q\rightarrow \infty} \chi_{(p,q)}(U)$  is oscillatory and not equal to zero for any $U$ where the irreps are enumerated using $p, q \in \mathbb{Z}$.  }
Since $|c^M_r(g^2)|=1$ is also precisely the condition required for unitarity of a real-time LGT transfer matrix, it is clear that this divergence is a generic feature of the character expansions for unitary real-time LGT actions.
A simple proof that this character expansion diverges for any real-time LGT action with kinetic and potential energy densities that only depend locally on timelike and spacelike plaquettes, respectively, is presented in Appendix~\ref{sec:divergent-actions}.

Path integral contour deformations can be used to improve this convergence as discussed in Sec.~\ref{sec:contour}, where a simple contour deformation is obtained for which the HFK action for $U(1)$ LGT is represented by an absolutely convergent character expansion.
Obtaining an analogous contour deformation providing an absolutely convergent representation of the $SU(N)$ HFK action suitable for numerical calculations is challenging.
For this reason, alternate real-time LGT actions are introduced below for which the kinetic-energy evolution operator takes a simpler form for both $G=U(1)$ and $G=SU(N)$, and a contour deformation that leads to an absolutely convergent path integral representation can be constructed in Sec.~\ref{sec:contour}.

\subsection{The imaginary-time heat-kernel action}\label{sec:heatkernelE}

The central role of the kinetic-energy evolution operator in establishing unitarity of the real-time transfer matrix suggests that it may be easier to construct unitary real-time LGT actions in the eigenbasis of the LGT kinetic energy operator.
A kinetic energy operator that generalizes the Laplacian operator for non-compact coordinates to compact variables including the quantum rotator as well as $SU(N)$ gauge fields was used by Kogut and Susskind to construct a lattice gauge theory Hamiltonian~\cite{Kogut:1974ag},
\begin{equation}
\begin{split}
    \hat{H} &= -\sum_{\vec{x},k}\left[ \frac{g^2}{2a} \hat{\Delta}_{\vec{x},k} \right] + V_W(\hat{U}) \\
    &\equiv \hat{K} + \hat{V}_W(\hat{U}),
    \end{split}
    \label{eq:Hdef}
\end{equation}
which includes the same potential energy term as the Euclidean Wilson action but a different kinetic energy term that is expected to be equivalent to the Wilson action kinetic term in the continuous-time limit.
Defining operators $\hat{L}^A_{\vec{x},k}$ by the commutation relation
\begin{equation}
    [\hat{L}^A_{\vec{x},k},\hat{U}_{\vec{x},k}] = t^A \hat{U}_{\vec{x},k},
\end{equation}
where the $t^A$ are Hermitian generators of $\mathfrak{g}$ normalized by $\Tr(t^A t^B) = \frac{1}{2}\delta^{AB}$, and the operator $\hat{\Delta}$ appearing in Eq.~\eqref{eq:Hdef} is defined by
\begin{equation}
    \hat{\Delta}_{\vec{x},k} = -\sum_A \hat{L}^A_{\vec{x},k}\hat{L}^A_{\vec{x},k}.
\end{equation}
This can be recognized as the quadratic Casimir operator (up to a sign) and for gauge groups $U(1)$ and $SU(N)$ acts on functions of the gauge field by $\hat{\Delta}_{\vec{x},k}f(U_{\vec{x},k}) = \Delta_{\vec{x},k} f(U_{\vec{x},k})$ where $\Delta_{\vec{x},k}$ is the Laplace-Beltrami differential operator as shown in Ref.~\cite{Menotti:1981ry}; see also Refs.~\cite{Dowker:1970vu,Montvay:1994cy}.
Denoting the character expansion for $f(U_{\vec{x},k})$ by $f(U_{\vec{x},k}) = \sum_r d_r f_r \chi_r(U_{\vec{x},k})$, the action of the Laplace-Beltrami operator on $f(U_{\vec{x},k})$ is given by
\begin{equation}
    \Delta_{\vec{x},k} f(U_{\vec{x},k}) = -\sum_r d_r C_r^{(2)} f_r \chi_r(U_{\vec{x},k}), \label{eq:Deltachar}
\end{equation}
where $C_r^{(2)}$ is the eigenvalue of the quadratic Casimir operator for representation $r$ of $G$.

Ref.~\cite{Menotti:1981ry} constructs eigenfunctions of the kinetic-energy time-evolution operator $e^{-a\hat{K}}$ by solving the diffusion or ``heat-kernel'' equation obtained by analytic continuation of the Schr{\"o}dinger equation $i \partial_t f(U)= \hat{K} f(U)$ to imaginary-time $\partial_{\tau} f(U)= - \hat{K} f(U)$.
Omitting the prefactor in Eq.~\eqref{eq:Hdef} for simplicity, the heat-kernel solution $\mathcal{K}_E(U,\tau)$ is defined as the solution to the differential equation 
\begin{equation}
    \partial_{\tau} \mathcal{K}_E(U,\tau) = \Delta \mathcal{K}_E(U,\tau),
    \label{eq:HK}
\end{equation}
with boundary condition $\mathcal{K}_E(U,0)=\delta(U,1)$. 
The heat-kernel solution provides an integral kernel for $\hat{\Delta}$,
\begin{equation}
\begin{split}
    \mbraket{U_{\vec{x},k} \otimes 1}{e^{\tau \hat{\Delta}_{\vec{x},k}}}{U'_{\vec{x},k} \otimes 1} &= \mbraket{1}{e^{\tau \hat{\Delta}_{\vec{x},k}}}{U'_{\vec{x},k} U^\dagger_{\vec{x},k} \otimes 1} \\
    &= \mathcal{K}_E\left( U'_{\vec{x},k} U^\dagger_{\vec{x},k}, \tau\right),
    \end{split}
\end{equation} 
where the state $\ket{U_{\vec{x},k} \otimes 1}$ is defined by assigning $U_{\vec{x},k}$ to link $(\vec{x},\vec{x}+\hat{k})$ and the identity to all other links and the first equality follows from commutativity of the Laplace-Beltrami operator with group multiplication~\cite{Menotti:1981ry}.
An integral representation for the (temporal-gauge) Kogut-Susskind kinetic-energy time-evolution operator is therefore given by
\begin{equation}
    \mbraket{U}{e^{-a\hat{K}}}{U'} = \prod_{\vec{x},k} \mathcal{K}_E\left( U'_{\vec{x},k} U^\dagger_{\vec{x},k}, \frac{g^2}{2} \right). \label{eq:HKEdef}
\end{equation}
The factor $U_{\vec{x},k}' U_{\vec{x},k}^\dagger $ can be identified with $P_{\vec{x},0k}$ in temporal gauge in order to construct a gauge-invariant expression. In Euclidean spacetime it is convenient to identify the kinetic and potential energy terms as identical functions of timelike and spacelike plaquettes respectively in order to obtain a LGT with $D$-dimensional rather than $(D-1)$-dimensional hypercubic symmetry. Such an isotropic Euclidean heat-kernel action is defined by~\cite{Menotti:1981ry} 
\begin{equation}
    e^{-S_{E,HK}(U)} = \prod_{x,\mu < \nu} \mathcal{K}_E\left( P_{x,\mu\nu}, \frac{g^2}{2} \right),
    \label{eq:SHKdef}
\end{equation}
which is gauge-invariant by the gauge-invariance of the Laplace-Beltrami operator (as well as more explicit arguments below).
The kinetic-energy piece of the transfer matrix associated with the heat-kernel action is given by construction as $\tkeucl{HK} = e^{-a \hat{K}}$.
Its character expansion in temporal gauge can be obtained from Eq.~\eqref{eq:Deltachar} and Eq.~\eqref{eq:HKEdef} as 
\begin{equation}
    \tkeucl{HK}(U,U') = \prod_{\vec{x},k} \left[\sum_r d_r e^{-g^2 C_r^{(2)}/2} \chi_r \left( U'_{\vec{x},k} U^\dagger_{\vec{x},k} \right) \right], \label{eq:EHKchar}
\end{equation} 
from which the character expansion coefficients can be identified as
\begin{equation}
\begin{split}
    c^{E,HK}_r(g^2) &=  e^{-g^2 C_r^{(2)}/2}.
    \label{eq:erHK}
    \end{split}
\end{equation}
Positivity of the quadratic Casimir eigenvalues $C_r^{(2)}$ gives $c^{E,HK}_r(g^2) > 0$, from which the positivity of $\tkeucl{HK}$ and $\teucl{HK}$ and formal existence of a unitary time-evolution operator $\hat{U}_{HK} = [\teucl{HK}]^i$ follows. 

The explicit construction of the Euclidean heat-kernel action requires an explicit form for $\mathcal{K}(U,\tau)$ for particular $G$.
For the non-compact group $G = \mathbb{R}$ the heat-kernel equation reduces to the usual diffusion equation 
\begin{equation}
\partial_\tau \mathcal{K}_{E,\mathbb{R}}(x,\tau) = -\frac{1}{2}\frac{\partial^2}{\partial x^2} \mathcal{K}_{E,\mathbb{R}}(x,\tau),
\end{equation}
which has the well-known solution
\begin{equation}
    \mathcal{K}_{E,\mathbb{R}}(x,\tau) = \mathcal{N} \exp\left[-\frac{x^2}{2\tau}\right],
    \label{eq:KR}
\end{equation}
with the normalization constant $\mathcal{N} = 1/\sqrt{2\pi \tau}$ fixed by the boundary condition $\mathcal{K}_{E,\mathbb{R}}(x,0) = \delta(x)$ at $\tau = 0$ and the evolution equation in Eq.~\eqref{eq:HK} for later $\tau$. In the heat-kernel solutions derived for various groups below, normalizing constants are not distinguished for conciseness; distinct normalizing constants apply to each case with the appropriate normalization clear from context.
For $G = U(1)$, the heat-kernel equation takes an analogous form
\begin{equation}
\partial_\tau \mathcal{K}_{E,U(1)}(e^{i\phi},\tau) = -\frac{1}{2}\frac{\partial^2}{\partial \phi^2} \mathcal{K}_{E,U(1)}(e^{i\phi},\tau),
\end{equation}
and has a solution given by a sum of Gaussian terms of the form in Eq.~\eqref{eq:KR} over coordinates $x = \phi + 2\pi n$ for all possible integers $n$.
Noting that the appropriate coupling constant normalization is given from Eq.~\eqref{eq:SHKdef} by $\tau = e^2$, the form of the $U(1)$ heat-kernel required for Euclidean LGT calculations is given by
\begin{equation}
\begin{split}
    \mathcal{K}_{E,U(1)}(e^{i\phi},e^2) &= \mathcal{N} \sum_{n=-\infty}^\infty  \exp\left[-\frac{1}{2e^2} (\phi + 2\pi n)^2 \right],
    \label{eq:KEU1}
    \end{split}
\end{equation}
with the normalization $\mathcal{N} = 1/\sqrt{2\pi e^2}$ fixed by $\mathcal{K}_{E,U(1)}(U,0) = \delta(U,1)$. This heat kernel weight corresponds to the Villain action~\cite{Villain:1974ir}.
For $G=SU(N)$, additional factors arise in coordinate descriptions of the heat-kernel equation from the non-trivial metric of the Riemannian manifold associated with $G$.
The construction of the heat-kernel for this case is detailed in Ref.~\cite{Menotti:1981ry}, and the solution is given in terms of the eigenvalues $e^{i\phi_1},\ldots,e^{i\phi_N}$ of $U$ by
\begin{widetext}
\begin{equation}
\begin{split}
    \mathcal{K}_{E,SU(N)}\left(U,\frac{g^2}{2}\right) &= \sum_{\{n\}} \mathcal{N}\, \mathcal{J}(\{\phi\},\{n\})\, \exp\left[-\frac{1}{g^2} \sum_A (\phi^A + 2\pi n^A)^2\right], \\
    \mathcal{J}(\{\phi\},\{n\}) &=  \prod_{A < B} \left( \frac{\phi^A - \phi^B + 2\pi(n^A - n^B)}{2\sin\left[ \frac{1}{2} \left( \phi^A - \phi^B + 2\pi(n^A - n^B) \right) \right]} \right) ,
\end{split}
\label{eq:KESUN}
\end{equation}
 where $\sum_{\{n\}}$ describes a sum of infinite sums $\sum_{n^A = -\infty}^{\infty}$ and $\mathcal{N}$ is fixed by $\mathcal{K}_{E,SU(N)}(U,0) = \delta(U,1)$ as in the $U(1)$ case. The integers $n^A, A=1,\ldots,N$ are subject to a constraint $\sum_A n^A = 0$ analogous to the eigenvalue phase constraint $\sum_A \phi^A = 0$ that ensures that $\det(U) = 1$.
Note that $\phi^A + 2\pi n^A$ is treated as a non-compact variable in the heat-kernel action in the sense that  $\sum_A \phi^A = 0$ and $\sum_A n^A = 0$ are enforced rather than $\sum_A \phi^A = 0 \mod 2\pi$~\cite{Menotti:1981ry}.
Eq.~\eqref{eq:KESUN} is invariant to permutation of eigenvalues $e^{i\phi_1}, \dots, e^{\phi_N}$ ensuring that the kernel does not depend on the (unphysical) ordering of eigenvalues. Gauge transformations act by matrix conjugation on $P_{x,\mu\nu}$ and therefore do not affect the unordered set of eigenvalues $e^{i\phi_1}, \dots, e^{i \phi_N}$, leaving Eq.~\eqref{eq:KESUN} invariant.

\end{widetext}

\subsection{The real-time heat-kernel action}\label{sec:heatkernelM}

The analog of the heat-kernel equation that describes the real-time evolution of $U \in G$  in the absence of a potential is the free Schr{\"o}dinger equation on $G$,
\begin{equation}
    i \partial_{t} \mathcal{K}_M(U,t) = - \Delta \mathcal{K}_M(U,t).
    \label{eq:HKi}
\end{equation}
This Schr{\"o}dinger equation is related to Eq.~\eqref{eq:HK} through analytic continuation using the identification $\tau = it$.
The solution to the  Schr{\"o}dinger equation  in Eq.~\eqref{eq:HKi} can therefore be immediately obtained through analytic continuation of the solution to the Euclidean heat-kernel equation,
\begin{equation}
    \mathcal{K}_M(U,t) = \mbraket{1}{e^{i t \hat{\Delta}_{\vec{x},k}}}{U_{\vec{x},k} \otimes 1} = \mathcal{K}_E(U,i\tau).
\end{equation}
Analytic continuation of Eq.~\eqref{eq:HKEdef} gives
\begin{equation}
\begin{split}
    \mbraket{U}{e^{-i a\hat{K}}}{U'} &= 
    \prod_{\vec{x},k}  \mbraket{1}{e^{ \frac{i g^2}{2} \hat{\Delta}_{\vec{x},k}}}{U'_{\vec{x},k} U^\dagger_{\vec{x},k} \otimes 1} \\
    &= \prod_{\vec{x},k} \mathcal{K}_M\left( U'_{\vec{x},k} U^\dagger_{\vec{x},k}, \frac{g^2}{2} \right).
    \end{split}
\end{equation}
The Minkowski heat-kernel action defined by analytic continuation of Eq.~\eqref{eq:SHKdef},
\begin{equation}
\begin{split}
    e^{iS_{M,HK}} &= \prod_{x,k} \mathcal{K}_M\left( P_{x,0k}, \frac{g^2}{2} \right) \\
    &\hspace{20pt} \times \prod_{x,i < j} \mathcal{K}_M\left( P_{x,ij}, -\frac{g^2}{2} \right),
    \label{eq:SHKMdef}
    \end{split}
\end{equation}
therefore has a kinetic-energy time-evolution operator with integral kernel 
\begin{equation}
\begin{split}
    \tkmink{HK}(U,U') &= \prod_{\vec{x},k} \mathcal{K}_M\left( U_{\vec{x},k}' U_{\vec{x},k}^\dagger , \frac{g^2}{2} \right),
    \label{eq:MHK}
    \end{split}
\end{equation}
which therefore satisfies $\tkmink{HK} = e^{-i a \hat{K}}$.
The unitarity of $\tkmink{HK}$ follows immediately from the Hermiticity of the Laplace-Beltrami operator and $\hat{K}$.

Unitarity can also be directly verified through analytic continuation of the Euclidean heat-kernel character expansion in Eq.~\eqref{eq:EHKchar}, which gives
\begin{equation}
\begin{split}
    \mathcal{K}_M\left(U,\frac{g^2}{2}\right) &= \mathcal{K}_E\left(U,\frac{ig^2}{2}\right) \\
    &= \sum_r d_r \chi_r(U) e^{-i g^2 C_r^{(2)} / 2}.
    \end{split}
\end{equation}
The coefficients $c^{M,HK}_r(g^2)$ of the character expansion of $\tkmink{HK}$ analogous to Eq.~\eqref{eq:mrWilson} are given by
\begin{equation}
\begin{split}
    c^{M,HK}_r(g^2) &= \left( \int \mathcal{D}U \prod_{\vec{x},k}  \left[ \frac{1}{d_r} \chi_r(U_{\vec{x},k})^* \right] \tkmink{HK}(1,U) \right)^{1/N_L} \\
    &=   e^{-i g^2 C_r^{(2)} / 2}.
    \end{split}
    \label{eq:mrHK}
\end{equation}
It follows that $|c^{M,HK}_r(g^2)| = 1$, which implies that $\tkmink{HK}$ is unitary.
Unitarity of $\tmink{}$ follows for any choice of real potential from Eq.~\eqref{eq:WWdag}.
In particular, this implies that the Minkowski heat-kernel action defined in Eq.~\eqref{eq:SHKMdef} leads to a unitarity LGT time-evolution operator $\tmink{HK}$.

Comparing Eq.~\eqref{eq:erHK} and Eq.~\eqref{eq:mrHK}, the Minkowski and Euclidean heat-kernel actions are further seen to satisfy 
\begin{equation}
  c^{M,HK}_r(g^2) = [c^{E,HK}_r(g^2)]^i.
\end{equation}
It follows from this that the real- and imaginary-time kinetic-energy evolution operators associated with the heat-kernel action satisfy $\tkmink{HK} = e^{-i a\hat{K}} =  \tkeucl{HK}^i$.
The corresponding real- and imaginary-time transfer matrices can therefore be expected to satisfy the commutative diagram shown in Fig.~\ref{fig:limits}.
Since the Euclidean heat-kernel and Wilson actions are expected to be equivalent in the continuum limit, the Minkowski heat-kernel action is a unitary real-time LGT action that should be equivalent to the HFK action in the continuum limit.

For the $U(1)$ gauge group, the explicit form of the Minkowski heat-kernel solution is given by
\begin{equation}
    \mathcal{K}_{M,U(1)}(e^{i\phi},e^2) = \mathcal{N}\sum_{n=-\infty}^\infty  \exp\left[\frac{i}{2e^2} (\phi + 2\pi n)^2 \right],
    \label{eq:KMU1}
\end{equation}
where the normalizing constant $\mathcal{N}$ is related to the Euclidean normalizing constant by the analytic continuation $e^2 \rightarrow ie^2$. For the $SU(N)$ gauge group, the Minkowski heat-kernel solution is
\begin{equation}
\begin{split}
    \mathcal{K}_{M,SU(N)}\left(U,\frac{g^2}{2}\right) &= \sum_{\{n\}}  \exp\left[\frac{i}{g^2} \sum_A (\phi^A + 2\pi n^A)^2\right] \\
    &\hspace{20pt} \times \mathcal{N}\, \mathcal{J}(\{\phi\},\{n\}),
    \end{split}
    \label{eq:KMSUN}
\end{equation}
where $ \mathcal{J}(\{\phi\},\{n\})$ is given in Eq.~\eqref{eq:KESUN} and $\{n\}$ again denotes a set of integers $n^A$ with $A = 1,\ldots N$ subject to the constraint $\sum_A n^A = 0$.
The $SU(N)$ normalizing constant is similarly related to the Euclidean case by the analytic continuation $g^2 \rightarrow ig^2$.

The infinite sums in Eqs.~\eqref{eq:KMU1}--\eqref{eq:KMSUN} are divergent since the summand does not vanish in the $n\rightarrow \pm \infty$ limit. This non-convergence of oscillating sums defining the path integral weights mirrors the problems faced in the HFK action. Motivated by a saddle point expansion, a possible approach to this problem is explored in Appendix~\ref{sec:truncated} in which sums over $n^A$ are truncated to $n^A = 0$. This truncation breaks unitarity at non-zero lattice spacing, but the dominance of the $n^A = 0$ terms in the saddle point approximation might suggest that this unitarity breaking is removed in the continuum limit. However, it is shown in the appendix that unitarity is not in fact recovered in the analytically solvable case of $(1+1)$D $U(1)$ LGT, indicating that this truncated version of the heat-kernel approach is not a useful starting point for real-time LGT and underscoring the importance of preserving unitarity in real-time LGT actions.

\subsection{The modified real-time heat-kernel action} \label{sec:heatkernelbarM}

The convergence issues of the HFK and Minkowski heat-kernel actions and the lack of a scaling limit for the truncated heat-kernel action motivate the definition of a real-time modified heat-kernel action that includes the heat-kernel kinetic term and the Wilson potential term,
\begin{equation}
\begin{split}
    e^{iS_{M,\HKbar}(U)} &= \prod_{t} e^{-i a V_W(U_t)}  \prod_{x,k} \mathcal{K}_M\left(P_{x,0k},  \frac{g^2}{2} \right)  \\
    &= \sum_{\{n\}} \prod_{x,k} \mathcal{N}\, \mathcal{J}(\{\phi_{x,0k}\},\{n_{x,k}\}) \\
    &\hspace{20pt} \times e^{-\frac{i}{g^2}\sum_{x,i<j}\Tr(2 - P_{x,ij} - P_{x,ij}^{-1})} \\
    &\hspace{20pt} \times    e^{\frac{i}{g^2} \sum_{x,k,A}  (\phi_{x,0k}^A + 2\pi n_{x,k}^A)^2 } ,
    \end{split}
        \label{eq:SHKbardef}
\end{equation}
where for the $U(1)$ case  $\mathcal{J}(\{\phi_{x,0k}\},\{n_{x,k}\})$ is replaced by unity, and $P_{x,ij}^\dagger = P_{x,ij}^{-1}$ has been used to express Eq.~\eqref{eq:SHKbardef} in a form suitable for discussions of contour deformations in Sec.~\ref{sec:contour}.
Since the kinetic term is the usual heat-kernel one, $\tkmink{\HKbar} = \tkmink{HK} = e^{-i a \hat{K}}$ is unitary and regardless of the different choice of potential in $S_{HK}(U)$ and $S_{\HKbar}(U)$ both are therefore unitary real-time LGT actions.
The real-time transfer matrix associated with $S_{\HKbar}(U)$ is given by
\begin{equation}
\begin{split}
    \tmink{\HKbar} &= e^{-i a \hat{V}_{W}/2} e^{-ia\hat{K}} e^{-i a \hat{V}_{W}/2} ,
    \end{split}
\end{equation}
and is manifestly unitary.
This is the expected form for a discretized real-time evolution operator and by the Lie-Trotter product formula $\tmink{\HKbar}^{t/a}$ should converge to $e^{-it(\hat{K}+\hat{V}_W)}$ in the continuous-time limit.
Since $T_W^{\tau/a}$ is expected to converge to $e^{-\tau(\hat{K}+\hat{V}_W)}$ in the continuous-time limit, $\teucl{W}$ and $\tmink{\HKbar}$ are expected to satisfy the commutative diagram shown in Fig.~\ref{fig:limits}.
It can also be seen at the level of the action that the real-time $\HKbar$ action is equivalent to the (truncated) Minkowski heat-kernel action in the $g^2 \rightarrow 0$ limit where in the stationary phase approximation the Wilson potential approaches $1 - \cos(\phi_{x,ij}^A) \approx \frac{1}{2}(\phi_{x,ij}^A)^2$.
The real-time $\HKbar$ action therefore provides another unitary real-time LGT action with the same naive continuum limit as the Minkowski heat-kernel and HFK actions.

It is noteworthy that all of the Minkowski actions above include at least a sign difference between kinetic and potential terms arising from Eq.~\eqref{eq:SMdef} and are therefore isotropic in $(D-1)$ spatial dimensions but not $D$ dimensions.
There is no subgroup of the Lorentz group that provides a symmetry of any of the Minkowski LGT actions described above, and the real-time $\HKbar$ action shares the same symmetries as the Minkowski heat-kernel and HFK actions.
Although the real-time $\HKbar$ action also shares the same downside as the HFK and Minkowski heat-kernel actions --- a definition involving a divergent sum --- 
it is demonstrated below in Sec.~\ref{sec:contour} that path integral contour deformations can be used to construct an alternative representation of $S_{M,\HKbar}$ in which the sum defining the kinetic energy is absolutely convergent.
This representation of the real-time $\HKbar$ action provides a well-defined starting point for numerical investigations of real-time LGT with unitary time-evolution at non-zero lattice spacing.

\subsection{Exact results in (1+1)D}\label{sec:exact}

In $(1+1)$D with open boundary conditions (OBCs), both the Wilson and heat-kernel Euclidean actions do not include potential terms and $SU(N)$ gauge theory is analytically solvable ~\cite{Gross:1980he,Wadia:2012fr,Menotti:1981ry}.
These solutions can be straightforwardly analytically continued in order to compare results for observables constructed using Minkowski actions leading to non-unitary and unitary time evolution.

A simple, non-trivial observable in $(1+1)$D is the Wilson loop,
\begin{equation}
   \begin{split}
      \wloop_{\mathcal{A}} = \frac{1}{N} \Tr \left( \prod_{x,\mu \in \partial \mathcal{A}} U_{x,\mu} \right),
   \end{split}\label{eq:Wdef}
\end{equation}
where $\prod_{x,\mu \in \partial \mathcal{A}} U_{x,\mu}$ denotes an ordered product
along the boundary of the two-dimensional rectangular region $\mathcal{A}$ with spatial extent $L$ and temporal extent denoted $\tau$ in Euclidean spacetime and $t$ in Minkowski spacetime.
Wilson loops can be interpreted as propagators for static quark-antiquark pairs separated by a distance $L$. In Euclidean spacetime, for a lattice gauge theory with positive-definite transfer matrix, Wilson loops satisfy the spectral representation
\begin{equation}
    \left< \wloop_{\mathcal{A}} \right>_E = \sum_n |Z_n|^2\ e^{-\tau E_n  },
    \label{eq:Espec}
\end{equation}
where $E_n > 0$ is the energy of the $n$-th energy eigenstate with appropriate quantum numbers for describing the static quark-antiquark system and $Z_n$ is the overlap factor onto the $n$-th energy eigenstate.
An ideal spectral representation for the corresponding Minkowski theory is obtained by analytic continuation with $\tau = it$,
\begin{equation}
    \left< \wloop_{\mathcal{A}} \right>_M = \sum_n |Z_n|^2\ e^{- i t E_n  }.
    \label{eq:Mspec}
\end{equation}
A suitable real-time LGT action should give rise to a spectral representation of the form Eq.~\eqref{eq:Mspec} with energies and overlap factors that may differ at non-zero lattice spacing but should agree in the continuum limit for states with energies much below the lattice cutoff.

With a local action $S_E = \sum_x \mathcal{L}_E(P_{x})$ that depends only on the plaquettes $P_x \equiv P_{x,01}$, Wilson loop expectation values in $(1+1)$D with OBCs take a simple factorized form~\cite{Gross:1980he,Wadia:2012fr} 
\begin{equation}
  \begin{split}
    \left< \wloop_{\mathcal{A}} \right>_{E} &= \left[ \frac{ \frac{1}{N} \int \mathcal{D}P \ \Tr(P) \ e^{-\mathcal{L}_E(P)} }{ \int \mathcal{D}P \ e^{-\mathcal{L}_E(P)} } \right]^{\tau L}.
    \label{eq:WEfac}
  \end{split}
\end{equation} 
This can be simplified using the character expansion of $e^{-\mathcal{L}_E(P)}$,
\begin{equation}
  \begin{split}
    \left< \wloop_{\mathcal{A}} \right>_{E} &= \left[ \frac{ \frac{1}{N} \int \mathcal{D}P \ \Tr(P) \ \sum_r d_r c^E_r(g^2) \chi_r(P) }{ \int \mathcal{D}P \ \sum_r d_r c^E_r(g^2) \chi_r(P) } \right]^{\tau L} \\
    &= \left[ \frac{ c_{\overline{f}}^{E}(g^2) }{ c_{0}^{E}(g^2) } \right]^{\tau L},
    \label{eq:WEchar}
  \end{split}
\end{equation}
where $r=\overline{f}$ denotes the antifundamental representation and has dimension $d_{\overline{f}} = N$.
The condition $c^E_r(g^2) > 0$ required for positivity of $\teucl{}$ guarantees that this can be expressed as
\begin{equation}
  \left< \wloop_{\mathcal{A}} \right>_{E} = e^{-L \tau \sigma }, \hspace{20pt} \sigma = \ln\left( \frac{ c^E_0(g^2) }{ c^E_{\overline{f}}(g^2) } \right),
\end{equation}
which is the expected spectral representation for a single state with energy $E_0 = \sigma L$.
An analogous relation holds in Minkowski spacetime,
\begin{equation}
  \begin{split}
    \left< \wloop_{\mathcal{A}} \right>_{M} &= \left[ \frac{ \int \mathcal{D}P \ \Tr(P) \ e^{i\mathcal{L}_M(P)} }{ \int \mathcal{D}P \ e^{i\mathcal{L}_M(P)} } \right]^{t L} \\
    &= \left[ \frac{ \frac{1}{N} \int \mathcal{D}P \ \Tr(P) \ \sum_r d_r c^M_r(g^2) \chi_r(P) }{ \int \mathcal{D}P \ \sum_r d_r c^M_r(g^2) \chi_r(P) } \right]^{t L} \\
    &= \left[ \frac{ c_{\overline{f}}^{M}(g^2) }{ c_{0}^{M}(g^2) } \right]^{t L}.
    \label{eq:WMchar}
  \end{split}
\end{equation}
The condition $c^M_r(g^2) = [c^E_r(g^2)]^i$ is sufficient for Eq.~\eqref{eq:WMchar} to assume the form of the desired Minkowski spectral representation Eq.~\eqref{eq:Mspec} with the same energy $E_0 = \sigma L$ as in the Euclidean case.
This condition is satisfied for the HFK and real-time (modified) heat-kernel actions; however, it is not satisfied for the real-time Wilson action for which the character expansion coefficients in real and imaginary time are given in Eq.~\eqref{eq:mrWilson} and Eq.~\eqref{eq:erWilson} and satisfy $c^{M,W}_r(g^2) = c^{E,W}_r(ig^2) \neq [c^{E,W}_r(g^2)]^i$.

\begin{figure}
    \centering
    \includegraphics[width=\columnwidth]{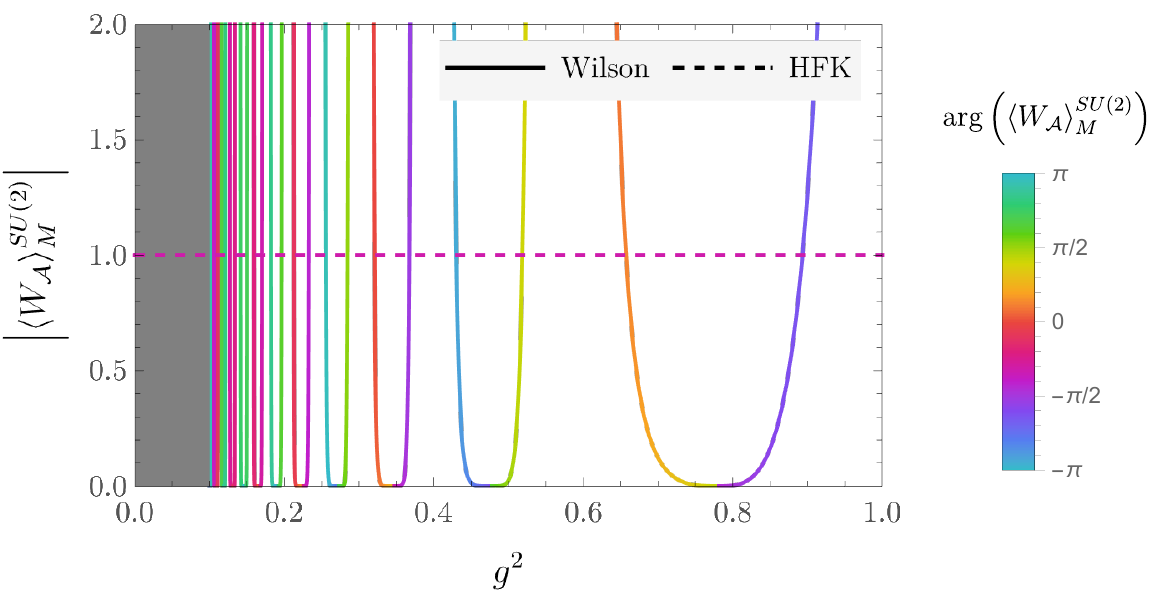}
    \caption{Analytically computed value of the Wilson loop $\wloop_{\mathcal{A}}$ in $SU(2)$ real-time LGT in $(1+1)$D as a function of $g^2$ with the area fixed to one in units of the (Euclidean) string tension, $A = 1/\sigma_{E,W}^{SU(2)}(g^2)$. The solid (dashed) lines show the magnitude of the Wilson loop expectation value computed using the real-time Wilson action (HFK action) with the corresponding phases of the Wilson loop expectation  value shown with the color of the corresponding line. The Wilson action result includes an infinite number of singularities that accumulate as $g^2 \rightarrow 0$ and is replaced by a gray background for $g^2 < 0.1$.
    \label{fig:wilson-action-wilson-loop} }
\end{figure}

For the gauge groups $U(1)$ and $SU(2)$, analytic results for $c^{M,W}_r(g^2)$ in Eqs.~\eqref{eq:mrWilsonU1}--\eqref{eq:mrWilsonSU2} and the relation $c^{E,W}_r(g^2) = c^{M,W}_r(-ig^2)$ can be used to derive the explicit forms of the character expansion coefficients appearing in these $(1+1)$D Wilson loops.
The string tension for the Euclidean Wilson action is given by
\begin{equation}
    \sigma_{E,W}^{SU(2)} = \ln \left( \frac{I_1(4/g^2)}{I_2(4/g^2)} \right),
\end{equation}
while the real-time Wilson action expectation value becomes
\begin{equation}
    \left< \wloop_{\mathcal{A}} \right>_{M,W}^{SU(2)} = \left( \frac{I_1(-4i/g^2)}{I_2(-4i/g^2)} \right)^{-Lt}.
    \label{eq:WMWilson}
\end{equation}
This does not correspond to a real-time spectral representation of the form Eq.~\eqref{eq:Mspec} for any value of $g^2$.
In particular $\lim_{g^2 \rightarrow 0} \left< \wloop_{\mathcal{A}} \right>_{M,W}^{SU(2)}$ does not exist, as shown in Fig.~\ref{fig:wilson-action-wilson-loop}.
This demonstrates that even though the $g^2 \rightarrow 0$ limit of the real-time transfer matrix exists in the $SU(2)$ case, well-defined continuum limits for real-time observables do not necessarily exist.
Similar results can be obtained for $SU(N)$ gauge groups in $(1+1)$D, and again the lack of a continuum limit where the real-time transfer matrix becomes unitary is associated with the lack of a well-defined continuum limit for Wilson loop expectation values.
Conversely, the HFK action by definition has a character expansion matching the Wick rotated Wilson action, $c^{M,HFK}_r(g^2) = [c^{E,W}_r(g^2)]^i$, and therefore
\begin{equation}
  \begin{split}
    \left< \wloop_{\mathcal{A}} \right>_{M,HFK}^{SU(2)} &= \left( \frac{I_1(4/g^2)}{I_2(4/g^2)} \right)^{-iLt} \\
    &= e^{-iLt \sigma_{E,W}^{SU(2)} },
    \label{eq:WMHFK}
  \end{split}
\end{equation}
demonstrating that the HFK action leads to unitary results that correspond to the $\tau = i t$ analytic continuation of the corresponding Euclidean Wilson action LGT results.
In higher dimensions where LGT potentials are non-trivial, this exact $t = i\tau$ correspondence will not hold in LGT but should emerge in the continuum limit.

The Euclidean heat-kernel action leads to a different form for the $(1+1)$D LGT string tension~\cite{Menotti:1981ry}
\begin{equation}
     \sigma_{E,HK}^{SU(N)} = \frac{1}{4}g^2 N \left( 1 - \frac{1}{N^2} \right).
\end{equation}
It can be explicitly seen that
\begin{equation}
    \lim_{g^2 \rightarrow 0} \sigma_{E,HK}^{SU(2)} / \sigma_{E,W}^{SU(2)} = 1
\end{equation} 
verifying that the heat-kernel action and Wilson action results agree in the continuum limit.
The heat-kernel action satisfies $c^{M,HK}_r(g^2) = [c^{E,HK}_r(g^2)]^i$, and it therefore follows from Eqs.~\eqref{eq:WEchar}--\eqref{eq:WMchar} that
\begin{equation}
    \left< \wloop_{\mathcal{A}} \right>_{M,HK} = e^{- i L t \sigma_{E,HK} },
    \label{eq:HKMspec}
\end{equation}
demonstrating that $ \left< \wloop_{\mathcal{A}} \right>_{M,HK}$  has the expected spectral representation associated with unitary time evolution in LGT.
This demonstrates that lattice artifacts leading to differences between the energies appearing in the Minkowski and Euclidean spectral representations are also absent for the heat-kernel action in $(1+1)$D, a feature which does not persist in higher dimensional LGT where potential operators are present.
The same result can be derived directly by inserting the character expansion coefficients from Eq.~\eqref{eq:mrHK} into Eq.~\eqref{eq:WMchar} and using $C_{\overline{f}}^{(2)} = \frac{N}{2}(1 - 1/N^2)$.
In the $U(1)$ case $C_{\overline{f}}^{(2)} = 1$ and it can be derived analogously that Eq.~\eqref{eq:HKMspec} holds with
\begin{equation}
     \sigma_{HK}^{U(1)} = \frac{e^2}{2}.
\end{equation}

The real-time $\HKbar$ action $S_{M,\HKbar}$ defined in Eq.~\eqref{eq:SHKbardef} uses the Wilson potential and heat-kernel kinetic terms.
Since there is no potential term present for LGT in $(1+1)$D, $S_{M,\HKbar}$  coincides with the Minkowski heat-kernel action $S_{M,HK}$  and leads to Wilson loop results identical to Eq.~\eqref{eq:HKMspec}.
In higher dimensions $S_{M,\HKbar}$ will not coincide with the $S_{M,HK}$ exactly, but $S_{M,\HKbar}$ still leads to a unitary time-evolution operator and therefore Wilson loop expectation values with a unitary spectral representation.
``Lattice artifacts'' take the form of modifications to the energy spectrum and overlap factors that vanish as $g^2 \rightarrow 0$ and are not expected to spoil the scaling properties of the continuum limit (unlike the non-unitary truncated heat-kernel action investigated in Appendix~\ref{sec:truncated}).

\section{Real-time LGT path integral contour deformation}\label{sec:contour}

Real-time LGT path integrals in more that two dimensions cannot be calculated analytically with current techniques.
In order to perform Monte Carlo calculations of real-time LGT path integrals, convergent representations of path integrands involving the real-time actions under study must be constructed.
Path integral contour deformation techniques previously used to tame sign problems are used here to construct such convergent representations of real-time LGT actions.
Path integral contour deformation techniques are reviewed in Sec.~\ref{sec:contourGen}.
These techniques are then applied to construct absolutely convergent representations of path integrals involving the HFK action for $U(1)$ real-time LGT in Sec.~\ref{sec:HFKconvergence} and the modified heat-kernel action for $U(1)$ and $SU(N)$ gauge theory in Sec.~\ref{sec:HKconvergence}.
An absolutely convergent representation of Schwinger-Keldysh path integrals using the modified heat-kernel action is discussed in Sec.~\ref{sec:SK}.
The absolutely convergent representations of unitary real-time LGT actions are applied in proof-of-principle Monte Carlo calculations of $U(1)$ real-time LGT in Sec.~\ref{sec:U1MC} and of $SU(3)$ real-time LGT in Sec.~\ref{sec:SU3MC}.

\subsection{Sign problems and path integral contour deformations in real time}\label{sec:contourGen}

For a compact Lie group the absolute value of the Minkowski path integral weight and measure,
\begin{equation}
    |e^{iS_M(U)}|\mathcal{D}U = \mathcal{D}U,
\end{equation}
provides a well-defined probability measure for performing Monte Carlo sampling, unlike in the case of non-compact scalar fields in real time discussed in Refs.~\cite{Tanizaki:2014xba,Alexandru:2016gsd,Alexandru:2017lqr,Mou:2019tck,Lawrence:2021izu}. 
However, the ``reweighting'' approach defined by sampling with respect to $\mathcal{D}U$ generically leads to an exponential sign-to-noise (StN) problem: the variance of Monte Carlo estimates of the average phase factor $e^{iS_M}$ required to determine the full partition function grows exponentially as the size of the system is increased.
Standard arguments for sign problems in Euclidean spacetime compare the ``phase-quenched'' partition function, defined by the integral of the absolute value of the path integrand, to the full partition function; if ignoring phase fluctuations leads to a free energy $f_Q$ distinct from the free energy $f$ in the full theory, then the variance of estimates of the partition function will be given for large volumes by $e^{-2f_Q V \beta} - e^{-2f V \beta}$ and positivity of the variance requires $f_Q \leq f$.
The signal-to-noise (StN) ratio associated with computing the ratio of phase-quenched to full partition function will therefore scale as $e^{-(f-f_Q)V \beta}$ and will decrease exponentially with increasing Euclidean spacetime volume~\cite{Gibbs:1986ut,Cohen:2003ut,Cohen:2003kd,Splittorff:2006fu,Splittorff:2007ck}.

Analogous arguments can be extended to Schwinger-Keldysh correlation functions~\cite{Lawrence:2021izu}.
For the case of Schwinger-Keldysh LGT in particular, the phase-quenched theories describing the Minkowski segments are defined by trivial path integral weights and can be given a thermodynamic interpretation as a Euclidean action that is exactly zero, corresponding to the strong coupling limit.
The phase-quenched Schwinger-Keldysh partition functions will thus scale with volume as $e^{-f' V L_T^E} e^{-f_Q' V L_T^M}$, where $L_T^M$ is the total length of the time contour with Minkowski signature, $L_T^E$ is the length of the time contour with Euclidean signature, and $f'$ and $f'_Q$ are the relevant free energies in the Euclidean and phase-quenched Minkowski regions. The corresponding full partition function is equal to $e^{-f' V L_T^E}$, where there is no dependence on $L_T^M$ because the amplitudes from forward and reverse Minkowski time evolution cancel by unitarity.
Positivity of the variance requires $f_Q' \leq 0$, and Schwinger-Keldysh LGT path integrals therefore face exponentially severe StN problems with StN ratios proportional to $e^{-|f_Q'| V L_T^M}$.
Analogous StN problems can be expected for purely Minkowski path integrals, although the details in this case will depend on the temporal boundary conditions.

Numerical calculations using the unitary Minkowski actions studied above --- the HFK, real-time heat-kernel (HK), and modified real-time heat-kernel ($\HKbar$) actions --- face the additional challenge that in all cases $e^{iS_M(U)}$ is formally defined by an infinite sum that does not converge for fixed $U$ and its average over the distribution $\mathcal{D}U$ cannot be calculated using Monte Carlo methods even in principle.
In the $(1+1)$D examples discussed above, where exact results are available, it is clear that performing the gauge field integral before the infinite sums provides convergent results that match the desired spectral representations obtained by
analytic continuation of Euclidean results.
One could imagine ordering the summation outside integration and explicitly performing Monte Carlo integration over $e^{i S_M(U)}$ for each combination of 
terms in the HFK character expansion or integers in the heat-kernel sum below a specific cutoff,
but the need to perform $O(V L_T)$ Monte Carlo calculations and the systematic uncertainties in the combined result arising from truncating the infinite sums make this approach undesirable.
If the combined sum-integral defining the path integral over $e^{iS_M}$ for a unitary action could be rendered absolutely convergent, it would instead be possible to
perform the sums and integrals together in a joint Monte Carlo calculation.
This is achieved below using path integral contour deformation methods.

The foundation of path integral contour deformations is the complex analysis result that for holomorphic integrands $\mathcal{O} e^{iS_M}$, the integration contour of the path integral can be deformed in order to affect the StN properties of the integral without modifying the total integral value.
Previously, path integral contour deformations have been used to improve the sign and associated StN problems affecting real-time $(0+1)$D quantum mechanics models~\cite{Tanizaki:2014xba,Alexandru:2016gsd,Alexandru:2017lqr,Mou:2019tck,Lawrence:2021izu}, as well as imaginary-time theories of scalars and fermions~\cite{Cristoforetti:2012su,Cristoforetti:2013wha,Fujii:2013sra,Aarts:2013fpa,Tanizaki:2014tua,Mukherjee:2014hsa,Cristoforetti:2014gsa,Kanazawa:2014qma,Fujii:2015vha,Fujii:2015bua,Tanizaki:2015rda,Alexandru:2015sua,Alexandru:2015xva,Alexandru:2016ejd,Alexandru:2017czx,Ulybyshev:2017hbs,Mori:2017zyl,Alexandru:2017oyw,Alexandru:2018fqp,Alexandru:2018ddf,Ohnishi:2018jjw,Ulybyshev:2019fte,Fukuma:2019uot,Fukuma:2019wbv,Wynen:2020uzx}, $U(1)$ gauge theory~\cite{Mukherjee:2013aga,Alexandru:2018ngw,Detmold:2020ncp,Kashiwa:2020brj,Pawlowski:2021bbu,Pawlowski:2020kok}, dimensionally reduced (single- or few-variable) non-Abelian gauge theory~\cite{Tanizaki:2015pua,Schmidt:2017gvu,Ohnishi:2018jjw,Zambello:2018ibq,Bluecher:2018sgj,Kashiwa:2019lkv,Mori:2019tux}, and recently large Wilson loops in $(1+1)$D $SU(N)$ gauge theory~\cite{Detmold:2021ulb}.
By modifying the integrand magnitude and phase, contour deformations also have the potential to improve the convergence problems highlighted above.

A lattice gauge theory path integral can be interpreted as an iterated integral over a set of compact, group-valued variables. For the $U(1)$ gauge group, these integrals can be written in terms of one angular variable $\phi \in [0,2\pi]$ per $U(1)$ gauge link. For $SU(N)$ gauge groups, deformations can be defined in terms of an angular parameterization of each $SU(N)$ gauge link, given by a set $\Omega = \{\phi^a,\theta^b\}$ of azimuthal angles $\phi^a \in [0,2\pi]$ and zenith angles $\theta^b \in [0,\pi/2]$. To be a valid contour deformation, endpoints must be handled properly for both $\phi$ and $\theta$ angles: for periodic $\phi$ angles (appearing in both the $U(1)$ and $SU(N)$ parameterizations) any deformation that keeps the endpoints identified will be valid, while for non-periodic $\theta$ angles the endpoints must be held fixed. Further details on angular parameters and deformations of $SU(N)$ variables can be found in Ref.~\cite{Detmold:2021ulb}. The Haar measure appearing in the path integral can be related to the natural measure on the relevant angular coordinates by $\mathcal{D}U = \prod_a d\phi^a \prod_b d\theta^b\ H(\Omega) = \mathcal{D}\Omega\ H(\Omega)$, where $H(\Omega)$ can be straightforwardly computed for particular angular parameterizations~\cite{Detmold:2021ulb,Bronzan:1988wa}.

After rewriting the path integral in terms of the chosen coordinates, i.e.~replacing the measure as above and replacing instances of $\mathcal{U}$ with $\mathcal{U}(\Omega)$,
contour deformation can be directly applied to the compact integration paths of each real-valued angular variable, potentially conditioned on other angular variables.
For a valid deformation $\Omega \rightarrow \widetilde{\Omega}(\Omega)$ describing integration on a new manifold $\mathcal{M}$, a generic integral can be deformed as
\begin{equation}
  \begin{split}
    \int \mathcal{D}U \ f(U) &= \int \mathcal{D}\Omega \ H(\Omega)\ f(U(\Omega)) \\
    &= \int_{\mathcal{M}} \mathcal{D}\widetilde{\Omega} \ H(\widetilde{\Omega})\ f(U(\widetilde{\Omega}))  \\
    &= \int \mathcal{D}\Omega \ H(\Omega)\ \left[ J(\Omega) \frac{ H(\widetilde{\Omega}) }{ H(\Omega) } \right]  f(U(\widetilde{\Omega})) \\
    &= \int \mathcal{D}U \ J(U) \  f(\widetilde{U}),
  \end{split}
  \label{eq:SUNCauchy}
\end{equation}
where Cauchy's theorem gives the equality between the first and second line, the Jacobian $J(U) = J(\Omega) H(\widetilde{\Omega}) / H(\Omega)$ accounts for the change in Haar measure arising from the deformation, and the deformed gauge field $\tilde{U} \equiv U(\widetilde{\Omega}(\Omega))$ is a member of the complexified group.\footnote{For example, $\widetilde{U} \in SL(N,\mathbb{C})$ for $U \in SU(N)$.}
To be valid, the deformation map must also be continuously connected to the identity map $\widetilde{\Omega}_{\text{Id}}(\Omega) \equiv \Omega$.
By deforming and then writing integration on the deformed manifold in terms of coordinates $\Omega$ in the same domain as the original integration, the net effect of contour deformation is to replace the integrand $f(U)$ with $J(U) f(\widetilde{U})$ without modifying the integral value.

Applying Eq.~\eqref{eq:SUNCauchy} to $f(U) = \mathcal{O}(U) e^{i S_M(U)}$ under the assumption that $\mathcal{O}(U)$ and $S_M(U)$ are holomorphic functions of (a coordinate description of) $U$ allows deforming both the numerator and denominator of real-time expectation values to give
\begin{widetext}
\begin{equation}
\begin{split}
    \left< \mathcal{O} \right> &= \frac{\int \mathcal{D}U\ \mathcal{O}(\widetilde{U}(U)) \ J(U)\, e^{iS_M(\widetilde{U}(U))}}{\int \mathcal{D}U\ J(U)\, e^{iS_M(\widetilde{U}(U))}} \\
    &= \left( \frac{\int \mathcal{D}U\ \left\{ \mathcal{O}(\widetilde{U}(U)) \ e^{i \text{Arg}[J(U)]} e^{i\Re[S_M(\widetilde{U}(U))]} \right\} \ 
    |J(U)| e^{-\Im[S_M(\widetilde{U}(U))]}}{\int \mathcal{D}U\ \ |J(U)| \, e^{-\Im[S_M(\widetilde{U}(U))]}} \right) \\
    &\hspace{20pt} \times \left( \frac{\int \mathcal{D}U\ \left\{ e^{i \text{Arg}[J(U)]} e^{i\Re[S_M(\widetilde{U}(U))]} \right\} \ 
    |J(U)|\, e^{-\Im[S_M(\widetilde{U}(U))]}}{\int \mathcal{D}U\ \ |J(U)| \, e^{-\Im[S_M(\widetilde{U}(U))]}} \right)^{-1}.
    \end{split}
\end{equation}
\end{widetext}
This can be expressed as a ratio of expectation values
\begin{equation}
\begin{split}
    \left< \mathcal{O} \right> &= \frac{ \left<\mathcal{O}(\widetilde{U}(U)) e^{i \text{Arg}[J(U)]} e^{i\Re[S_M(\widetilde{U}(U))]} \right>_{S'} }{ \left< e^{i \text{Arg}[J(U)]} e^{i\Re[S_M(\widetilde{U}(U))]} \right>_{S'} },
    \end{split}
    \label{eq:contoursign}
\end{equation}
where $\left< \cdot  \right>_{S'}$ denotes an average with respect to a probability distribution proportional to $e^{-S'(U)} \equiv |J(U)| e^{-\Im[S_M(\widetilde{U}(U))]}$.
If fluctuations in $\text{Arg}[J(U)] + \Re[S_M(\widetilde{U}(U))]$ are reduced compared to $S_M(U)$ then the sign and associated StN problem of the denominator in Eq.~\eqref{eq:contoursign} will be corresponding improved. The sign problem in this denominator is highly correlated with the numerator for local observables, and thus the StN ratios of estimates of observables $\left< \mathcal{O} \right>$ are expected to be improved overall.

A similar approach can be applied when the path integral is defined on a Schwinger-Keldysh contour consisting of both Euclidean and Minkowski spacetime regions. Sign problems in such path integrals arise from the weights associated with the Minkowski region, suggesting that useful deformations can therefore be largely restricted to this region, up to boundary effects.
Path integral contour deformations can therefore be used to improve the sign and StN problems associated with calculations of $e^{iS_M(U)}$ in real-time and Schwinger-Keldysh LGT in complete analogy to previous applications to quantum mechanical models.

\subsection{Convergent U(1) HFK path integrals}\label{sec:HFKconvergence}

By changing the integrand magnitude, contour deformations can also render a divergent sum-integral absolutely convergent in certain cases, giving a prescription for evaluation. In particular, if integrating first then summing gives a well-defined result, it may be possible to apply contour deformations to each integral within the sum such that the joint sum-integral becomes absolutely convergent. Doing so preserves the value defined by the integrate-then-sum order of operations while enabling reordering or joint integration/summation.
However, because the endpoints of the integration contour must remain fixed to define a valid contour deformation, there will always be a neighborhood of the endpoint that has similar convergence properties to the original integration contour.
In order to define an absolutely convergent prescription for jointly performing sum-integrals below, an additional regularization of the infinite sum appearing in the HFK kinetic term analogous to a continuum Wick rotation is introduced below.
In particular, the phases of $U(1)$ gauge fields and plaquettes posses a $2\pi$ shift symmetry that is used to enforce cancellations between integral contributions from segments near both endpoints before taking the limit in which the Wick rotation regularization is removed. The resulting sum-integral is absolutely convergent for all points on the deformed contour after this cancellation is enforced and the limit is subsequently taken. This can be thought of as a rigorous coordinate-based approach to dealing with the identification of points related by the $2\pi$ shift symmetry, which is sufficient to avoid these endpoint singularities.

A simple contour deformation can be constructed that in this sense provides an absolutely convergent representation of the HFK action for $G=U(1)$.
Real-time LGT path integrals involving the HFK action can be represented as
\begin{equation}
\begin{split}
    \left< \mathcal{O} \right>_{M,HFK} &=  \int \mathcal{D}U\ \sum_{\{r\}} \mathcal{O}(U) \frac{e^{iS_{M,HFK}(U,r)}}{Z_{M,HFK}}
    \end{split}
    \label{eq:OHFK}
\end{equation}
where $Z_{M,HFK} = \int \mathcal{D}U\ \sum_{\{r\}} e^{iS_{M,HFK}(U,r)}$ and the indices $\{ r\} \equiv \{ r_{x,k} : x \in \text{sites}, k \in \{1, \dots, D-1\} \}$ label the representations appearing in the character expansion for the HFK kinetic term involving temporal plaquettes $P_{x,0k} = e^{i \phi_{x,0k}}$. These indices can be considered auxiliary variables in the path integral, with a local action $S_{M,HFK}(U,r)$ defining the weights over $U$ and $r$ simultaneously. For $G=U(1)$ this action is given by
\begin{equation}
\begin{split}
e^{i S_{M,HFK}^{U(1)}(U,r)} &= \prod_{x,k} \left[ e^{i/e^2}  [I_{r_{x,k}}(1/e^2)]^i  e^{i  r_{x,k} \phi_{x,0k}} \right] \\
&\hspace{20pt} \times e^{-\frac{i}{2e^2}\sum_x \sum_{i<j} \left(2 - P_{x,ij} - P_{x,ij}^{-1} \right) },
    \end{split}
    \label{eq:HFKraction}
\end{equation}
which is a convergent function of $U_{x,\mu}$ and $r_{x,k}$. As discussed in Sec.~\ref{sec:HFK}, however, the path integral in Eq.~\eqref{eq:OHFK} involves sums over $r_{x,k}$ and when these are evaluated inside the integral (i.e.~holding the gauge configuration $U$ fixed) the large-$r$ behavior is divergent.
This prevents the sum-integrals defining $\left<\mathcal{O}\right>_{M,HFK}$ from being evaluated using Monte Carlo methods because the absolute value of the weights do not give a well-defined probability measure.

On the other hand, exchanging the order of summation and integration in Eq.~\eqref{eq:OHFK} does produce a well-defined prescription for evaluating $\left< \mathcal{O} \right>_{M,HFK}$,
\begin{equation}
\begin{split}
    \left< \mathcal{O} \right>_{M,HFK} &= \sum_{\{r\}} \int \mathcal{D}U\ \mathcal{O}(U) \frac{e^{iS_{M,HFK}(U,r)}}{Z_{M,HFK}}.
    \end{split}
    \label{eq:OHFK-switched}
\end{equation}
In the $G=U(1)$ action above, integration over the gauge variables produces strong cancellation in the term $e^{i r_{x,k} \phi_{x,0k}}$ when any of the $r_{x,k}$ are taken large, suppressing these terms in the sum. A similar suppression occurs for $G = SU(N)$, though it takes a more complex form due to the non-Abelian characters $\chi_r$ appearing in the integrand. Although this doesn't enable joint Monte Carlo sampling of $U$ and $r$ (the ordering of summation/integration would be lost), it does provide a starting point for contour deformations.

In order to define an absolutely convergent representation of real-time LGT path integrals involving the HFK action, we first regularize the divergent sums by replacing the factor of $[c_r^{E,W}(g^2)]^i$ appearing in the HFK action with $[c_r^{E,W}(g^2)]^{e^{i\theta}}$ where $\theta \in [0,\pi/2]$ parameterizes a transformation between the Euclidean Wilson action with $\theta=0$ and the Minkowski HFK action with $\theta = \pi/2$.
For generic $\theta$, the kinetic term appearing in the $U(1)$ HFK action becomes
\begin{equation}
\begin{split}
    &\prod_{x,k}\left[  \left[I_{r_{x,k}}(1/e^2) \right]^{\cos(\theta)} \left[I_{r_{x,k}}(1/e^2)\right]^{i\sin(\theta)} \right. \\
    &\hspace{30pt}\left. \times e^{-e^{-i\theta}/e^2}  e^{ir_{x,k} \phi_{x,0k}} \right].
    \end{split}
\end{equation}
For $\theta \in [0, \pi/2)$ the first factor $\left[ I_{r_{x,k}}(1/e^2) \right]^{\cos(\theta)}$ vanishes faster than exponentially at large $|r_{x,k}|$~\cite{olver2010nist1035}, leading to absolute convergence of path integrals of the form Eq.~\eqref{eq:OHFK-switched} involving this transformed kinetic term.
It is only exactly at the Minkowski limit corresponding to $\theta=\pi/2$ that this term is equal to unity and absolute convergence is lost.

Before taking the Minkowski limit, we perform an $r_{x,k}$-dependent deformation of the $\phi_{x,0k}$ integration contour defined by the map
\begin{equation}
    \phi_{x,0k} \rightarrow \widetilde{\phi}_{x,0k} = \phi_{x,0k} + i \sign(r_{x,k}), \label{eq:HFKshift}
\end{equation}
which can be interpreted as a distinct contour deformation for each term in the sum in Eq.~\eqref{eq:OHFK-switched}.
``Constant vertical deformations'' analogous to this one have previously been used to reduce sign problems in several applications, see for example Refs.~\cite{Alexandru:2018ddf,Alexandru:2018fqp}, typically including one or more free parameters that define the imaginary shift and can be optimized to determine an integration contour with minimal phase fluctuations.\footnote{Free parameters of this and more general forms could be included in our choice of deformation and then optimized, but such explorations of how to optimally tame the sign problem and efficiently compute real-time LGT path integrals are left to future work; our goal here is just to obtain contour deformations that lead to absolutely convergent summands/integrands so that Monte Carlo methods can be applied in principle.}
Constant vertical deformations like Eq.~\eqref{eq:HFKshift} are only valid for periodic compact variables: the contours parallel to the imaginary axis required to complete a closed path involving the original and deformed integration contours differ by a shift equal to the domain of periodicity and have opposite orientation, causing integrals along these contours to cancel~\cite{Detmold:2021ulb}.
For $\theta \in [0, \pi/2)$, the absolute convergence discussed above allows this cancellation to be performed.
After enforcing the cancellation of the perpendicular contours to leave only the integral along the shifted contour corresponding to Eq.~\eqref{eq:HFKshift}, the Minkowski limit can be taken.
For all points on the shifted contour, exponential damping factors appear in the transformed kinetic energy term from $e^{i\widetilde{\phi}_{x,0k}r_{x,k}} = e^{i\phi_{x,0k}r_{x,k}} e^{-|r_{x,k}|}$.
As detailed below, these exponential damping factors are sufficient to render path integrals involving the HFK action absolutely convergent for all points on the shifted contour, even in the Minkowski limit.

In order to apply the transformation defined in Eq.~\eqref{eq:HFKshift} for timelike plaquette phases, it is necessary to first perform a change of variables from the original set of path integral variables $\{U_{x,\mu}\}$ to a set of variables that includes the timelike plaquette phases. Although it is not possible to perform a one-to-one change of variables from links to plaquettes, it is possible to transform to the set of variables $\{U_{x,0},P_{x,0k}\}$ and an additional fixed gauge field $U_{(L_T,\vec{x}),k}$ acting as a boundary condition in the real-time direction.
For the OBC case used in Sec.~\ref{sec:exact} as well as Secs.~\ref{sec:U1MC}--\ref{sec:SU3MC} below, $U_{(L_T,\vec{x}),k} = 1$.
Periodic boundary conditions in real-time can be viewed as infinite-temperature Schwinger-Keldysh contours and include singular observables not suitable for numerical simulation.\footnote{For example the $(1+1)$D results for real-time Wilson loops at finite $Lt$ are replaced by divergent series with periodic boundary conditions in real time.}
For Schwinger-Keldysh contours with non-trivial Euclidean extent, $U_{(L_T,\vec{x}),k}$ can be taken to be the first link on the Euclidean side of the boundary with each Minkowski region as detailed in Sec.~\ref{sec:SK}.
For $U(1)$ gauge theory, it is convenient to represent the boundary field as $U_{(L_T,\vec{x}),k} = e^{i b_{\vec{x},k}}$, the gauge field as $U_{x,\mu} = e^{i a_{x,\mu}}$, and the plaquette as $P_{x,\mu\nu} = e^{i\phi_{x,\mu\nu}}$ as above.
The relation
\begin{equation}
    \phi_{x,\mu\nu} = A_{x,\mu} + A_{x+a\hat{\mu},\nu} - A_{x+a\hat{\nu},\mu} - A_{x,\nu}, \label{eq:U1linktoplaq}
\end{equation}
can be used to solve for the spacelike components of the gauge field phase as
\begin{equation}
\begin{split}
   A_{x,k} &= b_{\vec{x},k} - \sum_{t/a=0}^{(L_T - x^0 - a)/a} \left( \phi_{x+t\hat{0},0k} \right. \\
    & \hspace{20pt}  \left. + A_{x+a\hat{k}+t\hat{0},0} - A_{x+t\hat{0},0} \right).
    \end{split}\label{eq:U1plaqtolink}
\end{equation}
The spacelike plaquette phase can then be expressed as a function  $\phi_{x,ij}(\phi_{x,0k},A_{x,0},b_{\vec{x},k})$ by inserting Eq.~\eqref{eq:U1plaqtolink} into Eq.~\eqref{eq:U1linktoplaq} and the full spacelike plaquette further obtained as $P_{x,ij}(\phi_{x,0k},A_{x,0},b_{\vec{x},k}) = e^{i \phi_{x,ij}(\phi_{x,0k},A_{x,0},b_{\vec{x},k})}$.
Path integrands of the form in Eq.~\eqref{eq:OHFK-switched} can therefore be transformed from base coordinates where the measure is proportional to $\prod_{x,\mu} dA_{x,\mu}$ into coordinates with measure proportional to $\prod_{x,k} d\phi_{x,0k} dA_{x,0}$ through a linear transformation with unit Jacobian.
In this basis, the contour deformation defined by Eq.~\eqref{eq:HFKdeform} can be simply applied to $\phi_{x,0k}$ with $A_{x,0}$ held fixed.
The spacelike plaquette phase for the deformed contour is therefore given (with $\theta = \pi/2$ taken after enforcing cancellation of perpendicular contours) by
\begin{equation}
\begin{split}
    &\phi_{x,ij}(\widetilde{\phi}_{x,0k},A_{x,0},b_{\vec{x},k}) = \phi_{x,ij}(\phi_{x,0k},A_{x,0},b_{\vec{x},k}) \\
    &\hspace{20pt} - i\sum_{t/a=0}^{(L_T - x^0 - a)/a} \left[ \sign(r_{x+t\hat{0},0i}) + \sign(r_{x+t\hat{0}+\hat{i},0j})    \right. \\
    & \hspace{20pt}  \left.  - \sign(r_{x+t\hat{0}+\hat{j},0i}) - \sign(r_{x+t\hat{0},0i}) \right].
    \end{split}
\end{equation}
The HFK action evaluated for the transformed gauge field can therefore be expressed using $\widetilde{P}_{x,ij} \equiv e^{i \phi_{x,ij}(\widetilde{\phi}_{x,0k},A_{x,0},b_{\vec{x},k})}$ and $\widetilde{P}_{x,ij}^{-1} \equiv e^{-i \phi_{x,ij}(\widetilde{\phi}_{x,0k},A_{x,0},b_{\vec{x},k})}$  as
\begin{equation}
\begin{split}
e^{i S_{M,HFK}^{U(1)}(\widetilde{U},r)} &= \prod_{x,k} \left[ e^{i/e^2}  [I_{r_{x,k}}(1/e^2)]^i  e^{i  r_{x,k} \phi_{x,0k}} e^{-|r_{x,k}|} \right] \\
&\hspace{20pt} \times e^{-\frac{i}{2e^2}\sum_x \sum_{i<j} \left(2 - \widetilde{P}_{x,ij} - \widetilde{P}_{x,ij}^{-1} \right) }.
    \end{split}
    \label{eq:HFKdeform}
\end{equation}
The $r$-dependent kinetic term $[I_{r_{x,k}}(1/e^2)]^i e^{i r_{x,k} \phi_{x,0k}}$ has unit magnitude. The factors of $\widetilde{P}_{x,ij}$ in the potential term only depend on $\{ \sign(r_{x,k}) \}$ and cannot have asymptotically diverging magnitude in the large-$r$ limit. Any observable that is a function of gauge fields similarly can only depend on $\{ \sign(r_{x,k}) \}$. The summand in Eq.~\eqref{eq:OHFK-switched} thus takes the form of $\prod_{x,k} e^{-|r_{x,k}|}$ multiplied by a bounded integral over the compact space of gauge configurations. 
The sum-integrals appearing in Eq.~\eqref{eq:OHFK-switched} are absolutely convergent using this contour deformation and can be used to calculate expectation values involving the $U(1)$ HFK action by simultaneously evaluating the sum and integral in
\begin{equation}
\begin{split}
    \left< \mathcal{O} \right>_{M,HFK} &= \sum_{\{r\}} \int \mathcal{D}U\  \mathcal{O}(\widetilde{U}(U,r))  \frac{e^{iS_{M,HFK}(\widetilde{U}(U,r),r)}}{Z_{M,HFK}}.
    \end{split}
    \label{eq:OHFK-deform}
\end{equation}
The absolute value weight $|\frac{e^{iS_{M,HFK}(\widetilde{U}(U,r),r)}}{Z_{M,HFK}}|$ now defines a valid probability measure over the combined space $(U,r)$ and Monte Carlo sampling techniques can be applied.

For $G = SU(2)$, the sum over representations can be converted to a sum over integers $r_{x,k}^A$ with $A=1,2$ satisfying $\sum_A r_{x,k}^A =0$ of the form Eq.~\eqref{eq:OHFK-deform}.
The undeformed action in this case is given by
\begin{equation}
\begin{split}
& e^{iS_{M,HFK}^{SU(2)}(U,r)} = e^{-\frac{i}{g^2}\sum_{x}\sum_{i<j}\Tr\left(2 - P_{x,ij} - P_{x,ij}^{-1}\right)}\prod_{x,k}  \\
    &\hspace{15pt}  \left[  (r_{x,k}+1)  e^{-4i/g^2} [I_{r_{x,k}}(4/g^2)]^i  \frac{\sin((r_{x,k}+1)\phi^1_{x,0k})}{\sin(\phi^1_{x,0k})} \right],
    \end{split}
\end{equation}
where $\phi^1_{x,0k} = -\phi^2_{x,0k}$ are the phases of the eigenvalues $e^{i\phi^A_{x,0k}}$ of $P_{x,0k}$ and $r_{x,k} \equiv r_{x,k}^1$.
These weights cannot be brought to an absolutely convergent form through either a constant vertical deformation or through the more general affine transformations explored in Sec.~\ref{sec:HKconvergence} below.
The exploration of more sophisticated contour deformation to transform the HFK action to an absolutely convergent representation for $G=SU(N)$ is left to future work.

\subsection{Convergent SU(N) and U(1) \texorpdfstring{$\HKbar$}{HKbar} path integrals}\label{sec:HKconvergence}

We now turn to the construction of a path integral contour deformation that provides absolutely convergent representations of path integrals using the real-time $\HKbar$ LGT action $S_{M,\HKbar}(U)$ that are suitable for numerical calculations using Monte Carlo techniques. Deformations for the case of both a $U(1)$ and $SU(N)$ gauge group are explored.

Since the divergent sums appear in the heat-kernel kinetic term, which is expressed in terms of the eigenvalues of the plaquette, it is convenient to begin with a change of path integral variables to a set that includes $P_{x,0k}$.
For real-time systems, one is generally interested in fixed or open boundary conditions in time or Schwinger-Keldysh contours as discussed in Sec.~\ref{sec:SK}
rather than periodic boundary conditions in real time.
As discussed above in Sec.~\ref{sec:HFKconvergence}, a boundary field configuration $U_{(\vec{x},L_T),k}$ is available for each of these cases that be used to perform a change of variables from spatial links to timelike plaquettes.
For both $U(1)$ and $SU(N)$ gauge theory, it is possible to invert the relation
\begin{equation}
    P_{x,0k} = U_{x,0} U_{x+a\hat{0},k} U_{x+a\hat{k},0}^{-1} U_{x,k}^{-1}
\end{equation}
and solve for the spatial gauge field in terms of the plaquettes, boundary field configuration, and temporal links as
\begin{equation}
\begin{split}
    U_{x,k} &= \left[ \prod_{n=0}^{(L_T-x^0-a)/a} P_{x+an\hat{0},0k}^{-1} U_{x+an\hat{0},0} \right] U_{(L_T,\vec{x}),k} \\
    &\hspace{20pt} \times\left[ \prod_{n=0}^{(L_T-x^0-a)/a} U_{x+a\hat{k}+an\hat{0},0}^{-1}  \right].
    \end{split}
    \label{eq:UPdef}
\end{equation}
It follows that once $U_{(L_T,\vec{x}),k}$ is specified the path integral variables $U_{x,\mu}$ are in one-to-one correspondence with the set of variables $\{ P_{x,0k},\ U_{x,0} \}$.
Further, the Jacobian for performing the change of variables to $\{ P_{x,0k},\ U_{x,0} \}$ is unity by the group multiplication invariance of the Haar measure.
The plaquettes $P_{x,0k}$ can be diagonalized and written in terms of eigenvalues $e^{i\phi_{x,0k}^A}$ and corresponding unitary eigenvector matrices $V_{x,0k}$ satisfying
\begin{equation}
    P_{x,0k} = V_{x,0k}^{\dagger} \text{diag}\left(e^{i\phi_{x,0k}^A}\right) V_{x,0k} \label{eq:Pdiag}
\end{equation}
where $\phi_{x,0k}^A \in [-\pi,\pi]$ and, for $G=SU(N)$, $A=1,\ldots,N$ with $\sum_A \phi_{x,0k}^A = 0 \mod 2\pi$.
Together these provide a new set of path integral variables
\begin{equation}
    \{ \phi_{x,0k}^A,\ V_{x,0k},\ U_{x,0} \}, \label{eq:phivar}
\end{equation}
where $V_{x,0k}$ and $U_{x,0}$ are parameterized according to some coordinate description of $G$. These variables are overcomplete in comparison to $\{ P_{x,0k},\ U_{x,0} \}$ since $V_{x,0k}$ can be multiplied by any unitary diagonal matrix without changing $P_{x,0k}$, but this redundancy only affects the normalization of the path integral measure since the unitary phases can be ``integrated in'' without changing the integral value. In general an integral over an $SU(N)$ plaquette can thus be rewritten using these variables as~\cite[\S3.1]{meckes_2019_p74}
\begin{equation}
\begin{split}
    &\int d P_{x,0k}\ f(P_{x,0k}) \\
    &= \int \frac{1}{N!}\prod_{A=1}^N \left[ \frac{d\phi_{x,0k}^A}{2\pi} \prod_{B<A} \left| e^{i\phi_{x,0k}^A} - e^{i\phi_{x,0k}^B} \right|^2 \right]  \\
    &\hspace{20pt} \times 
    \sum_n \delta\left(2\pi n - \sum_A \phi_{x,0k}^A\right) d V_{x,0k} \\
    &\hspace{20pt} \times f( V_{x,0k}^\dagger \text{diag}(e^{i\phi_{x,0k}^A}) V_{x,0k} ),
    \end{split} \label{eq:meas_phivar}
\end{equation}
where $dV_{x,0k}$ is the Haar measure for $U(N)$.
Using the variables in Eq.~\eqref{eq:phivar} as independent path integral variables with the measure appearing on the right-hand-side Eq.~\eqref{eq:meas_phivar} permits the use of deformations of the $\phi_{x,0k}^A$ integration contours that leave $V_{x,0k}$ and $U_{x,0}$ unaffected.

\begin{figure*}[t]
    \centering
    \includegraphics[width=\textwidth]{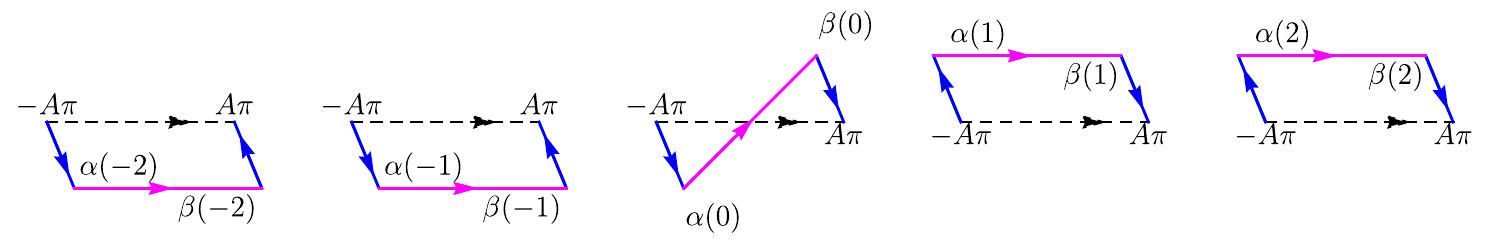}
    \caption{The deformed path integral contour for $\widetilde{\psi}_{x,k}^A$ is shown as a function of $m_{x,k}^A \in \{-2, -1, 0,  1, 2\}$ from left to right. The dashed line shows the original contour $[-A \pi, A \pi]$, while the solid lines show the segments of the deformed contour described by the coordinates $\widetilde{\psi}^{(0),A}_{x,0k}$ in magenta and $\widetilde{\psi}^{(1),A}_{x,0k}$ in blue. The points $\alpha(m_{x,k}^A)$ and $\beta(m_{x,k}^A)$ where the deformed contour segments intersect are defined in Eq.~\eqref{eq:alphabeta}. This deformation leads to a convergent real-time $\HKbar$ path integral after the outer contour segments $[-A \pi, \alpha(m_{x,k}^A)]$ and $[\beta(m_{x,k}^A), A \pi]$ are exactly cancelled and the limit to the Minkowski signature is analytically evaluated.
   \label{fig:contour}}
\end{figure*}

In analyzing the convergence of $\HKbar$ path integrals, it will be useful to consider the limit as the prefactor in the kinetic energy term of the $\HKbar$ action is rotated from $-1$ (Euclidean) to $i$ (Minkowski), and for intervening points we use the prefactor $-e^{-i \theta}$, with $\theta \in [0,\pi/2]$, and represent a generic heat-kernel kinetic term for link $(x,k)$ as
\begin{equation}
    \mathcal{G}_{x,k} = \prod_{A=1}^N e^{-\frac{1}{g^2} e^{-i\theta} (\phi_{x,0k}^A + 2 \pi n_{x,k}^A)^2}.
\end{equation}
For the gauge group $U(1)$, this term is already in a suitable form for deforming as below; however, for the gauge group $SU(N)$ further rewriting of this term is helpful.

The determinant constraint for $SU(N)$ implies that $\phi_{x,0k}^N$ and $n_{x,0k}^N$ are not independent variables that are integrated over, and instead can be related to the other variables by $\phi_{x,0k}^N = - \sum_{A=1}^{N-1} \phi_{x,0k}^A$ and $n_{x,k}^N = -\sum_{A=1}^{N-1} n_{x,k}^A$.
In terms of the $N-1$ independent phases and integers, $\mathcal{G}_{x,k}$ is a correlated Gaussian,
\begin{equation}
\begin{split}
    \mathcal{G}_{x,k} &= \exp\Big( - \frac{e^{-i\theta}}{g^2} \times \\
    &\sum_{A,B=1}^{N-1} ({\phi}_{x,0k}^A+2\pi n_{x,k}^A) \, \sigma^{AB} \, (\phi_{x,0k}^B + 2\pi n_{x,k}^B) \Big),
    \end{split}
\end{equation}
where $\sigma^{AB}$ equals 2 if $A=B$ and 1 if $A\neq B$.
The eigenvalues of $\sigma$ are $\rho^A = (N,1,\ldots,1)$ corresponding to an orthogonal (not unit normalized) basis of eigenvectors $S^{AB}$ satisfying
\begin{equation}
    \sigma^{AB} =  \sum_C  S^{AC} \rho^C (S^{-1})^{CB}, \label{eq:sigmadiag}
\end{equation}
where
\begin{equation}
    (S^{-1})^{AC} = \begin{cases}
    (1, 1, \dots, 1)_A & C = 1 \\
    \vphantom{(1,)}\smash{(\underbrace{1, 1, 1, \dots, 1}_{\text{$C-1$ elements}}, -(C-1), 0, \dots)_A} & C > 1
    \end{cases}.
\label{eq:Spsiphi}
\end{equation}
New variables can be defined for this eigenbasis by
\begin{equation}
\begin{split}
    \psi_{x,0k}^A &= \sum_{B=1}^{N-1} (S^{-1})^{AB} \phi_{x,0k}^B \\
    m_{x,k}^A &= \sum_{B=1}^{N-1} (S^{-1})^{AB} n_{x,k}^B.
    \end{split}\label{eq:psiphi}
\end{equation}
This permits changing variables from $\phi_{x,0k}^A$ to the newly introduced $\psi_{x,0k}^A$ with a Jacobian given by $|\det(S)| = (N-1)!$.
The heat-kernel term $\mathcal{G}_{x,k}$ can thus be expressed as an uncorrelated Gaussian,
\begin{equation}
    \mathcal{G}_{x,k} = \prod_{A=1}^{N-1} e^{ - \frac{1}{g^2}e^{-i\theta} \rho^A (\psi_{x,0k}^A + 2\pi m_{x,k}^A)^2 }.
    \label{eq:HKGdiag}
\end{equation}

To change variables in the path integral, the particular linear combinations involved in the definitions of $m_{x,k}^A$ and $\psi_{x,k}^A$ affect the domain of summation/integration. Unlike the set of $n_{x,k}^A$, for $m_{x,k}^A$ it is necessary to write an iterated summation $\sum_{m_{x,k}^1} \sum_{m_{x,k}^2} \dots$ with bounds and increments determined by the linear combinations involved in Eq.~\eqref{eq:psiphi}. A convenient choice of order is to sum over $m_{x,k}^1$ outermost, then successively sum over $m_{x,k}^{N-1}, \dots, m_{x,k}^{2}$ in decreasing order.  In this order, each $m_{x,k}^A$ can be related to variables in outer sums by the relations
\begin{equation}
    m_{x,k}^A = \begin{cases}
    \sum_{B=1}^{N-1} n_{x,k}^B & A = 1\\
    m_{x,k}^1 - (N-1) n_{x,k}^{N-1} & A = N-1\\
    m_{x,k}^{A+1} + A(n_{x,k}^{A+1} - n_{x,k}^{A}) & 1 < A < N-1
    \end{cases}.
\label{eq:HKm-relations}
\end{equation}
The domain of the $m_{x,k}^1$ sum is thus $\mathbb{Z}$, the domain of the $m_{x,k}^{N-1}$ sum is $m_{x,k}^1 + (N-1)\mathbb{Z}$, and the domain of the remaining $m_{x,k}^{A}$ sums are $m_{x,k}^{A+1} + A \mathbb{Z}$, indicating that the increment of the sum over $m_{x,k}^A$ is always $A$ though the offsets are generally functions of variables in outer sums. A similar iterated integral is required to properly cover the original domain when integrating over the new $\psi_{x,0k}^A$. These variables satisfy relations analogous to  Eq.~\eqref{eq:HKm-relations},
\begin{equation}
    \psi_{x,k}^A = \begin{cases}
    \sum_{B=1}^{N-1} \phi_{x,k}^B & A = 1\\
    \psi_{x,k}^1 - (N-1) \phi_{x,k}^{N-1} & A = N-1\\
    \psi_{x,k}^{A+1} + A(\phi_{x,k}^{A+1} - \phi_{x,k}^{A}) & 1 < A < N-1
    \end{cases},
\end{equation}
and from these relations we find that each $\psi_{x,k}^A$ should be integrated over a domain of width $2\pi A$, with offsets determined by outer integration variables. Based on the form of the diagonalized heat-kernel term in Eq.~\eqref{eq:HKGdiag} and the domains of the sums above, the pair $(\psi_{x,k}^A, m_{x,k}^A)$ satisfy a shift symmetry $(\psi_{x,k}^A, m_{x,k}^A) \rightarrow (\psi_{x,k}^A + 2\pi A, m_{x,k}^A - A)$. The shift in $m_{x,k}^A$ can be absorbed into the infinite summation, and thus we are free to use this symmetry to shift the domain of integration of each $\psi_{x,k}^A$ to the simple range $[-A \pi, A \pi]$.

In terms of these new variables, Fig.~\ref{fig:contour} shows the $m_{x,k}^A$-dependent deformation of $\psi_{x,0k}^A$ that is used to construct a convergent representation of $\HKbar$ path integrals.\footnote{For the gauge group $U(1)$, the trivial identification $\psi_{x,0k}^1 = \phi_{x,0k}^1$ and $m_{x,0k}^1 = n_{x,0k}^1$ should be used to determine the following deformation.}
The contour for each value of $m_{x,k}^A$ and a fixed angle $\theta$ consists of the union of three segments. The middle segment is defined to be the interval $[ \alpha(m_{x,k}^A),\beta(m_{x,k}^A)]$ where
\begin{equation}
    \begin{split}
       \alpha(m_{x,k}^A) &= \begin{cases} -A \pi e^{i \theta / 2}  & m_{x,k}^A = 0 \\ 
       -A \pi + (e^{i \theta / 2}-1)A \pi \sign(m_{x,k}^A) & m_{x,k}^A \neq 0 \end{cases}, \\
       \beta(m_{x,k}^A) &= \begin{cases} A \pi e^{i \theta / 2}  & m_{x,k}^A = 0 \\ 
      A \pi + (e^{i \theta / 2}-1)A \pi \sign(m_{x,k}^A) & m_{x,k}^A \neq 0 \end{cases}.
    \end{split}\label{eq:alphabeta}
\end{equation}
The sign function is defined by $\sign(x) = x/|x|$ with the midpoint convention $\sign(0) = 0$.
The two outer segments of the contour are defined to be the linear intervals $[-A\pi, \alpha(m_{x,k}^A)]$ and $[\beta(m_{x,k}^A),A\pi]$ in the complex plane. To understand the convergence properties of this deformed path integral, it is useful to split each integration over $\widetilde{\psi}_{x,k}^A$ into an integral over a coordinate parameterization of the middle segment and a second integral over a coordinate parameterization of the outer segments. The first integral can be defined using a coordinate transformation $\widetilde{\psi}_{x,0k}^{(0),A}(\psi_{x,0k}^A, m_{x,k}^A)$ relating the base coordinate interval $[-A\pi,A\pi]$ to the middle segment of the deformed contour,
\begin{equation}
\begin{split}
     \widetilde{\psi}_{x,0k}^{(0),A} &= \begin{cases} \psi_{x,0k}^A e^{i \theta / 2}  & m_{x,k}^A = 0 \\ 
      \psi_{x,0k}^A + (e^{i \theta / 2}-1)A \pi \sign(m_{x,k}^A) & m_{x,k}^A \neq 0 \end{cases}. 
\end{split}
\label{eq:HKdeform0}
\end{equation}
The second integral can be defined using a similar coordinate transformation $\widetilde{\psi}_{x,0k}^{(1), A}(\psi_{x,0k}^A, m_{x,k}^A)$ relating the base coordinate interval $[-A\pi,A\pi]$ to the union of the two outer deformed contour segments,
\begin{equation}
\begin{split}
    \widetilde{\psi}_{x,0k}^{(1), A} &= \begin{cases}
     \alpha(m_{x,k}^A) + \left( 1 + \frac{\alpha(m_{x,k}^A)}{A \pi}\right)\psi_{x,k}^A, & \psi_{x,0k}^A < 0 \\
     \beta(m_{x,k}^A) + \left( 1 - \frac{\beta(m_{x,k}^A)}{A \pi}\right)\psi_{x,k}^A,  & \psi_{x,0k}^A \geq 0 
    \end{cases} .
    \end{split}
    \label{eq:HKdeform1}
\end{equation}
The transformed coordinates $\widetilde{\psi}_{x,0k}^{(c),A}$ are indexed by $c_{x,0k}^A \in \{0,1\}$, abbreviated $c$ in the $\widetilde{\psi}$ label, which in the full deformed path integral must be summed over.

The eigenvalue phases of each deformed temporal plaquette can then be obtained by inverting Eq.~\eqref{eq:psiphi},
\begin{equation}
    \begin{split}
        \widetilde{\phi}_{x,0k}^{(c),A} &=  \sum_{B=1}^{N-1}  S^{AB} \widetilde{\psi}_{x,0k}^{(c),B}, \\
        n_{x,k}^A &=  \sum_{B=1}^{N-1}  S^{AB} m_{x,k}^B,
    \end{split}\label{eq:phipsi}
\end{equation}
with the value of $\widetilde{\phi}_{x,0k}^{(c),N}$ and $n_{x,k}^N$ obtained by solving the constraints $\widetilde{\phi}_{x,0k}^{(c),N} = -\sum_{A=1}^{N-1} \widetilde{\phi}_{x,0k}^{(c),A}$ and $n_{x,k}^N = -\sum_{A=1}^{N-1} n_{x,k}^A$.
These deformed eigenvalue phases can be further used to define deformed timelike plaquettes as
\begin{equation}
    \widetilde{P}_{x,0k}^{(c)} = V_{x,0k}^\dagger \text{diag}\ \left(e^{i\widetilde{\phi}_{x,0k}^{(c),A}}\right)\ V_{x,0k},
    \label{eq:Ptildedef}
\end{equation}
and from these the deformed gauge fields using Eq.~\eqref{eq:UPdef},
\begin{equation}
\begin{split}
     \widetilde{U}^{(c)}_{x,k} &= \left[ \prod_{t/a=0}^{(L_T-x^0-a)/a} \left( \widetilde{P}^{(c)}_{x+t\hat{0},0k} \right)^{-1} U_{x+t\hat{0},0} \right] U_{(L_T,\vec{x}),k} \\
    &\hspace{20pt} \times\left[ \prod_{t/a=0}^{(L_T-x^0-a)/a} U_{x+a\hat{k}+t\hat{0},0}^{-1}  \right].
    \end{split}
    \label{eq:UPtilde}
\end{equation}
This contour deformation therefore defines a map $U_{x,\mu} \rightarrow \widetilde{U}_{x,\mu}^{(c)}(U,n)$ that depends non-locally on the gauge field as well as the integers $n_{x,\mu}^A$ ensuring $2\pi$ shift invariance of the heat-kernel action.
Transformations of functions of $U_{x,\mu}$ can be obtained immediately from the definition of $\widetilde{U}_{x,\mu}^{(c)}(U,n)$, for example transformed spacelike plaquettes are given by
\begin{equation}
    \widetilde{P}_{x,ij}^{(c)} = \widetilde{U}_{x,i}^{(c)} \widetilde{U}_{x+a\hat{i},j}^{(c)} \left(\widetilde{U}_{x+a\hat{j},i}^{(c)}\right)^{-1} \left( \widetilde{U}_{x,j}^{(c)} \right)^{-1}.
    \label{eq:Pijtildedef}
\end{equation}

This transformation does not leave the Haar measure invariant.
The coordinate piece of the Jacobian is given by
\begin{equation}
    \begin{split}
        j_{x,k}^{(c)} &= \det\left( \frac{\partial  \widetilde{\phi}_{x,0k}^{(c),A}}{\partial \phi_{x,0k}^{(c),B}}\right) \\
        &= \det \left( \sum_{C,D} \frac{\partial \widetilde{\phi}_{x,0k}^{(c),A}}{\partial \widetilde{\psi}_{x,0k}^{(c),C}}  \frac{\partial \widetilde{\psi}_{x,0k}^{(c),C}}{\partial \psi_{x,0k}^{(c),D}} \frac{\partial \psi_{x,0k}^{(c),D}}{\partial \phi_{x,0k}^{(c),B}} \right) \\
        &= \det\left( \frac{\partial \widetilde{\psi}_{x,0k}^{(c),A}}{\partial \psi_{x,0k}^{(c),B}} \right) \\
        &= \begin{cases} e^{i\theta/2} & c_{x,k}^A = 0,\ m_{x,k}^A = 0 \\
        1 & c_{x,k}^A = 0,\ m_{x,k}^A \neq 0 \\
        1 + \frac{\alpha\left(m_{x,k}^A\right)}{A\pi} & c_{x,k}^A = 1,\ \psi_{x,0k}^A < 0 \\ 
        1 - \frac{\beta\left(m_{x,k}^A\right)}{A\pi} & c_{x,k}^A = 1,\ \psi_{x,0k}^A \geq 0\end{cases}.
    \end{split}
\end{equation}
Including the explicit change of the Haar measure in the coordinates Eq.~\eqref{eq:meas_phivar}, the full Jacobian  $J^{(c)}$ for each deformed contour segment is given by
\begin{equation}
\begin{split}
    J^{(c)}(U,n) &= \prod_{x,k} j_{x,k}^{(c)} h_{x,k}^{(c)}, \\
    h_{x,k}^{(c)}  &= \left[ \frac{\prod_{A<B} \left| e^{i\widetilde{\phi}_{x,0k}^{(c),A}} - e^{i\widetilde{\phi}_{x,0k}^{(c),B}} \right|^2}{\prod_{A<B} \left| e^{i\phi_{x,0k}^A} - e^{i\phi_{x,0k}^B} \right|^2} \right]. \\
    \end{split}
    \label{eq:Jdeform}
\end{equation}
For $G=U(1)$ the explicit change of the Haar measure is trivial and $h_{x,k}^{(c)}$ should be replaced by unity in Eq.~\eqref{eq:Jdeform}.

The $\HKbar$ action given in Eq.~\eqref{eq:SHKbardef} is a holomorphic function of $\phi_{x,0k}^A$, with the independent variables $V_{x,0k}$ and $U_{x,0}$ left undeformed. The spacelike plaquettes appearing in the action are functions of these variables, as defined in Eq.~\eqref{eq:Pijtildedef}, and are also holomorphic in $\phi_{x,0k}^A$. These eigenvalue phases $\phi_{x,0k}^{A}$ are themselves holomorphic functions of the linear combinations $\psi_{x,0k}^{A}$ introduced in Eq.~\eqref{eq:psiphi}, and results for expectation values of observables that are similarly holomorphic functions of $\psi_{x,0k}^A$ (including in particular Wilson loops and other polynomial functions of $U_{x,\mu}$) are therefore unchanged by deformations of the $\psi_{x,0k}^A$ integration contours, as long as endpoints of each contour are held fixed and the unit determinant constraint for $G=SU(N)$ is not violated.
Expectation values using the deformation defined by Eq.~\eqref{eq:HKdeform0}--\eqref{eq:HKdeform1} of the $\HKbar$ path integral can therefore be expressed as
\begin{equation}
\begin{split}
    \left< \mathcal{O} \right>_{M,\HKbar} &= \frac{1}{Z_{M,\HKbar}} \int \mathcal{D}U\ \mathcal{O}(U)\ e^{iS_{M,\HKbar}(U)} \\
    &= \frac{1}{Z_{M,\HKbar}} \sum_{\{n\}} \sum_{\{c\}}   \int \mathcal{D}U\ J^{(c)}(U,n)\ \\
    &\hspace{10pt} \times \mathcal{O}(\widetilde{U}^{(c)}(U,n))\ e^{iS_{M,\HKbar}(\widetilde{U}^{(c)}(U,n),n)}, 
    \end{split}
    \label{eq:OHKtilde}
\end{equation}
where $Z_{M,\HKbar} = \int \mathcal{D}U\ \sum_{\{n\}}  e^{iS_{M,\HKbar}(U,n)}$ and $\sum_{\{c\}}$ denotes a collection of sums $\sum_{c_{x,k}^A = 0}^1$.
Using the general prefactor $-e^{-i\theta}$, the modified heat-kernel action applied to the transformed variables takes the form
\begin{equation}
\begin{split}
e^{iS_{\theta,\HKbar}(\widetilde{U}^{(c)},n)} &=  \prod_{x,k}  \mathcal{N}\, \mathcal{J}(\{\widetilde{\phi}_{x,0k}\},\{n_{x,k}\}) \\
&\hspace{10pt} \times \prod_{x,k,A} e^{ -\frac{1}{g^2} e^{-i\theta} \left(\widetilde{\phi}_{x,0k}^{(c),A}  + 2\pi n_{x,k}^A\right)^2 } \\
    &\hspace{10pt} \times \prod_{x,i<j} e^{ - \frac{i}{g^2}\Tr \left(2 - \widetilde{P}_{x,ij}^{(c)} - \left(\widetilde{P}_{x,ij}^{(c)} \right)^{-1} \right)} ,
    \end{split}
    \label{eq:SHKbardeform}
\end{equation}
where $\mathcal{J}$ is the $SU(N)$ heat-kernel factor defined in Eq.~\eqref{eq:KESUN} applied to the transformed coordinates $\widetilde{\phi}_{x,0k}^A$, 
\begin{equation}
\begin{split}
    &\mathcal{J}(\{\widetilde{\phi}\},\{n\}) =   \\ 
    &\hspace{5pt} \prod_{A < B} \left( \frac{\widetilde{\phi}^A - \widetilde{\phi}^B + 2\pi(n^A - n^B)}{2\sin\left[ \frac{1}{2} \left( \widetilde{\phi}^A - \widetilde{\phi}^B + 2\pi(n^A - n^B) \right) \right]} \right).
\label{eq:HKJdeform}
\end{split}
\end{equation}
This factor can be naturally analytically continued using the complex $\sin$ function. The poles at $\frac{1}{2} \left( \widetilde{\phi}^A - \widetilde{\phi}^B + 2\pi(n^A - n^B) \right) = \pi k$ for $k \neq 0$ appear to cause problems with holomorphy, but these are exactly the cases when $\widetilde{\phi}^A = \widetilde{\phi}^B \mod 2\pi$ and these poles are cancelled by zeros of the deformed Haar measure appearing as the numerator of Eq.~\eqref{eq:Jdeform}.

To analyze absolute convergence, we consider the phase-quenched path integral given for general rotation angle $\theta$ by
\begin{equation}
\begin{split}
    Z^{pq}_{\theta, \HKbar} &= \sum_{\{n\}} \sum_{\{c\}} \int \mathcal{D}U\ |J^{(c)}(U,n)| \\
    &\hspace{20pt} \times\prod_{x,k} \left| \mathcal{N}\,  \mathcal{J}(\{ \widetilde{\phi}_{x,0k} \}, \{ n_{x,k} \}) \right| \\
    &\hspace{20pt} \times\prod_{x,k,A} e^{- \Re\left[ \frac{1}{g^2} e^{-i\theta} \left(\widetilde{\phi}_{x,0k}^{(c),A}  + 2\pi n_{x,k}^A\right)^2 \right]} \\
    &\hspace{20pt} \times \prod_{x,i<j} e^{ - \Re\left[\frac{i}{g^2}\Tr \left(2 - \widetilde{P}_{x,ij}^{(c)} - \left(\widetilde{P}_{x,ij}^{(c)} \right)^{-1} \right) \right]}.
\end{split}
\label{eq:HKbarZpq}
\end{equation}
The summation $\sum_{\{c\}}$ is over a finite set and the integral $\int \mathcal{D}U$ is over a compact domain, so showing boundedness of the integrand and absolute convergence of the sum $\sum_{\{n\}}$ is sufficient for convergence of the absolute sum-integral.
The term in Eq.~\eqref{eq:HKbarZpq} involving the spatial plaquettes $\widetilde{P}_{x,ij}^{(c)}$ and the term $\mathcal{D}U|J^{(c)}(U,n)|$ determining the deformed Haar measure are functions of only the variables $\{\sign(m_{x,k}^A)\}$ and therefore do not diverge asymptotically. To understand the large-$n$ behavior we focus on the remaining terms. The heat-kernel factor $\mathcal{J}$ on the second line grows polynomially with $n_{x,k}^A$ (the singularities in the denominator are cancelled by the Haar measure, as discussed above). The kinetic-energy term on the third line contains non-trivial dependence on $n_{x,k}^A$, and it is helpful to expand the kinetic-energy term for a general decomposition into real and imaginary components of the deformed Gaussian eigenbasis variables $\widetilde{\psi}_{x,0k}^{(c),A} \equiv y_{x,0k}^{(c),A} + i z_{x,0k}^{(c),A}$,
\begin{equation}
\begin{split}
    \ln|\mathcal{G}_{x,k}| &=\Re\left[ -\frac{1}{g^2}\rho^A e^{-i\theta}  \left(\widetilde{\psi}_{x,0k}^{(c),A}  + 2\pi m_{x,k}^A\right)^2 \right] \\
    &= -\frac{1}{g^2}\rho^A \cos{\theta}  \left[ ( y_{x,0k}^{(c),A} + 2\pi m_{x,k}^A )^2 - (z_{x,0k}^{(c),A})^2 \right] \\
    &\hspace{20pt} -\frac{1}{g^2} \rho^A \sin{\theta} \left[ 2 (y_{x,k}^{(c),A} + 2\pi m_{x,k}^A)z_{x,k}^{(c),A} \right].
\end{split}
\label{eq:HKkineticDeform}
\end{equation}
For $\theta < \pi/2$, the first term provides a Gaussian cutoff $e^{-\frac{4\pi^2 \rho^A}{g^2}\cos\theta (m_{x,k}^A)^2}$ for large $m_{x,k}^A$, which equivalently corresponds to large-$n$ behavior. This provides absolute convergence for Wick rotated path integrals with $\theta < \pi/2$ for all points on all contour segments.

In the desired $\theta \rightarrow \pi/2$ limit, only the last term survives. If $\lim_{m\rightarrow \pm \infty} \sign(z_{x,k}^A) = \sign(m_{x,k}^A)$, that is if the sign of the imaginary part of $\widetilde{\phi}_{x,0k}^{(c),A}$ matches the sign of $m_{x,k}^A$ asymptotically, then this surviving term provides an exponential cutoff for large $m_{x,k}^A$ sufficient to overcome the polynomial growth in $\mathcal{J}$ and provide convergence.\footnote{If the heat-kernel potential is used in place of the Wilson potential, this argument remains valid but additional divergent infinite sums appear for the space-space plaquette eigenvalues that are not rendered convergent by this contour deformation. If the truncated heat-kernel potential is used in place of the Wilson potential, holomorphy of the potential as a function of $\phi_{x,0k}^A$ is no longer manifest because path integrals employing the truncated heat-kernel action do not possess a $\phi_{x,0k}^A \rightarrow \phi_{x,0k}^A + 2\pi$ shift symmetry and are therefore sensitive to branch cuts in $\text{arg}(e^{i\phi_{x,0k}^A})$ that could cause further complications.} This is true of all points on the middle contour segments in Fig.~\ref{fig:contour} identified by $c_{x,k}^A = 0$. However, this does not hold for the endpoints of the outer segments identified by $c_{x,k}^A = 1$.

The deformation employed here is therefore convergent everywhere except a set of measure zero for the action and deformation with $\theta = \pi/2$, but this is not sufficient to guarantee absolute convergence. In fact, these singularities do cause the phase-quenched partition function $Z_{\theta,\HKbar}^{pq}$ to diverge for $\theta = \pi/2$. Fundamentally, the fixed endpoints of the deformation contour prevent convergence at these points for any choice of contour deformation. This problem can be resolved, and the path integral rendered absolutely convergent, by noting that in the full path integral these outer segments can be identified and cancelled using $2\pi$ shift symmetry. As discussed above, path integrals using the $\HKbar$ action are invariant under the transformation $(\psi_{x,0k}^A ,m_{x,k}^A) \rightarrow (\psi_{x,0k}^A + 2\pi A , m_{x,k}^A - A)$.
This symmetry relates the outer contour endpoints $\beta(m_{x,k}^A) = \alpha(m_{x,k}^A + A) + 2\pi A$ for integrals along the blue contours shown in Fig.~\ref{fig:contour} for $\widetilde{\psi}_{x,0k}^A$ integrals with $m_{x,k}^A$ and $m_{x,k}^A + A$.
The other endpoints of these contours at $\pm A \pi$ are also identical because of this shift symmetry, and therefore the blue contour segments in Fig.~\ref{fig:contour} describe oppositely oriented contour integrals of holomorphic functions between identified endpoints and therefore cancel. 
Strictly, this identification and cancellation should only be considered valid for an absolutely convergent integral. 
For all $\theta < \pi/2$, the term $\cos{\theta} (2\pi m_{x,k}^A)^2$ gives absolute convergence for the deformation under study, and for all $\theta < \pi/2$ the outer contour segments in the neighborhood of the endpoints can be identified and exactly cancelled.
This defines a limiting procedure for defining $\HKbar$ path integrals in which integrals are defined for a generic Wick rotation angle $\theta$ and the cancellation of the blue contours in Fig.~\ref{fig:contour} is performed before taking the $\theta \rightarrow \pi/2$ Minkowski limit.

Finally, any observable $\mathcal{O}(U)$ without explicit $n$-dependence leads to an $\mathcal{O}(\widetilde{U}^{(c)}(U,n))$ whose only $n$-dependence can be described as dependence on $\sign(m_{x,k}^A)$ and is therefore finite in the $n_{x,k}^A \rightarrow \infty$ limit.
This demonstrates that the exponential convergence provided by the kinetic term is not spoiled and that $\HKbar$ path integrals including observables are absolutely convergent.

It is noteworthy that in addition to providing an exponential convergence factor for the integer sums in the heat-kernel kinetic term, this contour deformation also removes the sign problem associated with the Gaussian factors $e^{i (\phi_{x,0k}^A)^2}$ in the heat-kernel kinetic term by transforming them to positive weights $e^{- (\phi_{x,0k}^A)^2}$.
The magnitude fluctuations of $e^{- (\phi_{x,0k}^A)^2}$ are amenable to importance sampling, and in particular are the same weights that appear in the Euclidean heat-kernel action.
Phase fluctuations arise from terms with $n_{x,k}^A \neq 0$; however, the contributions from these terms are exponentially suppressed for small $g^2$.
This contour deformation can therefore be expected to significantly improve the sign problem associated with the real-time heat-kernel kinetic term for sufficiently weak couplings in addition to providing absolute convergence.

\begin{figure}[t!]
    \centering
    \includegraphics[width=.93\linewidth]{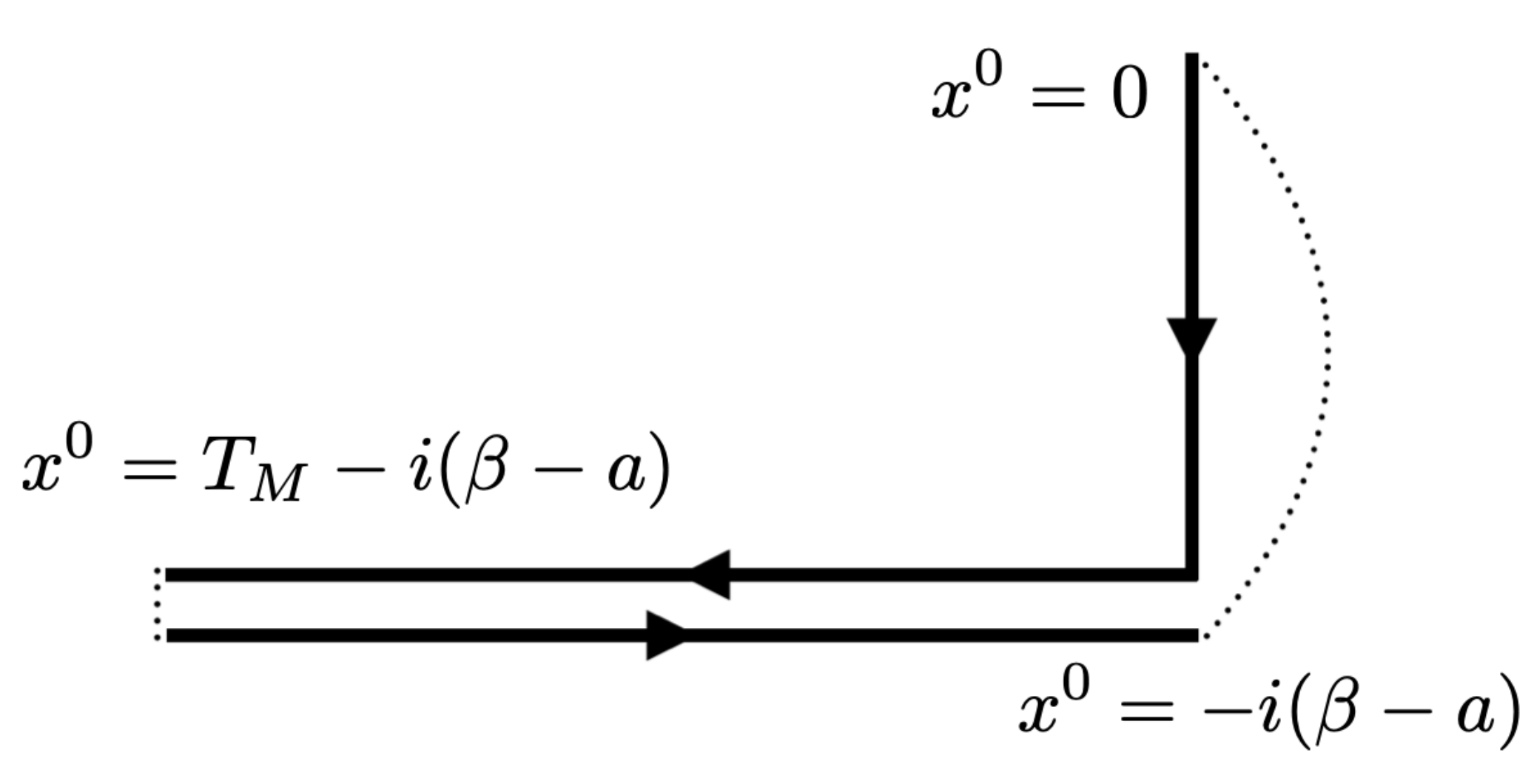}
    \caption{The $x^0$ coordinate Schwinger-Keldysh contour for evaluation of out-of-equilibrium observables. Spatial directions of the lattice are omitted for clarity. Subsets $\mathcal{T}_E$, $\mathcal{T}_{M^+}$, and $\mathcal{T}_{M^-}$ correspond to regions of the contour with Euclidean time evolution $x^0\in [0,-i(\beta-a)]$, forward Minkowski time evolution $x^0\in [-i(\beta-a),T_M-i(\beta-a)]$, and reverse Minkowski time evolution $x^0\in [T_M-i(\beta-a),-i(\beta-a)]$, respectively. Dashed lines indicate that the imaginary parts of $x^0$ on $\mathcal{T}_M^+$ and $\mathcal{T}_M^-$ are identical and displaced for visual clarity and also that gauge fields at $x^0 = -i\beta$ are identified with gauge fields at $x^0 = 0$. \label{fig:SK-contour}}
\end{figure}

\subsection{Convergent LGT Schwinger-Keldysh path integrals}\label{sec:SK}

For applications involving the Schwinger-Keldysh formalism, an action $S_{SK}$ must be constructed for the time integration contour $x^0 \in \{\mathcal{T}_E,\ \mathcal{T}_M^+,\ \mathcal{T}_M^-\}$ consisting of a Euclidean time evolution segment $\mathcal{T}_E$, a forward Minkowski time evolution segment $\mathcal{T}_M^+$, and a reverse Minkowski time evolution segment $\mathcal{T}_M^-$, as depicted in Fig.~\ref{fig:SK-contour}. The length of $\mathcal{T}_E$ is denoted $\beta$, the length of $\mathcal{T}_M^\pm$ is denoted $T_M$, and the total extent of the time direction with Minkowski signature is $L_T^M = 2T_M$.
A suitable generalization of the real-time $\HKbar$ action for Schwinger-Keldysh contours is given by
\begin{widetext}
\begin{equation}
\begin{split}
   e^{iS_{SK,\HKbar}(U)} &=  \prod_{x^0 \in \mathcal{T}_E} \left[  e^{-\frac{1}{g^2}\sum_{\vec{x},\mu<\nu} \Tr(2 - P_{x,\mu\nu} - P_{x,\mu\nu}^{-1})} \right] \\
    &\hspace{20pt} \times \prod_{x^0 \in \mathcal{T}_M^+} \left[ \prod_{\vec{x},k} \mathcal{K}_M\left( P_{x,0k}, \frac{g^2}{2} \right)  e^{-\frac{i}{g^2}\sum_{\vec{x},i<j} \Tr(2 - P_{x,ij} - P_{x,ij}^{-1})} \right] \\
     &\hspace{20pt} \times \prod_{x^0 \in \mathcal{T}_M^-} \left[ \prod_{\vec{x},k} \mathcal{K}_M\left( P_{x,0k}, -\frac{g^2}{2} \right) e^{\frac{i}{g^2}\sum_{\vec{x},i<j} \Tr(2 - P_{x,ij} - P_{x,ij}^{-1})} \right].
    \label{eq:SHKSKdef}
    \end{split}
\end{equation}
The real-time transfer matrices for $x^0 \in \mathcal{T}_M^+$ and $x^0 \in \mathcal{T}_M^-$ correspond to $\hat{T}_{M,\HKbar}$ and $\hat{T}_{M,\HKbar}^\dagger$ respectively and are therefore unitary.
The imaginary-time transfer matrix for $x \in \mathcal{T}_E$ corresponds to $\hat{T}_{E,W}$ and is therefore positive.
With the contour ordering shown in Fig.~\ref{fig:contour}, the Schwinger-Keldysh partition function associated with $S_{SK,\HKbar}$ is equal to the Euclidean Wilson partition function (even before taking the continuum limit),
\begin{equation}
    Z_{SK,\HKbar} = \Tr\left( (\hat{T}_{M,\HKbar}^\dagger)^{T_M} \hat{T}_{M,\HKbar}^{T_M} \hat{T}_W^\beta \right) = \Tr\left( \hat{T}_W^\beta \right) = Z_{E,W}.
\end{equation}
It is noteworthy that either the Wilson or heat-kernel kinetic or potential terms could be used for $x^0 \in \mathcal{T}_E$, and in all cases the imaginary-time transfer matrix for $x^0\in \mathcal{T}_E$ would be positive.\footnote{It is also possible to construct a more general action with the same symmetries by introducing independent gauge couplings $g_1$ for the Wilson action for $x^0 \in \mathcal{T}_E$ and $g_2$ and $g_3$ for the kinetic and potential terms, respectively, of the $\HKbar$ action for $x^0 \in \mathcal{T}_M^\pm$. Choosing a different trajectory in this three-dimensional coupling space besides the choice $g_1 = g_2 = g_3$ in Eq.~\eqref{eq:SHKSKdef} would modify the LGT spectrum in the Euclidean and Minkowski regions (which in general differ by lattice artifacts) and the approach to the continuum limit defined by $\lim_{g_1,g_2,g_3 \rightarrow 0}$. The investigation of lattice artifacts and the existence of a continuum limit for the trajectory $g_1=g_2=g_3$ is deferred to future work. }
However, the choice of Wilson kinetic term for $x^0 \in \mathcal{T}_E$, as well as the choice of Wilson potential term for $x^0 \in \mathcal{T}_M^\pm$, will be seen below to be important for establishing convergence of Schwinger-Keldysh path integrals using path integral contour deformations analogous to those in Sec.~\ref{sec:HKconvergence}.

In order to apply a generalization of this contour deformation to Schwinger-Keldysh path integrals, auxiliary variables $n_{x,k}^A$ corresponding to the indices for all of the sums in the heat-kernel kinetic terms appearing Eq.~\eqref{eq:SHKSKdef} can be introduced and an $n$-dependent action can be defined as
\begin{equation}
\begin{split}
   e^{iS_{SK,\HKbar}(U,n)} &=  \prod_{x^0 \in \mathcal{T}_E} \left[  e^{-\frac{1}{g^2}\sum_{\vec{x},\mu<\nu} \Tr(2 - P_{x,\mu\nu} - P_{x,\mu\nu}^{-1})} \right] \\
    &\hspace{20pt} \times \prod_{x^0 \in \mathcal{T}_M^+} \left[ \prod_{\vec{x},k} \mathcal{N}\, \mathcal{J}(\{\phi_{x,0k},n_{x,k}\})\  e^{\frac{i}{g^2}\sum_{\vec{x},k,A}\left(\phi_{x,0k}^A + 2\pi n_{x,k}^A\right)^2 }  e^{-\frac{i}{g^2}\sum_{\vec{x},i<j} \Tr(2 - P_{x,ij} - P_{x,ij}^{-1})} \right] \\
     &\hspace{20pt} \times \prod_{x^0 \in \mathcal{T}_M^-} \left[ \prod_{\vec{x},k} \mathcal{N}\, \mathcal{J}(\{\phi_{x,0k},n_{x,k}\})\  e^{-\frac{i}{g^2}\sum_{\vec{x},k,A}\left(\phi_{x,0k}^A + 2\pi n_{x,k}^A\right)^2 } e^{\frac{i}{g^2}\sum_{\vec{x},i<j} \Tr(2 - P_{x,ij} - P_{x,ij}^{-1})} \right],
    \end{split}\label{eq:SKact2}
\end{equation}
\end{widetext}
with $\mathcal{J}(\{\phi_{x,0k},n_{x,k}\})$ defined in Eq.~\eqref{eq:KESUN}.
Non-equilibrium LGT observables with Schwinger-Keldysh contour path integral representations can be defined using the $\HKbar$ action as
\begin{equation}
   \left< \mathcal{O} \right>_{SK,\HKbar} = \frac{1}{Z_{SK,\HKbar}} \sum_{\{n\}} \int \mathcal{D}U \mathcal{O}(U)\ e^{iS_{SK,\HKbar}(U,n)}, \label{eq:SKHKbardef}
\end{equation}
where $Z_{SK,\HKbar} = \sum_{\{n\}} \int \mathcal{D}U e^{iS_{SK,\HKbar}(U,n)}$.
A straightforward generalization of the contour deformation introduced in Sec.~\ref{sec:HKconvergence} can be used to provide an absolutely convergent representation of $\left< \mathcal{O} \right>_{SK,\HKbar}$ as described below.

For Schwinger-Keldysh applications the path integrals involving gauge fields on $\mathcal{T}_E$ are convergent before applying contour deformations and do not have sign problems arising from the action. The gauge field variables on $\mathcal{T}_E$ can therefore be held fixed, and only the gauge field variables on $\mathcal{T}_M^\pm$ need to be actively deformed (induced transformations to plaquettes on $\mathcal{T}_E$ near the boundaries with $\mathcal{T}_M^\pm$ are discussed below). The fixed gauge field variables on $\mathcal{T}_E$ can then be used as boundary field configurations to perform a change of variables from $\{U_{x,\mu}\}$ to $\{P_{x,0k},U_{x,0}\}$ for $x^0 \in \mathcal{T}_M^\pm$.
The original gauge field variables $\{U_{x,\mu}\}$ will continue to be used a path integral variables for $x^0 \in \mathcal{T}_E$.
For $x^0\in\mathcal{T}_M^+$, this change of variables is given by the function $U_{x,\mu}(P_{x,0k},U_{x,0})$ defined as
\begin{equation}
\begin{split}
    U_{x,k} &= \left[ \prod_{n=0}^{(x^0-\eta-a)/a} U_{x-(n+1)a\hat{0},0}^{-1} P_{x-a\hat{0},0k}  \right] U_{(\eta,\vec{x}),k} \\
    &\hspace{20pt} \times\left[ \prod_{n=0}^{(x^0-\eta-a)/a} U_{x+a\hat{k}-(n+1)a\hat{0},0} \right],
    \end{split}
    \label{eq:UPdef+}
\end{equation}
where $\eta = -i(\beta - a)$ denotes the imaginary part of $x^0$ for $x^0 \in \mathcal{T}_M^\pm$.
An analogous change of variables valid for $x^0 \in \mathcal{T}_M^-$ is defined by
\begin{equation}
\begin{split}
    U_{x,k} &= \left[ \prod_{n=0}^{(L_T+\eta-x^0-a)/a} P_{x+an\hat{0},0k}^{-1} U_{x+an\hat{0},0} \right] U_{(0,\vec{x}),k} \\
    &\hspace{20pt} \times\left[ \prod_{n=0}^{(L_T+\eta-x^0-a)/a} U_{x+a\hat{k}+an\hat{0},0}^{-1}  \right].
    \end{split}
    \label{eq:UPdef-}
\end{equation}
Further introducing the timelike plaquette eigenvalues $e^{i\phi_{x,0k}^A}$ and eigenvalues $V_{x,0k}$ defined by Eq.~\eqref{eq:Pdiag} as well as the variables $\psi_{x,0k}^A$ with $A=1,\ldots,N-1$ defined by Eq.~\eqref{eq:psiphi} for which $SU(N)$ heat kernel takes an uncorrelated Gaussian form, the variables $\{\psi_{x,0k}^A, V_{x,0k}, U_{x,0}\}$ can be used as a set of independent path integral variables on $\mathcal{T}_M^\pm$. The redundancy introduced by treating $V_{x,0k}$ as independent variables is irrelevant by Eq.~\eqref{eq:meas_phivar}.
Deformed contours for $\psi_{x,0k}^A$ can then be introduced that are equal to the contour shown in Fig.~\ref{fig:contour} for $\mathcal{T}_M^+$ and equal to its reflection about the horizontal axis for $\mathcal{T}_M^-$.
The central segments of these contours correspond to the coordinate transformation $\psi_{x,0k}^A \rightarrow \widetilde{\phi}_{x,0k}^A(\phi_{x,0k}^A,n_{x,k}^A)$ given for $x^0 \in \mathcal{T}_M^\pm$ by
\begin{equation}
\begin{split}
     \widetilde{\psi}_{x,0k}^{A} &= \begin{cases} \psi_{x,0k}^A e^{\pm i \pi/4}  & m_{x,k}^A = 0 \\ 
      \psi_{x,0k}^A + (e^{\pm i \pi/4}-1)\pi \sign(m_{x,k}^A) & m_{x,k}^A \neq 0 \end{cases},
\end{split} \label{eq:SKtransform}
\end{equation}
where the $\pm$ signs in $e^{\pm i \pi/4}$ correspond to $x^0 \in \mathcal{T}_M^\pm$.
Defining $\widetilde{\phi}_{x,0k}^A$ in terms of $\widetilde{\psi}_{x,0k}^A$ and $m_{x,0k}^A$ using Eq.~\eqref{eq:phipsi}, this transformation turns the $n_{x,k}^A$ dependent oscillatory factors in Eq.~\eqref{eq:SKact2} into exponential damping factors that render the sums in Eq.~\eqref{eq:SKHKbardef} exponentially convergent as in Sec.~\ref{sec:HKconvergence}.
Integrals over the additional contour segments required to complete the contours in Fig.~\ref{fig:contour} cancel between terms with different $m_{x,k}^A$ after regularizing the sums with transformations of kinetic prefactor exactly equivalent to the ``Wick rotation'' procedure described in Sec.~\ref{sec:HKconvergence}  for $x^0 \in \mathcal{T}_M^+$ and with $\theta\rightarrow - \theta$ for $x^0 \in \mathcal{T}_M^-$.
Formally, Schwinger-Keldysh path integrals can then be defined by performing this cancellation with a generic Wick rotation angle and subsequently taking the Minkowski limit.

A noteworthy difference between Schwinger-Keldysh contour deformations as compared with the purely real-time case discussed in Sec.~\ref{sec:HKconvergence} is that the contour deformation associated with Eq.~\eqref{eq:SKtransform} induces a non-trivial transformation of the timelike plaquette $P_{x,0k}$ with $x^0 = -i(\beta - a) \in \mathcal{T}_E$.
If a heat-kernel kinetic term were used for the Euclidean segments, this induced transformation could introduce a divergence to the Euclidean heat-kernel sum for this boundary plaquette; however, the choice of the Wilson kinetic term for $x^0 \in \mathcal{T}_E$ ensures that the only infinite sums appearing are associated with $x^0 \in \mathcal{T}_M^\pm$  and therefore that the transformations of fields on $\mathcal{T}_E$ cannot spoil convergence. 

An absolutely convergent representation for Schwinger-Keldysh path integrals with the $\HKbar$ action is therefore given by
\begin{equation}
\begin{split}
    \left< \mathcal{O} \right>_{SK,\HKbar} &= \frac{1}{Z_{SK,\HKbar}} \sum_{\{n\}}  \int \mathcal{D}U\  J(U,n)\ \\
    &\hspace{10pt} \times \mathcal{O}(\widetilde{U}(U,n))\ e^{iS_{SK,\HKbar}(\widetilde{U}(U,n),n)},
    \end{split}
    \label{eq:OSKHKbartilde}
\end{equation}
where the full Jacobian $J(U,n)$ is given in analogy to Eq.~\eqref{eq:Jdeform} by
\begin{equation}
\begin{split}
    J(U,n) &= \left( \prod_{x \in \mathcal{T}_M^+,k} j_{x,k}  \right) \left(  \prod_{x \in \mathcal{T}_M^-,k}  j_{x,k} \right),
      \end{split}
    \label{eq:SKJdeform}
\end{equation}
where the contributions from each link are given by
\begin{equation}
    h_{x,k}  = \left[ \frac{\prod_{A<B} \left| e^{i\widetilde{\phi}_{x,0k}^{A}} - e^{i\widetilde{\phi}_{x,0k}^{B}} \right|^2}{\prod_{A<B} \left| e^{i\phi_{x,0k}^A} - e^{i\phi_{x,0k}^B} \right|^2} \right],
    \end{equation}
and
\begin{equation}
    \begin{split}
        j_{x,k} &= \begin{cases} e^{\pm i \pi/4} & m_{x,k}^A = 0,\ x^0 \in \mathcal{T}_M^\pm \\
        1 & m_{x,k}^A \neq 0 
        \end{cases}.
    \end{split}
\end{equation}

\subsection{Monte Carlo simulations of two-dimensional U(1) gauge theory}\label{sec:U1MC}

\begin{figure*}[t]
    \centering
    \includegraphics[width=.47\textwidth]{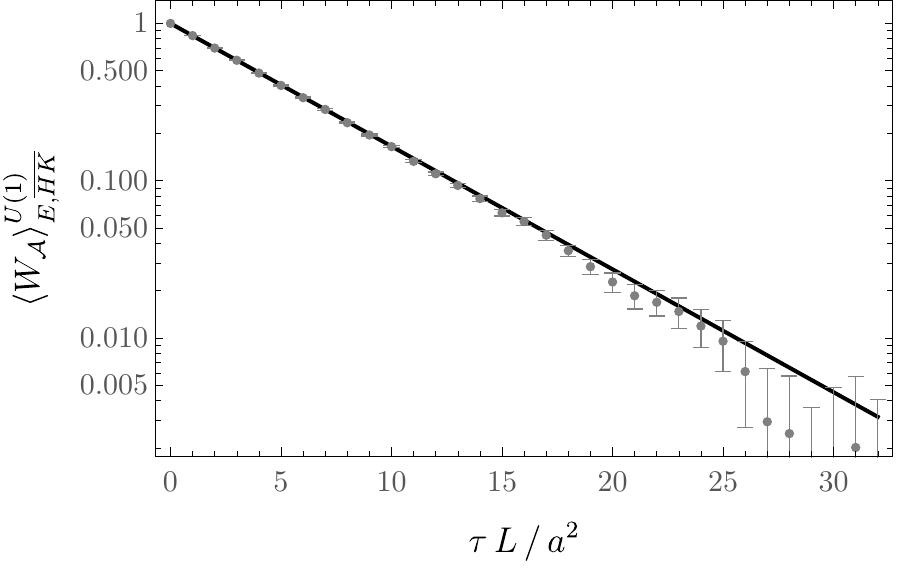}
    \includegraphics[width=.47\textwidth]{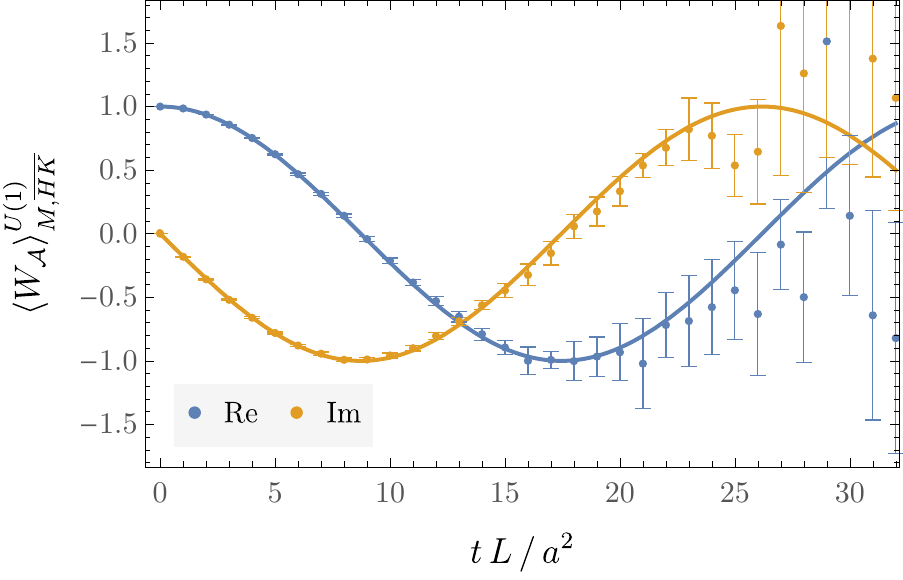}
    \caption{Wilson loop expectation values using the heat-kernel $U(1)$ LGT actions in $(1+1)$D Euclidean spacetime (left) and Minkowski spacetime (right). Solid curves show exact analytic results, while points with error bars show the central values and bootstrap $67\%$ confidence intervals for the numerical Monte Carlo calculations described in the main text.  \label{fig:U1}}
\end{figure*}

\begin{figure*}[t]
    \centering
    \includegraphics[width=.47\textwidth]{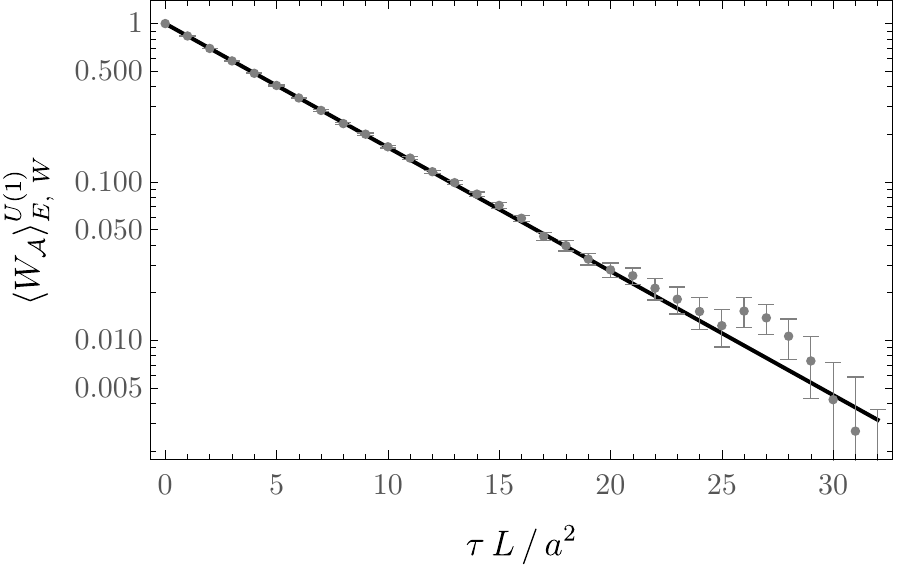}
    \includegraphics[width=.47\textwidth]{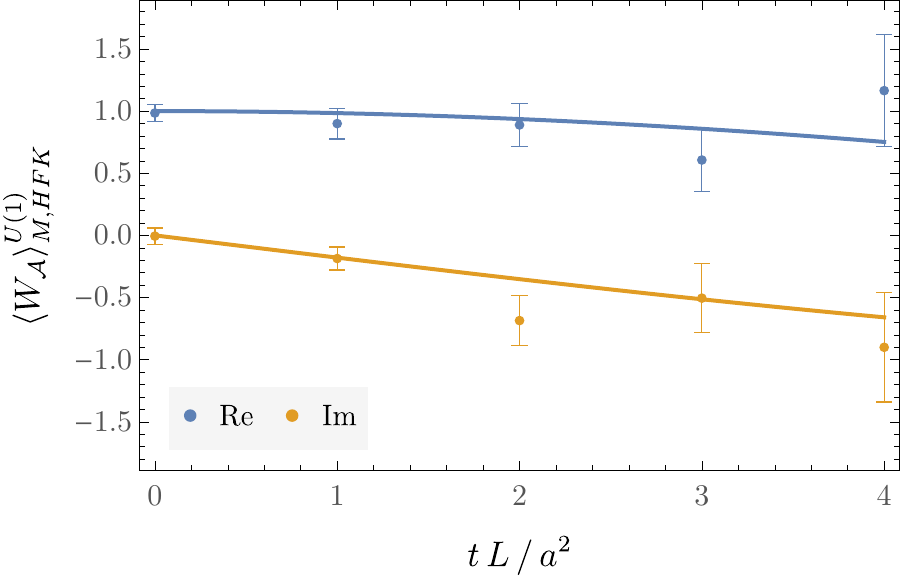}
    \caption{Wilson loop expectation values using the Wilson $U(1)$ LGT actions in $(1+1)$D Euclidean spacetime (left) and the HFK action in Minkowski spacetime (right) analogous to Fig.~\ref{fig:U1}. The sign problem for the HFK action using the deformed contour studied in this work is much more severe than the corresponding sign problem for the $\HKbar$ action and contour deformation, and identical sized statistical ensembles were used for the HFK action with a volume of $L_T L =4$ and the $\HKbar$ action with a volume of $L_T L = 32$. \label{fig:HFK}}
\end{figure*}

An exploratory study of the feasibility of Monte Carlo calculations of real-time LGT observables using the actions and deformations introduced above can be performed for $(1+1)$D $U(1)$ gauge theory with open boundary conditions, where the exact results obtained analytically in Sec.~\ref{sec:exact} can be used to verify the correctness of Monte Carlo results.
We specifically investigate the performance of the $\HKbar$ action and the HFK action, the only two actions for which the contour deformations above result in well-defined Monte Carlo sampling schemes.
In $(1+1)$D the remaining gauge links after gauge fixing are in one-to-one correspondence with the plaquettes $P_x \equiv P_{x,01}$.
The factorization of Wilson loop path integrals into products of one-dimensional integrals for each $P_x$ can be exploited to obtain a simple Monte Carlo simulation strategy in which plaquette variables $P_x$ and integer-valued auxiliary variables are drawn independently from the probability distribution
\begin{equation}
    p_{M,\HKbar}(P_x,n_x) \propto |J(P_x,n_x)| e^{-\Im[S_{M,\HKbar}(\widetilde{P}_x, n_x)]},
    \label{eq:pMHKdef}
\end{equation}
using the Metropolis algorithm.
In $(1+1)$D there is no potential term, and therefore $S_{M,\HKbar}$ is identical to $S_{M,HK}$; the former notation is used for consistency with the choice of the modified heat-kernel action as a unitary and convergent action in higher dimensions.
Expectation values using the deformed path integral, as defined in Eq.~\eqref{eq:OHKtilde}, are then computed from the sample mean of an ensemble of $N_{\rm cfg}$ plaquettes sampled independently for each lattice site,
\begin{equation}
\begin{split}
    \left< \mathcal{O} \right>_{M,\HKbar} &\approx  
    \frac{1}{\hat{Z}_{M,\HKbar}} \frac{1}{N_{\rm cfg}} \sum_{i=1}^{N_{\rm cfg}} \mathcal{O}(\widetilde{P}(P_x^i,n_x^i)) \\
    &\hspace{20pt} \times  e^{i\text{Arg}[J(P_x^i,n_x^i)]} e^{i \Re[S_{M,\HKbar}(\widetilde{P}_x^i, n_x^i)]}, \\
    \hat{Z}_{M,\HKbar} &\equiv \frac{1}{N_{\rm cfg}}    \sum_{i=1}^{N_{\rm cfg}}  e^{i\text{Arg}[J(P_x^i,n_x^i)]} e^{i \Re[S_{M,\HKbar}(\widetilde{P}_x^i, n_x^i)]}.
\end{split}
\label{eq:OHKtildeMC}
\end{equation}

Results for the expectation values of Minkowski Wilson loops as a function of area $t L$ are shown in Fig.~\ref{fig:U1} for an ensemble of $N_{\rm cfg} = 50,000$ configurations for a lattice with total area $L_T L = 32 a^2$  and gauge coupling $e = 0.6$. In $(1+1)$D with OBCs these results are only sensitive to the total area $L_T L$ and the total Wilson loop area $t L$, independent of the shape, due to factorization of the path integral.
Results for Euclidean Wilson loop expectation values obtained using the heat-kernel action for an ensemble with the same number of configurations, lattice area, and gauge coupling are also shown for comparison in Fig.~\ref{fig:U1}.
Good agreement between numerical Monte Carlo and exact analytic results is obtained in both Euclidean and Minkowski cases.
The inclusion of integer-valued auxiliary variables is not found to significantly increase autocorrelation times or introduce other undesirable numerical features in comparison to calculations with the Wilson action.

In this simple theory, similar statistical precision is achieved in the Euclidean and Minkowski cases.
This can be understood by noting that $e^{i S_{M,\HKbar}^{U(1)}(\widetilde{U})} = e^{-S_{E,\HKbar}^{U(1)}(U)}$ and therefore that the probability distribution
\begin{equation}
    p_{E,\HKbar}(P_x,n_x) \propto  e^{-S_{E,\HKbar}(P_x, n_x)},
        \label{eq:pEHKdef}
\end{equation}
is identical to $p_{M,\HKbar}(P_x,n_x)$ for the case of $G = U(1)$ in $(1+1)$D. In other words, the contour deformation identified for the $U(1)$ heat-kernel action completely removes phase fluctuations in $(1+1)$D,
and the reweighting factor multiplying the observable $\mathcal{O}$ in Eq.~\eqref{eq:OHKtildeMC} is exactly equal to unity.
The only difference between Euclidean and Minkowski results is the use of the original observable $\mathcal{O}(U)$ or the transformed observable $\mathcal{O}(\widetilde{U}(U,n))$.
Writing the Wilson loop in terms of the plaquette variables gives
\begin{equation}
\begin{split}
    \wloop_{\mathcal{A}}(P) &=  \prod_{x \in \mathcal{A}} P_x = \prod_{x \in \mathcal{A}} e^{i \phi_x} ,
    \end{split}
\end{equation}
from which $\wloop_{\mathcal{A}}(\widetilde{P}(P,n))$ can be explicitly constructed as
\begin{equation}
\begin{split}
   \wloop_{\mathcal{A}}(\widetilde{P}(P,n)) &= \prod_{x \in \mathcal{A}} \widetilde{P}_x(P,n)  \\
    &= \prod_{x \in \mathcal{A}} e^{ \sqrt{i} \phi_x + \sqrt{i}\sign(n_x)}.
    \end{split}
\end{equation}
In the Euclidean case, a StN problem arises because the average Wilson loop scales as $e^{-L\tau}$ while the variance is $O(1)$ for all $L\tau$.
In the Minkowski case, the magnitude of each term in the product fluctuates as well as the phase since $\Re[\sqrt{i}]=\Im[\sqrt{i}]=1/\sqrt{2}$ and the variance increases exponentially while the average Wilson loop is $O(1)$.
The result is similar exponential StN degradation for Minkowski and Euclidean Wilson loops as seen in Fig.~\ref{fig:U1}.
This level of precision for Minkowski observables is remarkable given the severity of the sign problem before applying the contour deformation in Eq.~\eqref{eq:HKdeform0}, but it should not necessarily be expected in higher dimensions, where potential terms are present and $\Re[S_{M,\HKbar}^{U(1)}(\widetilde{U})] \neq 0$.

An analogous Monte Carlo sampling strategy can be applied to calculations of observables using the HFK $U(1)$ action and the deformation in Eq.~\eqref{eq:HFKshift}. 
Plaquette variables $P_x$ and integer-valued auxiliary variables $r_x$ are sampled independently from the distribution
\begin{equation}
    p_{M,HFK}^{U(1)}(P_x,n_x) \propto e^{-\Im[S_{M,HFK}(\widetilde{P}_x,r_x)]}
\end{equation}
using the Metropolis algorithm, and Wilson loop results are computed from ensemble averages defined in analogy to Eq.~\eqref{eq:OHKtildeMC}.
These results can be contrasted against the equivalent Euclidean theory given by the Wilson gauge action, using identical coupling $e$. In both cases, the coupling is tuned to $e = 0.542947$ so that the exact string tension $\sigma_{E,W}^{U(1)}=\ln \left( \frac{I_0(1/e^2)}{I_1(1/e^2)} \right)$ is identical to the exact string tension $\sigma_{E,\HKbar}^{U(1)} = e^2 / 2$ for the heat-kernel action with $e=0.6$.
Results for Monte Carlo calculations using this gauge coupling and the same ensemble size, $N_{\rm cfg} = 50,000$, and lattice area as the heat-kernel calculations are shown in Fig.~\ref{fig:HFK}. Comparison to the Euclidean heat-kernel results in Fig.~\ref{fig:U1} shows that the two Euclidean actions achieve similar precision as well as good agreement with the corresponding (identical) exact results.
Significantly lower precision is achieved by the Minkowski HFK action in comparison to $\HKbar$ action calculations above, which can be attributed to the presence of a reweighting factor $e^{i\Re[S_{M,HFK}^{U(1)}(\widetilde{U},r)]}$ with severe phase fluctuations whose StN is observed to decrease exponentially with increasing lattice area $L_T L$.
The results shown in Fig.~\ref{fig:HFK} use a lattice with area $L_T L = 4 a^2$. A statistical ensemble many orders of magnitude larger would be required to extend these results to a lattice with area comparable to the $L_T L = 32a^2$ areas explored in the $\HKbar$ case.
Alternatively, a more sophisticated contour deformation than the simple shift in Eq.~\eqref{eq:HFKdeform} might be able to significantly improve the sign/StN problem associated with $e^{i\Re[S_{M,HFK}^{U(1)}(\widetilde{U},r)]}$ while maintaining the convergence properties. Studying the best choice of action and deformation for practical simulations is deferred to future work.

\subsection{Monte Carlo simulations of two-dimensional SU(3) gauge theory}\label{sec:SU3MC}

\begin{figure*}[t]
    \centering
    \includegraphics[width=.47\textwidth]{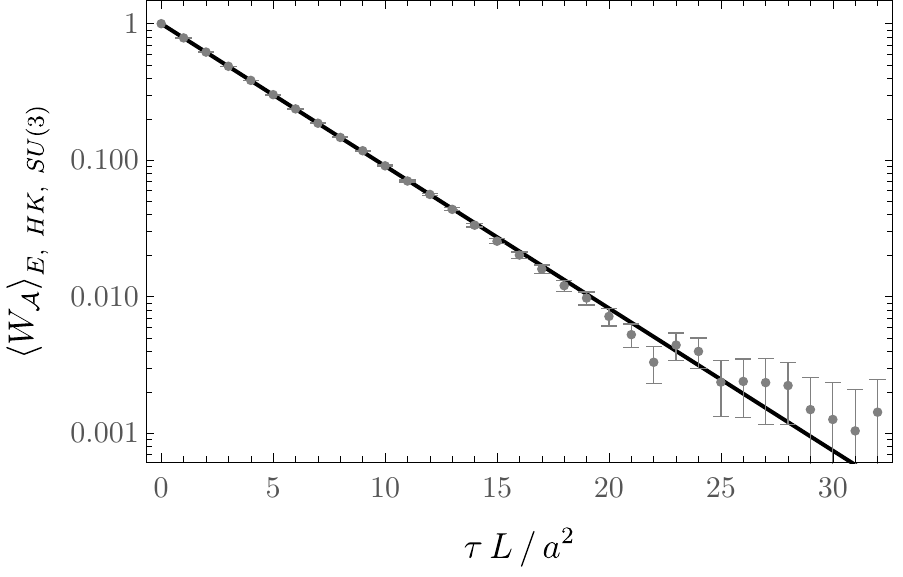}
    \includegraphics[width=.47\textwidth]{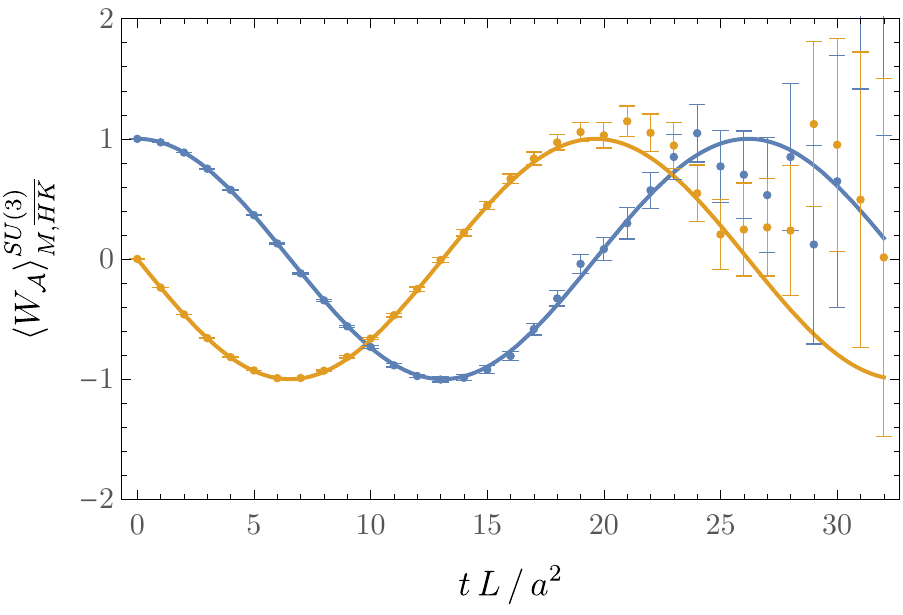}
    \caption{Wilson loop expectation values using the $\HKbar$ $SU(3)$ LGT actions in $(1+1)$D Euclidean spacetime (left) and Minkowski spacetime (right). Solid curves show exact analytic results, while points with error bars show the central values and bootstrap $67\%$ confidence intervals for the numerical Monte Carlo calculations described in the main text.  \label{fig:SU3}}
\end{figure*}

Exactly as in the $U(1)$ case, an exploratory study of the feasibility of numerical Monte Carlo calculations of real-time non-Abelian gauge theory observables can be performed in $(1+1)$D with OBCs and validated against exact results from Sec.~\ref{sec:exact}. The $\HKbar$ action is the only unitary $SU(3)$ LGT action studied in this work for which a contour deformation is obtained that results in a well-defined distribution for Monte Carlo sampling.
Factorization of path integrals into products of one-dimensional integrals is again exploited to obtain a Monte Carlo algorithm in which $P_x \in SU(3)$ is sampled along with integer-valued auxiliary variables $n_x^A$ satisfying the constraint $\sum_A n_x^A = 0$ from the probability distribution defined in Eq.~\eqref{eq:pMHKdef}.
Expectation values are then approximated using Monte Carlo ensembles averages defined by Eq.~\eqref{eq:OHKtildeMC}.
Wilson loops factorize as
\begin{equation}
    \wloop_{\mathcal{A}}(P) = \frac{1}{N} \Tr\left( \prod_{x\in A} P_x \right),
\end{equation}
and the corresponding transformed observables are given by
\begin{equation}
    \wloop_{\mathcal{A}}(\widetilde{P}(P,n)) = \frac{1}{N} \Tr\left( \prod_{x\in A} \widetilde{P}_x(P,n) \right),
\end{equation}
where $\widetilde{P}_x(P,n)$ is defined by Eq.~\eqref{eq:Ptildedef}.

Fig.~\ref{fig:SU3} shows results from Monte Carlo calculations using the $SU(3)$ $\HKbar$ action with an ensemble of $N_{\rm cfg} = 50,000$ gauge field configurations, using a lattice with area $L_T L = 32 a^2$, and gauge coupling $g = 0.6$.
Results for the $SU(3)$ Euclidean heat-kernel action for an ensemble with identical parameters are also shown in Fig.~\ref{fig:SU3} for comparison.
Good agreement between numerical Monte Carlo results and analytic exact results is seen for both Euclidean and Minkowski results.
Unlike the $U(1)$ case, $\Re[S_{M,\HKbar}^{SU(3)}(\widetilde{U},n)] \neq 0$ because $\mathcal{J}(\{\widetilde{\phi}\},\{n\})$ is complex, and further phase fluctuations arise from the contour deformation Jacobian.
These phase fluctuations are observed to be relatively mild in practice, and the average reweighting factor $e^{i\Re[S_{M,\HKbar}^{SU(3)}(\widetilde{U},n)]} e^{i \text{Arg}[J(P_x,n_x)]}$ appearing in the denominator of Eq.~\eqref{eq:OHKtildeMC} can be calculated with sub-percent-level precision for the lattice area and ensemble size used here.
The precision obtained for Minkowski and Euclidean $SU(3)$ Wilson loop results is similar, and the familiar exponential StN degradation with loop area is seen in both cases (note that this is distinct from, and much milder than, the expected sign problem extensive in total lattice volume associated with undeformed real-time simulation).
In higher dimensions, these phase fluctuations may become more severe and additional phase fluctuations will arise from potential terms in the action.
Further studies are needed to explore the feasibility of Monte Carlo calculations of real-time $SU(N)$ LGT in higher dimensions, and more sophisticated contour deformations or other approaches to improving the sign problem may be required to achieve precise results for four-dimensional real-time LGT observables.

\section{Conclusions and Outlook}\label{sec:concl}

Naive Wick rotation of the Euclidean Wilson LGT action leads to a Minkowski LGT with non-unitary time evolution, as pointed out in Ref.~\cite{Hoshina:2020gdy}.
It is demonstrated in this work that there is no way to recover a unitary time-evolution operator even in the continuum limit using the real-time Wilson LGT action.
Further, it is demonstrated that for exactly solvable examples in $(1+1)$D the weak-coupling limit associated with the continuum limit of Euclidean LGT does not exist using the real-time Wilson LGT action.
Real-time LGT calculations must therefore use an alternative Minkowski action.

One alternative action leading to unitary time-evolution for real-time LGT is provided by HFK in Ref.~\cite{Hoshina:2020gdy}.  The character expansion defining this action is divergent and requires regularization.
Here, a path integral contour deformation that renders the character expansion absolutely convergent is provided for the gauge group $U(1)$, making it possible to use Monte Carlo methods for this action. Exploratory Monte Carlo calculations of $U(1)$ Wilson loops in $(1+1)$D Minkowski spacetime were performed using the HFK action and found to be consistent with exact results obtained through analytic continuation of results for the Euclidean Wilson action. Finding a similar contour deformation is difficult for the $SU(N)$ HFK action because the character expansion  involves sums over irreducible representations of $SU(N)$, and understanding the convergence properties of these sums is non-trivial. However, no fundamental obstacle was encountered that would prevent the construction of such a deformation, and it may be an interesting subject of future work to find a convergent representation of the $SU(N)$ HFK path integral because of its closeness to the commonly employed Wilson gauge action.

Another class of actions leading to unitary time-evolution in Minkowski LGT is obtained in this work through analytic continuation of the Euclidean heat-kernel action.
Unitarity of the real-time transfer matrix depends only on using the heat-kernel or HFK kinetic term in the action and does not depend on the form of the potential term (besides the assumption that the potential term is real).
In particular, this permits the definition of a modified heat-kernel, or $\HKbar$, action using the potential term from the Wilson action.
The real-time $\HKbar$ action is defined by a divergent series required to ensure periodicity of angular variables, but by exploiting the Gaussian form of the heat kernel we obtain path integral contour deformations leading to absolutely convergent representations of $\HKbar$ path integrals for both $U(1)$ and $SU(N)$ real-time LGT.
Exploratory Monte Carlo calculations of $U(1)$ and $SU(3)$ Wilson loops in $(1+1)$D are performed and found to agree with exact results obtained through analytic continuation of results for the Euclidean heat-kernel action.

The contour deformations of the HFK and $\HKbar$ path integrals are crafted to provide a complete cancellation of the sign problem associated with the kinetic energy term in the weak coupling limit, beyond giving a convergent representation of these path integrals amenable to Monte Carlo sampling. In the $(1+1)$D proof-of-principle calculations, this is apparent as an exponential reduction in the sign problem associated with the reweighting factors $e^{i S_{M,HFK}}$ and $e^{i S_{M,\HKbar}}$, respectively. For the HFK action, the sign problem is not completely eliminated and the deformed theory still suffers from a signal-to-noise problem that grows exponentially with the lattice size, making numerical results less precise than in analogous Euclidean LGT calculations. On the other hand, the $\HKbar$ deformation completely removes the sign/StN problem for the $U(1)$ case and exponentially improves it for the $SU(3)$ case, in both cases resulting in comparable precision for Minkowski and Euclidean Wilson loop results. These results should however be considered specific to $(1+1)$D; gauge theories in $(1+1)$D do not include a potential term and therefore a careful analysis of the kinetic energy term is sufficient to directly find good contour deformations. It can be expected that sign/StN problems that are approximately or entirely solved in $(1+1)$D will be significantly worse in $(3+1)$D. Exploration of the contour deformations needed to practically tame sign/StN problems in higher dimensions is left to future work. It is possible that more sophisticated contour deformations must be constructed or that numerical optimization of a variational ansatz for the contour deformation will be necessary to make progress.

This work focused exclusively on pure gauge theory, but the inclusion of fermionic matter fields is not expected to spoil unitarity~\cite{Hoshina:2020gdy}.
Convergence of the $\HKbar$ action is not affected by the presence of fermionic matter since the fermion determinant does not explicitly depend on the heat-kernel sum index and therefore will only depend on $\sign(m_{x,k}^A)$ after applying our contour deformation.
Care should be taken with real-time fermion doublers and the continuum limits of lattice gauge theories including fermions, and a study of these subtleties is deferred to future work.

The absolutely convergent representation of $\HKbar$ path integrals defined here provides a suitable starting point in principle for Monte Carlo calculations of $SU(3)$ LGT in $(3+1)$D Minkowski spacetime and calculations of Schwinger-Keldysh path integrals in QCD.
If phase fluctuations arising from sign problems in $(3+1)$D can be tamed for some real-time LGT observables, significant strides could be made towards predicting real-time gauge theory observables relevant for phenomenology.

\acknowledgments{
The authors are grateful to Marcela Carena, William Detmold, Henry Lamm, Scott Lawrence, Yingying Li, Hersh Singh, Neill Warrington, Yikun Wang, and Uwe-Jens Wiese for helpful discussions. G.K.\ is supported in part by the U.S.\ DOE grant No.\ DE-SC0011090. This manuscript has been authored by Fermi Research Alliance, LLC under Contract No. DE-AC02-07CH11359 with the U.S. Department of Energy, Office of Science, Office of High Energy Physics. This work is supported by the National Science Foundation under Cooperative Agreement PHY-2019786 (The NSF AI Institute for Artificial Intelligence and Fundamental Interactions, \url{http://iaifi.org/}).
}

\appendix

\section{Scalar field theory real-time transfer matrix}\label{sec:scalar}
Lattice scalar field theory with an action of the form
\begin{equation}
    S_E(\varphi) = \sum_{\tau/a=0}^{N_T-1} \sum_{\mathbf{x}}  \left[ \frac{a^2(\varphi_{(\tau+a,\mathbf{x})}-\varphi_{(\tau,\mathbf{x})})^2}{2} + a V(\varphi_\tau) \right],
\end{equation}
where $V(\varphi)$ is the scalar field potential,
has a transfer matrix given by
\begin{equation}
   \teucl{}(\varphi,\varphi') = e^{-a V(\varphi)/2} \tkeucl{}(\varphi,\varphi') e^{-a V(\varphi)/2},
\end{equation}
where the kinetic term is included in
\begin{equation}
    \tkeucl{}(\varphi,\varphi') = \prod_{\mathbf{x}} e^{-(\varphi_\mathbf{x} - \varphi'_\mathbf{x})^2 a^2/2}.
\end{equation}
The Minkowski lattice scalar field theory defined by the action
\begin{equation}
    S_M(\varphi) = \sum_{t/a=0}^{N_T-1} \sum_{\mathbf{x}} \left[ \frac{a^2(\varphi_{(t+a,\mathbf{x})}-\varphi_{(t,\mathbf{x})})^2}{2} - a V(\varphi_t) \right],
\end{equation}
has a corresponding time-evolution operator
\begin{equation}
   \tmink{}(\varphi,\varphi') = e^{-iaV(\varphi)/2} \tkmink{}(\varphi,\varphi') e^{-iaV(\varphi')/2},
   \label{eq:Wscalar}
\end{equation}
where the kinetic term is included in
\begin{equation}
    \tkmink{}(\varphi,\varphi') = \prod_{\mathbf{x}} e^{i(\varphi_\mathbf{x} - \varphi'_\mathbf{x})^2 a^2/2}.
\end{equation}
Defining the measure $\mathcal{D}\mathcal{\varphi} = \prod_\mathbf{x}d \varphi_\mathbf{x}$ and $\delta(\varphi - \varphi') = \prod_\mathbf{x} \delta(\varphi_\mathbf{x} - \varphi_\mathbf{x}')$, it can be explicitly verified that $\tkmink{}$ is unitary after multiplying by a normalization factor of $\prod_{\mathbf{x}} \sqrt{ \frac{i a^2}{2\pi} }$,
\begin{equation}
    \begin{split}
    & \int \mathcal{D}\varphi^{\prime\prime} \;  \left[\prod_{\mathbf{x}}\frac{a^2}{2\pi}\right] \tkmink{}(\varphi,\varphi^{\prime\prime}) \tkmink{}^\dagger(\varphi^{\prime\prime},\varphi) \\
    &=  \prod_{\mathbf{x}}\frac{a^2}{2\pi} \int_{-\infty}^\infty d\varphi^{\prime\prime}_\mathbf{x}  \;  e^{i(\varphi_\mathbf{x} - \varphi^{\prime\prime}_\mathbf{x})^2 a^2/2} e^{-i(\varphi^{\prime\prime}_\mathbf{x} - \varphi^{\prime}_\mathbf{x})^2 a^2/2} \\
      &=  \prod_{\mathbf{x}} e^{i \varphi_\mathbf{x}^2 a^2/2} e^{-i \varphi^{\prime 2}_\mathbf{x} a^2/2} \left(\frac{a^2}{2\pi}\right) \int_{-\infty}^\infty d\varphi^{\prime\prime}_\mathbf{x}  \;   e^{ i\varphi^{\prime\prime}_\mathbf{x} (\varphi^{\prime}_\mathbf{x} - \varphi_\mathbf{x} ) a^2} \\
      &= \delta(\varphi - \varphi').
        \label{eq:scalarunitary}
    \end{split}
\end{equation}
The unitarity of $\tmink{}$ follows immediately from Eq.~\eqref{eq:Wscalar}.

\section{Non-unitarity of the real-time Wilson action}\label{sec:nonunitary-Wilson}

The character expansion coefficients for arbitrary $SU(N)$ groups can be calculated to demonstrate non-unitarity of the Wilson action as discussed for $U(1)$, $SU(2)$, and $SU(3)$ in Sec.~\ref{sec:wilson}. All ratios of character expansion coefficients $c_r^{M,SU(N)}(g^2) / c_{0}^{M,SU(N)}(g^2)$ must have unit norm for a given action to lead to unitary real-time evolution. In this appendix we numerically compute ratios $c_f^{M,W,SU(N)}(g^2) / c_0^{M,W,SU(N)}(g^2)$ to demonstrate non-unitarity of the Wilson action for $N \in \{2,\ldots,9\}$ and study the $g^2\rightarrow 0$ limit using a stationary phase expansion. Clear patterns are observed that suggest non-unitarity for larger $N$.

\begin{figure*}
    \centering
    \includegraphics[width=0.47\textwidth]{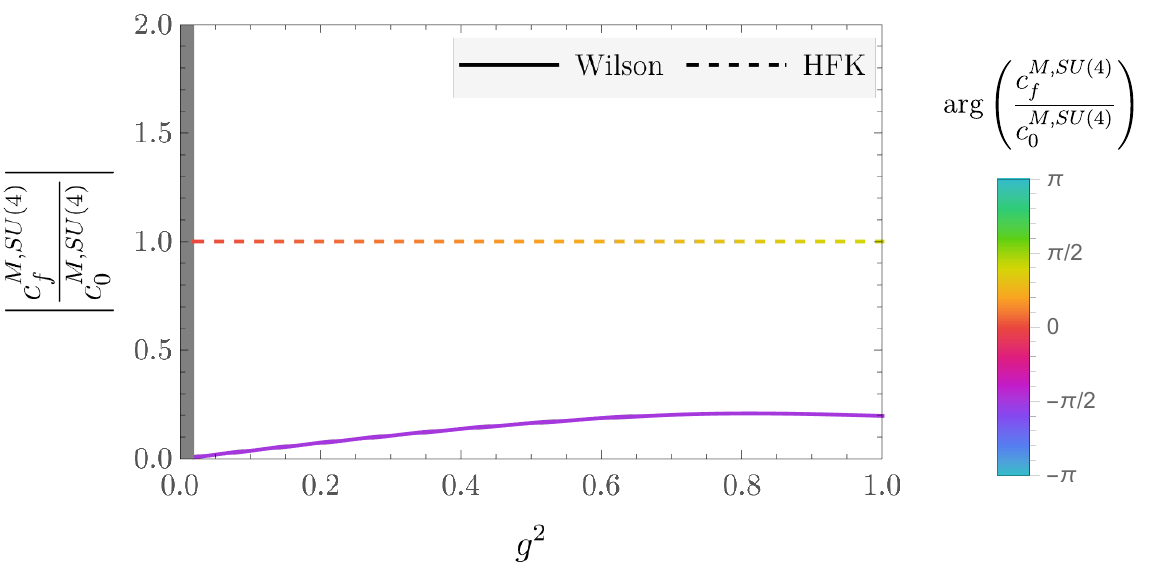} \hspace{25pt}
    \includegraphics[width=0.47\textwidth]{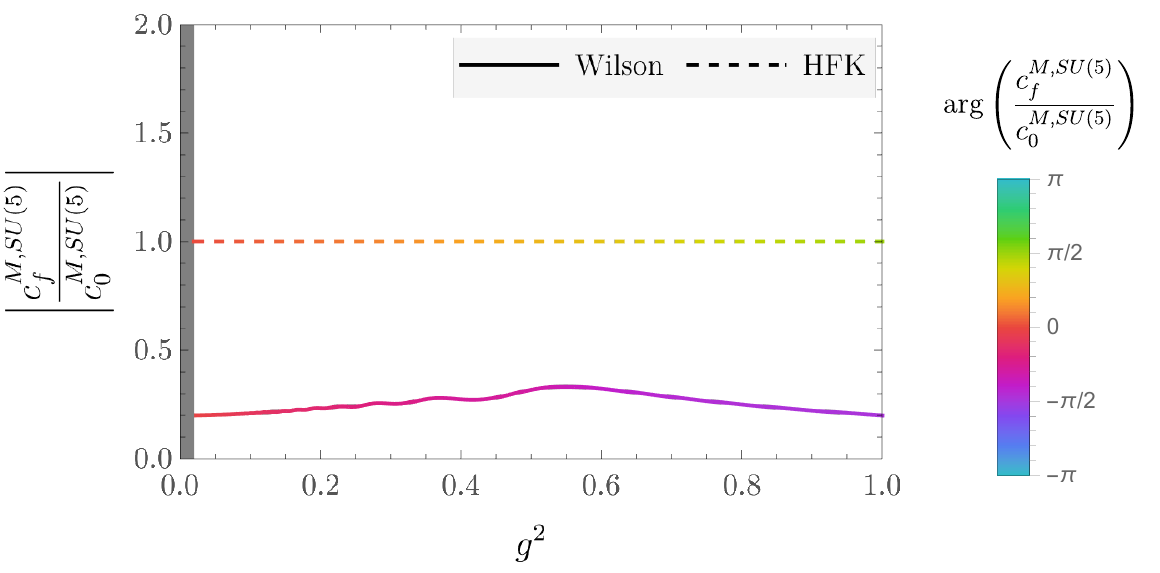} \\
    \includegraphics[width=0.47\textwidth]{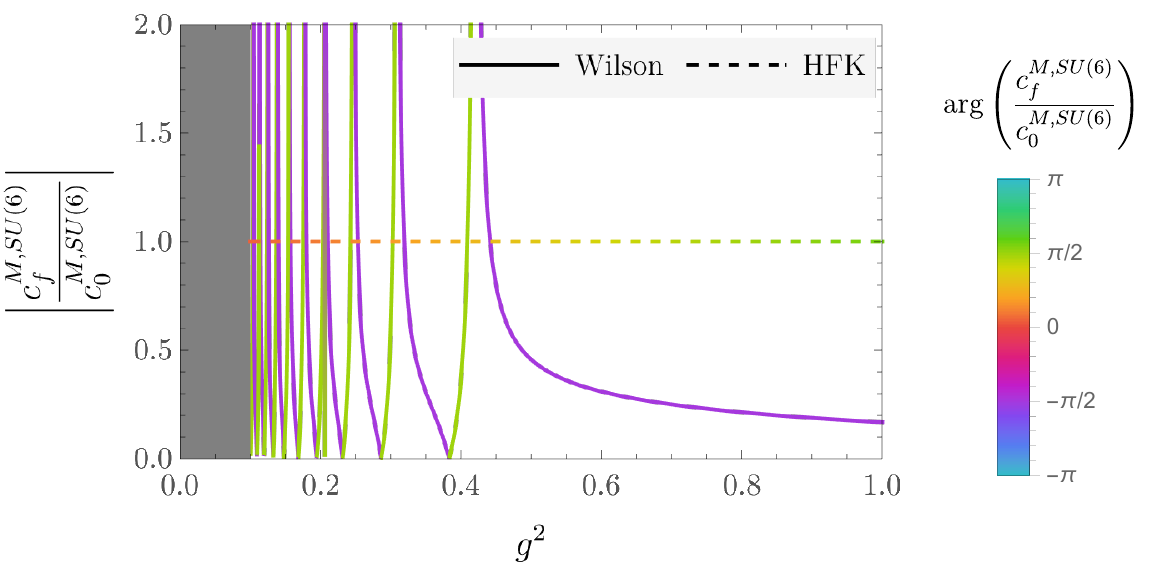} \hspace{25pt}
    \includegraphics[width=0.47\textwidth]{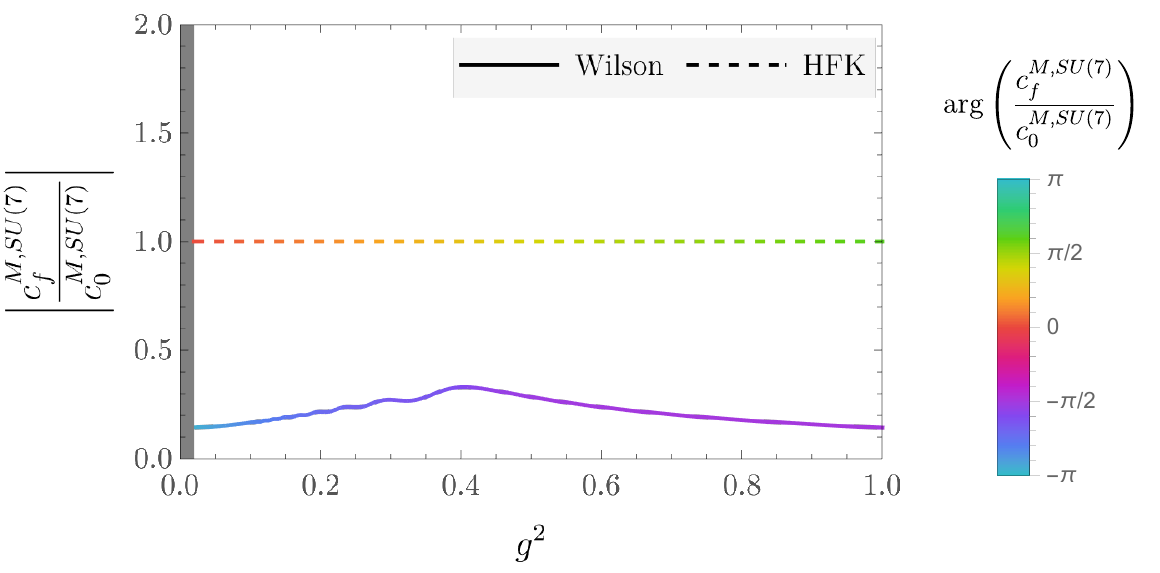} \\
    \includegraphics[width=0.47\textwidth]{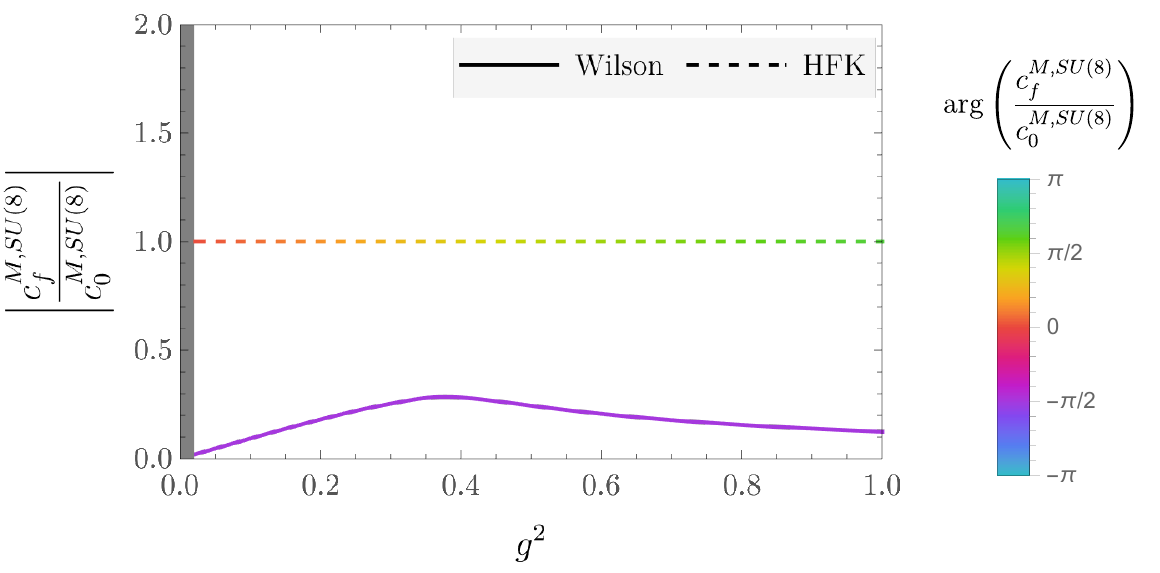} \hspace{25pt}
    \includegraphics[width=0.47\textwidth]{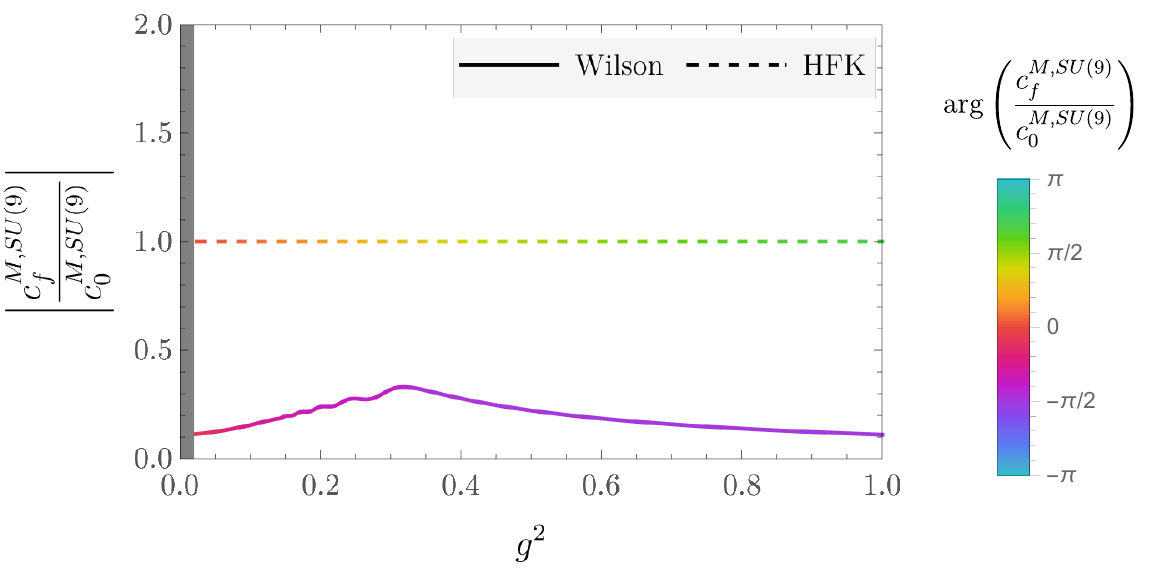}
    \caption{Ratios of the fundamental and trivial irrep character expansion coefficients for the Wilson and HFK actions for $SU(N)$ gauge theory with $N = 4,\ldots,9$. The Wilson action result for $SU(6)$ includes an infinite number of singularities that accumulate as $g^2 \rightarrow 0$ and is replaced by a gray background for $g^2 < 0.1$. Numerical precision limits the approach to $g^2 \rightarrow 0$ for other choices of $N$, and the results are replaced by a gray background for $g^2 < 0.02$. The Wilson action can be seen to have a non-unitary coefficient ratio almost everywhere in $g^2$ for all $N$. In the $SU(6)$ case, the limit $g^2 \rightarrow 0$ is ill-defined due to the repeated singularities, while for other $SU(N)$ there is an apparently well-defined limit which is non-unitary. The limit values are explained by a stationary phase expansion in the text. The HFK action has a unitary ratio for all choices of $g^2$ and $N$.
    \label{fig:Wchar49} }
\end{figure*}

The imaginary-time Wilson action character expansion coefficients are known for general $SU(N)$ gauge groups~\cite{Drouffe:1983fv},
\begin{equation}
   \begin{split}
      c_r^{E,W,SU(N)}(g^2) = \sum_{q=-\infty}^\infty \det(\mathcal{Z}^{q;r}(g^2)),
   \end{split}\label{eq:eigenHaar3}
\end{equation}
where the entries of the matrix $\mathcal{Z}^{q;r}$ are given by
\begin{equation}
   \begin{split}
      \mathcal{Z}^{q;r}_{IJ} &= I_{r_J + q+I-J}\left(\frac{2}{g^2}\right),
   \end{split}\label{eq:Zmat}
\end{equation}
where $I,J \in \{1,\ldots,N\}$ and the representation $r$ is labeled by a set of integers $r_J$ ordered so that $r_I \geq r_J$ for $I < J$ and $r_N = 0$ for $SU(N)$ groups~\cite{Drouffe:1983fv}.
The corresponding real-time Wilson action character expansion coefficients are simply obtained using Eq.~\eqref{eq:ermrWilson} as
\begin{equation}
   \begin{split}
      c_r^{M,W,SU(N)}(g^2) = \sum_{q=-\infty}^\infty \det(\mathcal{Z}^{q;r}(ig^2)).
   \end{split}\label{eq:eigenHaar3}
\end{equation}
The trivial representation $r=0$ corresponds to $r_J=0$ for all $J$, while the fundamental representation $r=f$ corresponds to $r_1 = 1$ and $r_J = 0$ for $J>1$, which permits calculation of character expansion ratios
\begin{equation}
    \frac{c_f^{M,W,SU(N)}}{c_0^{M,W,SU(N)}} = \frac{ \sum_{q=-\infty}^\infty \det(\mathcal{Z}^{q;f}(g^2)) } {\sum_{q=-\infty}^\infty \det(\mathcal{Z}^{q;0}(g^2)) }. \label{eq:charratio}
\end{equation}
In order to numerically evaluate Eq.~\eqref{eq:charratio}, the infinite sums can be replaced by finite sums $\sum_{q=-\Lambda}^\Lambda$ with exact results obtained in the limit $\Lambda \rightarrow \infty$.
Although the infinite sum converges for any fixed $g^2 > 0$, the truncation errors arising from using a fixed cutoff $\Lambda$ increase rapidly as $g^2$ is decreased towards zero.
Numerical results for $N\in \{2,3\}$ shown in Fig.~\ref{fig:Wchar23} and for $N\in \{4,\ldots,9\}$ shown in Fig.~\ref{fig:Wchar49} are computed using $\Lambda = 100$ for $N\geq 3$ and are verified to be indistinguishable with results obtained using lower values of $\Lambda$ for $g^2 \geq 0.1$ for $N \equiv 2 \pmod{4}$ and $g^2 \geq 0.02$ for other values of $N$.
For $N=2$ the equivalent expressions for $c_r^{M,W,SU(2)}$ without infinite series given in Eq.~\eqref{eq:mrWilsonSU2} are used for numerical calculations.

Non-unitarity of the Wilson action at a range of bare coupling $g^2$ corresponding to a range of non-zero lattice spacings would be acceptable if unitarity were recovered in the continuum limit. As can be clearly seen in Fig.~\ref{fig:Wchar23} and Fig.~\ref{fig:Wchar49}, a hypothetical unitary continuum limit is not approached at all, even for small $g^2$ that would correspond to very fine lattice spacings. For choices of $SU(N)$ with $N \equiv 2 \pmod{4}$, the $g^2 \rightarrow 0$ limit does not exist at all, due to repeated divergences in the ratio. For other choices of $N$, the continuum limit does appear to be well-defined, though non-unitary; in particular, for $N \equiv 0 \pmod{4}$ the ratio appears to approach $0$ in the continuum limit, while for other groups the ratio appears to approache a fixed non-zero value.

This analysis can be made precise by considering a stationary phase expansion about $g^2 = 0$ to determine the asymptotic values of ratios, when such asymptotic values exist. The character expansion coefficients for a single link variable are given by the integral
\begin{equation}
  c_r^{M,W,SU(N)} = \int dP \, \frac{\chi_r(P)^*}{d_r} \, e^{\frac{2 i}{g^2} \Re\Tr(1-P)},
\end{equation}
which is dominated by the local minima, maxima, and saddle points of the exponent $\frac{2 i}{g^2} \Re\Tr(1-P)$ in the limit $g^2 \rightarrow 0$. The stationary phase expansion provides a systematic means of expanding about $g^2 = 0$. To start, the kinetic integral kernel can be explicitly written in terms of the eigenvalues $\lambda_1, \dots, \lambda_N \equiv e^{i \phi_1}, \dots, e^{i \phi_N}$, and the eigenvector part of the measure can be integrated away to give a universal constant. The Haar measure over eigenvalues must also be included, giving in total
\begin{equation}
\begin{split}
  c_r^{M,W,SU(N)} &\propto \int \prod_{A} d\phi^A \left[ \prod_{A < B} \left| e^{i\phi^A} - e^{i\phi^B} \right|^2 \right] \, \\
  &\hspace{20pt} \times e^{\frac{2 i N}{g^2} - \frac{2 i}{g^2} \sum_A \Re(e^{i\phi^A})}.
  \end{split}
\label{eq:crMeigIntegral}
\end{equation}
The constant of proportionality is independent of the representation $r$, and will cancel from ratios.

Finding the stationary points of the action requires addressing the unit determinant constraint for the $N$ eigenvalues. One approach to including this constraint is through the use of a Lagrange multiplier term, $\alpha (2\pi m - \sum_A \phi^A)$ with arbitrary $m\in \mathbb{Z}$. The stationary points of the action with respect to the $N-1$ free directions are then given by the solutions to
\begin{equation}
\begin{aligned}
    0 &= \frac{\partial}{\partial \phi^A} \left[ \frac{2i}{g^2} \Re\sum_B e^{i\phi^B} + \alpha(2\pi m - \sum_B \phi^B) \right] \\
    &= -\frac{2i}{g^2} \Im e^{i \phi^A} - \alpha.
\end{aligned}
\end{equation}
This is solved by the group elements for which the imaginary component of every eigenvalue is identical; for example, this includes the identity element, which is also the global minimum of the Wilson action with Euclidean signature. For specific choices of $N$, other solutions are also possible. The Haar measure in square brackets in Eq.~\eqref{eq:crMeigIntegral} suppresses eigenvalue degeneracy, so the stationary points with lowest eigenvalue degeneracy will dominate the integral. Table~\ref{tab:char-stationary-points} gives these lowest-degeneracy stationary points for $N \in \{ 2, \dots, 9 \}$.

\begin{table*}[t!!]
    \centering
    \begin{tabular}{>{\centering\arraybackslash}p{1.2cm} @{\hspace{.7cm}} c @{\hspace{.7cm}} >{\centering\arraybackslash}p{1.75cm}}
    \toprule
         $N$ & Stationary points (lowest degeneracy) & $\chi_f$ \\
    \colrule
         2 & $(1,1)$ / $(-1,-1)$ & $2$ / $-2$ \\
         3 & $(1,-1,-1)$ & $-1$ \\
         4 & $(e^{i\theta}, e^{i\theta}, e^{i(\pi-\theta)}, e^{i(\pi-\theta)})$ & $4 i \sin(\theta)$ \\
         5 & $(1,1,1,-1,-1)$ & $1$ \\
         6 & $(1,1,1,1,-1,-1)$ / $(1,1,-1,-1,-1,-1)$ & $2$ / $-2$ \\
         7 & $(1,1,1,-1,-1,-1,-1)$ & $-1$ \\
         8 & $(e^{i\theta}, e^{i\theta}, e^{i\theta}, e^{i\theta}, e^{i(\pi-\theta)}, e^{i(\pi-\theta)}, e^{i(\pi-\theta)}, e^{i(\pi-\theta)})$ & $8 i \sin(\theta)$ \\
         9 & $(1,1,1,1,1,-1,-1,-1,-1)$ & $1$ \\
    \botrule
    \end{tabular}
    \caption{The group elements that are stationary points of the Wilson action for the group $SU(N)$ with $N = 2, \dots, 9$; only the elements with lowest eigenvalue degeneracy are listed, as these dominate the character coefficient integrals due to the suppression of degenerate eigenvalues by the Haar measure. Group elements are reported in terms of their spectra (i.e.~sets of eigenvalues) because the Wilson action is invariant under transformations of the eigenvectors alone. Stationary points always include the center elements of the group, but additional elements may also solve the constraints. The center elements have maximum eigenvalue degeneracy, so when other solutions exist they will dominate the integrals. The fundamental representation character is given for all cases with a single dominant stationary phase.}
    \label{tab:char-stationary-points}
\end{table*}

For a generic $N$, we can label the set of $k$ dominant stationary points $P_1, \dots, P_k \in SU(N)$. Expanding the evaluation of character expansion coefficients in the limit $g^2 \rightarrow 0$ gives
\begin{equation}
\begin{aligned}
  c_r^{M,W, SU(N)} &\approx \sum_{\ell=1}^{k} A_{\ell} \left[ \prod_{A<B} \left| e^{i \phi^A_\ell} - e^{i \phi^B_\ell} \right|^2 \right] \frac{\chi_r(P_{\ell})^*}{d_r} , \label{eq:stationaryphase}
\end{aligned}
\end{equation}
where the eigenvalues of $P_\ell$ are denoted by $e^{i\phi^A_\ell}$.
The Haar measure factor in brackets in Eq.~\eqref{eq:stationaryphase} vanishes exactly at the stationary points, and to be precise one should expand to higher order in the stationary phase approximation and then determine the limiting values of character expansion coefficient ratios using these higher-order results. However, these higher-order terms are determined by the structure of the Haar measure and the action about the stationary points, which is independent of the representation $r$, and coefficients of these terms will cancel in ratios of character expansion coefficients. Ratios of the approximate expressions in Eq.~\eqref{eq:stationaryphase} will therefore give correct $g^2\rightarrow 0$ results for ratios of character coefficients under the assumption that the stationary phase expansion has a non-trivial radius of convergence about $g^2 = 0$.

For some choices of $N$, it is apparent from Fig.~\ref{fig:Wchar23} and Fig.~\ref{fig:Wchar49} that the $g^2\rightarrow 0$ limits of character expansion coefficient ratios do not exist. These values align with the cases of $N \equiv 2 \pmod{4}$ where multiple stationary points contribute to the value of the integral as shown in Table~\ref{tab:char-stationary-points}, and one expects the divergences to correspond to cancellations between these contributions to $c_0^{M,W,SU(N)}(g^2)$ for particular values of $g^2$. 

For other $N$, the character expansion coefficients ratios shown in Fig.~\ref{fig:Wchar23} and Fig.~\ref{fig:Wchar49} appear to approach a limiting value as $g^2 \rightarrow 0$, suggesting that the stationary phase expansion can be reliably used to understand the $g^2 \rightarrow 0$ limit in these cases. For $N \equiv 0 \pmod{4}$ a class of group elements parameterized by an angle $\theta \in [-\pi/2,\pi/2]$ contributes to the value of the integral, as detailed in Table~\ref{tab:char-stationary-points}. For the ratio $c_f^{M,W,SU(N)} / c_0^{M,W,SU(N)}$ these group elements are weighted by the fundamental character, which is proportional to $\sin(\theta)$. Integrating over $\theta \in [-\pi/2,\pi/2]$ gives cancelling contributions from this set, and indeed the ratios can be seen to vanish as $g^2 \rightarrow 0$ for these cases in Fig.~\ref{fig:Wchar49}. 

In the remaining case of odd $N$, character expansion coefficient ratios are dominated by a single stationary point, and labelling the corresponding group element $P_1$, the would-be continuum limit of these ratios is given by
\begin{equation}
    \lim_{g^2 \rightarrow 0} \frac{c_f^{M,W,SU(N)}}{c_0^{M,W,SU(N)}} \approx \frac{\chi_f(P_1)^*}{N}.
\end{equation}
The fundamental character $\chi_f(P_1)^*$ in these cases corresponds to $\pm 1$, and the ratio therefore approaches $\pm 1/N$ with the sign depending on the particular choice of $N$. This behavior is observed for $SU(3)$ in Fig.~\ref{fig:Wchar23} as well as $SU(5)$, $SU(7)$, and $SU(9)$ in Fig.~\ref{fig:Wchar49}.

For every choice of $N \in \{2,\ldots,9\}$, the $g^2\rightarrow 0$ limit of the ratio $c_f^{M,W,SU(N)} / c_0^{M,W,SU(N)}$ either does not exist or does not approach a unit-norm value. The Wilson action thus does not give rise to a unitary real-time transfer matrix in the would-be continuum limit.

\section{Divergence of unitary real-time LGT actions}\label{sec:divergent-actions}

Suppose that a real-time LGT action has the form
\begin{equation}
    S_M(U) = a \sum_{t/a=0}^{N_T-1} \sum_{\vec{x}} [ k(P_{x,0k}) - v(P_{x,ij})],
\end{equation}
where $P_{x,\mu\nu}$ is the plaquette obtained from $U_{x,\mu}$. In temporal gauge $P_{x,0k} = U_{x,k} U_{x+a\hat{0},k}^\dagger$ and therefore the kinetic-energy evolution operator $\hat{R}_M$ defined by $\hat{R}_M(U_{t+a},U_t) = e^{i a \sum_{\vec{x}} k(U_{x,k} U_{x+a\hat{0},k}^\dagger)}$ has the character expansion
\begin{equation}
    \hat{R}_M(U,U') = \prod_{\vec{x},k} \left[ \sum_r d_r c_r(g^2) \chi_r(U_{\vec{x},k}' U_{\vec{x},k}^\dagger ) \right],
\end{equation}
with a corresponding gauge-invariant series definition for $e^{i S_M(U)}$ given by
\begin{equation}
    e^{i S_M(U)} = \prod_{x} e^{- i a v(P_{x,ij})} \prod_{\vec{x},k} \left[ \sum_r d_r c_r(g^2) \chi_r(P_{x,0k}) \right].\label{eq:gendivergence}
\end{equation}
If $S_M(U)$ is a unitary real-time LGT action, that is if $\hat{R}_M$ and therefore the corresponding real-time transfer matrix $\hat{T}_M$ are unitary, then it follows from the character decomposition in Eq.~\eqref{eq:Mchar} that $|c_r(g^2)|=1$.
The fact that each character is normalized by $\int dU |\chi_r(U)|^2 = 1$ implies that $\chi_r(P_{x,0k})$ is non-vanishing for a set of non-zero measure in path integrals involving $e^{i S_M(U)}$, and it therefore follows from  $|c_r(g^2)|=1$ that the $r$-th term in the sum in Eq.~\eqref{eq:gendivergence} using any enumeration of the representations of $G$ does not vanish as $r\rightarrow \infty$.
It immediately follows that the character expansion for $e^{iS_M(U)}$ in Eq.~\eqref{eq:gendivergence} diverges for a set of gauge fields with non-zero measure in real-time LGT path integrals.

\section{The truncated real-time heat kernel action}\label{sec:truncated}
In the classical limit, the real-time LGT path integral is dominated by the stationary phase solutions. Though the real-time heat kernel path integral introduced in Sec.~\ref{sec:heatkernelM} involves weights given by non-convergent sums over integers $n^A$, this classical limit is dominated by $n^A = 0$. This section explores the path integral defined by a truncation of these sums to $n^A = 0$, and demonstrates that ultimately it results in non-unitarity in the continuum limit in the analytically tractable case of $U(1)$ LGT in $(1+1)$D.

Explicitly, the truncated real-time heat-kernel action is defined by
\begin{equation}
\begin{split}
    e^{i S_{M,n=0}}(U) &= \prod_{x,k}\mathcal{N} \mathcal{J}(\{\phi_{x,0k}\},\{0\})\  e^{\frac{i}{g^2}\sum_{x,A}\sum_k (\phi_{x,0k}^A)^2} \\
    &\hspace{20pt} \times e^{-\frac{i}{g^2}\sum_{x,A}\sum_{i<j} (\phi_{x,ij}^A)^2},
    \end{split}
    \label{eq:SHKMtrunc}
\end{equation}
for $G=SU(N)$ with $\mathcal{J}(\{\phi_{x,0k}\},\{0\})$ replaced by 1 for $G=U(1)$. In the case of $G=U(1)$, the normalizing constant is explicitly $\mathcal{N} = 1 / \sqrt{2\pi i e^2}$.
The truncated real-time heat-kernel action corresponds to the $n=0$ term in the real-time heat-kernel action defined by Eqs.~\eqref{eq:SHKMdef}, \eqref{eq:KMU1}, \eqref{eq:KMSUN}.
The properties of the truncated real-time heat kernel action can be analyzed most simply for $G=U(1)$, and we specialize to this case below.
The kinetic-energy evolution operator associated with this action for $G=U(1)$ is given by
\begin{equation}
\begin{split}
    \tkmink{n=0}^{U(1)}(U,U') &= \prod_{\vec{x},k} \frac{1}{\sqrt{2 \pi i e^2}}\, e^{\frac{i}{e^2} (\phi_{\vec{x},k})^2},
\end{split}
\end{equation}
where $e^{i\phi_{\vec{x},k}} = U_{\vec{x},k}^\dagger U_{\vec{x},k}'$ as above.
The character expansion coefficients $c^{M,n=0,U(1)}_r(e^2)$ for the truncated heat-kernel kinetic operator $\tkmink{n=0,U(1)}$ can be explicitly computed as 
\begin{equation}
\begin{split}
    c^{M,n=0,U(1)}_r(e^2) &= \int_{-\pi}^{\pi} \frac{d\phi}{\sqrt{2 \pi i e^2}}  e^{i r \phi} e^{\frac{i}{2e^2} \phi^2} \\
    &=  e^{-i \left(\frac{e^2}{2}\right) r^2} \frac{1}{2}  \left[\text{erf}\left(\frac{-i\sqrt{i}(\pi - e^2 r)}{\sqrt{2} e}\right) \right. \\
    &\hspace{20pt} \left. + \text{erf}\left(\frac{-i\sqrt{i}(\pi + e^2 r)}{\sqrt{2} e}\right) \right]
    \end{split}
    \label{eq:mtruncU1}
\end{equation}
It follows that $|c^{M,n=0,U(1)}_r(e^2)| \neq 1$ for generic values of $e^2$ and therefore that the truncated heat-kernel kinetic term does not lead to a unitary time-evolution operator for $U(1)$ LGT.
From $\lim_{x\rightarrow \infty} \text{erf}(-i\sqrt{i} x) = 1$ we can derive
\begin{equation}
    \lim_{e^2 \rightarrow 0} \left|c^{M,n=0,U(1)}_r(e^2)\right| = 1,
\end{equation}
which establishes that unitarity is recovered for a fixed lattice volume in the $e^2 \rightarrow 0$ limit.
However, the scaling limit of interest to continuum QFT is defined by taking $e^2 \rightarrow 0$ with some dimensionful observable held fixed.
It is discussed in Ref.~\cite{Menotti:1981ry} and Sec.~\ref{sec:exact} that, in $(1+1)$D where LGT is analytically solvable, correlation lengths diverge $\propto 1/e$ in the $e^2 \rightarrow 0$ limit.
The appropriate scaling limit in $(1+1)$D is
\begin{equation}
    \lim_{e^2 \rightarrow 0} \left|c^{M,n=0,U(1)}_r(e^2)\right|^{1/e^2},
\end{equation}
which does not exist.
It is further shown below that the scaling limits of simple observables such as $U(1)$ Wilson loops in $(1+1)$D do not exist using the truncated Minkowski heat-kernel action, even though they do exist for a Euclidean analog of Eq.~\eqref{eq:SHKMtrunc}.
This suggests that the truncated Minkowski heat-kernel action is in fact not a useful starting point for real-time LGT and underscores the importance of preserving unitarity in LGT.

Wilson loop results in $(1+1)D$ using the truncated heat-kernel action analogous to those obtained in Sec.~\ref{sec:exact} can be easily obtained for the $U(1)$ truncated heat-kernel action by inserting the character expansion coefficients in Eq.~\eqref{eq:mtruncU1} into Eq.~\eqref{eq:WMchar},
\begin{equation}
\begin{split}
    &\left< \wloop_{\mathcal{A}} \right>_{M,n=0,U(1)} = e^{ - i \left(\frac{ e^2}{2} \right) L t} \Bigg[ \\
    &\hspace{50pt} \frac{\text{erf}\left( \frac{-i\sqrt{i}(\pi + e^2)}{\sqrt{2}e} \right) }{2 \text{erf} \left( \frac{-i\sqrt{i}\pi}{\sqrt{2}e} \right) } + \frac{\text{erf}\left( \frac{-i\sqrt{i}(\pi - e^2)}{\sqrt{2}e} \right)}{2 \text{erf} \left( \frac{-i\sqrt{i}\pi}{\sqrt{2}e} \right) } \Bigg]^{Lt}.
    \end{split}
\end{equation}
As $e^2 \rightarrow 0$, the factor in brackets approaches unity with corrections suppressed by $O(e)$ and the Minkowski heat-kernel result is recovered.
However, if $Lt \sigma_{HK}^{U(1)} = Lg e^2 / 2$ is kept fixed as $e^2 \rightarrow 0$, these $O(e)$ effects are magnified and the $e^2 \rightarrow 0$ limit of $\left< \wloop_{\mathcal{A}} \right>_{M,n=0}^{U(1)}$ at fixed $Lt \sigma_{HK}^{U(1)}$ does not exist.
This is in marked contrast to the situation in Euclidean spacetime, where the corresponding truncated Euclidean heat-kernel action gives rise to a result
\begin{equation}
\begin{split}
    &\left< \wloop_{\mathcal{A}} \right>_{E,n=0,U(1)} = e^{ - \left(\frac{ e^2}{2} \right) L t} \Bigg[ \\
    &\hspace{65pt} \frac{\text{erf}\left( \frac{\pi + e^2}{\sqrt{2}e} \right) }{2 \text{erf} \left( \frac{\pi}{\sqrt{2}e} \right) } + \frac{\text{erf}\left( \frac{\pi - e^2}{\sqrt{2}e} \right)}{2 \text{erf} \left( \frac{\pi}{\sqrt{2}e} \right) } \Bigg]^{Lt}.
    \end{split}
\end{equation}
The factor in brackets approaches unity exponentially faster than in the Minkowski case, and the truncated Euclidean heat-kernel action leads to $O(e^{-1/e^2})$ lattice artifacts.
Further, the limit of $\left< \wloop_{\mathcal{A}} \right>_{M,n=0,U(1)}$ at fixed $L\tau \sigma_{HK}^{U(1)}$ exists and equals the Euclidean heat-kernel result $e^{-L \tau \sigma_{HK}^{U(1)}}$ in the same limit.
This example demonstrates that ``lattice artifacts'' associated with non-unitary Minkowski actions can be difficult to remove, even if unitarity is restored in the would-be continuum limit.
It is unlikely that the truncated heat-kernel action provides a useful starting point for studying the continuum limits of higher-dimensional lattice gauge theories.
\vspace{1cm}

\bibliography{realtime}

\end{document}